\documentclass[a4paper, 12pt, openright, twoside, onecolumn,
final,spanish, english]{memoir}
 \settrimmedsize{297mm}{210mm}{*}
 \settypeblocksize{634pt}{448.13pt}{*}
 \setulmargins{4cm}{*}{*}
 \setlrmargins{*}{*}{1.5}
 \setmarginnotes{17pt}{51pt}{\onelineskip}
 \setheadfoot{\onelineskip}{2\onelineskip}
 \setheaderspaces{*}{2\onelineskip}{*}
\checkandfixthelayout
\fixpdflayout

\usepackage[T1]{fontenc}
\usepackage[utf8]{inputenc}
\usepackage{amssymb,amsmath}

\usepackage{fourier}

\usepackage{microtype}
\UseMicrotypeSet[protrusion]{basicmath} 

\usepackage{hyperref}
\usepackage{babel}

\usepackage{graphicx}

\usepackage[style=numeric-comp]{biblatex}
\providecommand{\tightlist}{%
  \setlength{\itemsep}{0pt}\setlength{\parskip}{0pt}}
\ifx\paragraph\undefined\else
\let\oldparagraph\paragraph
\renewcommand{\paragraph}[1]{\oldparagraph{#1}\mbox{}}
\fi
\ifx\subparagraph\undefined\else
\let\oldsubparagraph\subparagraph
\renewcommand{\subparagraph}[1]{\oldsubparagraph{#1}\mbox{}}
\fi
\usepackage{amsthm,physics}
\usepackage{cleveref}
\usepackage{enumerate}
\usepackage{aliases,mythm}

{
   \end{minipage}
   \vspace*{\stretch{3}}
   \clearpage
}

\title{Stability and area law for rapidly mixing quantum dissipative systems}
\author{Angelo Lucia}
\date{}

\def\englishtitle{
  Stability and area law  for rapidly mixing quantum dissipative systems}
\def\spanishtitle{
  Estabilidad y ley de área para sistemas cuánticos disipativos con equilibración rápida}
\def\author{Angelo Lucia}
\def\advisors{Prof. David P\'erez Garc\'ia \\ Prof. Toby S. Cubitt}
\def\ucmfac{Facultad de Ciencias Matem\'aticas}
\def\ucmdept{Departamento de An\'alisis Matem\'atico}
\def\theyear{2016}

\newcommand{\thetitlepage}{{%
   \clearpage
   \thispagestyle{empty}
   \begin{center}
   {\MakeUppercase{\Large Universidad Complutense de Madrid} \\
   \MakeUppercase{\ucmfac}\\
   \ucmdept
   }
   \vskip 2em%
   \includegraphics[width=150px]{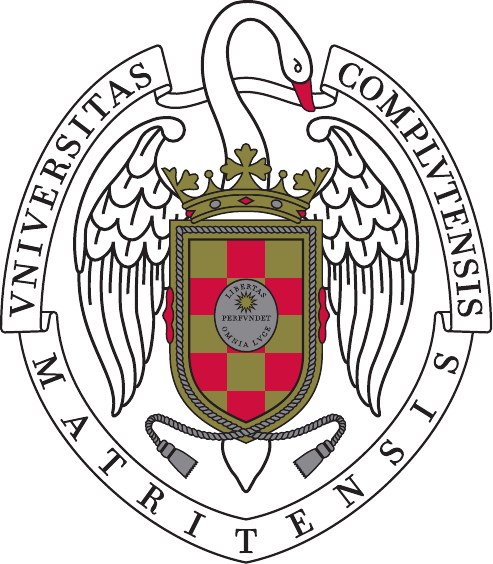}
   \vskip 1em%

   \begin{DoubleSpace*}
   {\LARGE
     \englishtitle
   }
   \end{DoubleSpace*}
   \vskip 0.7em%
   \begin{DoubleSpace*}
   {\LARGE
     \spanishtitle
   }
   \end{DoubleSpace*}

   \vfill

   {\SingleSpace
    \MakeUppercase{Memoria para optar al Grado de Doctor} \\
    \MakeUppercase{presentada por}\\
     \large{\author}
   }

   \vfill

   {\SingleSpace
     \MakeUppercase{Bajo la dirección de los doctores} \\
     \large{\advisors}
   \par}

   \vfill

   {\large \MakeUppercase{Madrid}, \theyear}
   \par
   \end{center}
   \clearpage}}

\begin{document}

\thetitlepage

\chapterstyle{veelo}
\pagestyle{ruled}

\chapter*{Acknowledgments}
Having gone through almost 5 years of Ph.D. studies in Madrid,
the list of people I would like to thank in a way or the other
would quickly exceed the length of this thesis. The ones who will
be missing should know that it is so for this reason and none other.
\vskip 1em \noindent
I would like to deeply thank my advisors David and Toby for having
introduced me into this exciting world of research. They have
supported me in countless different ways, and I am glad to have had
the opportunity to work with them.
\vskip 1em \noindent
I would also like to thank my collaborators and co-authors Spiros and
Fernando which have contributed invaluably to the papers in this
thesis. Together with them I would also thank Michael Wolf and
Johannes Bausch. Working with them has been a pleasure which
I hope will continue in the future.
\vskip 1em \noindent
Some people have made this journey smoother, and for this I am
grateful to Fernando Pastawski, Kristan Temme, Michael Beverland,
Jed Kaniewski, Cecilia Lancíen, Gemma De las Cuevas, William Matthews,
Ugo Siciliani, Gabriele Benedetti and dozens of others.
\vskip 1em \noindent
Me gustaría agradecer al Departamento de Análisis Matemático de la
Universidad Complutense por el trato siempre cariñoso que
he recibido, y a los miembros presentes y pasados del grupo de
Matemáticas y Información Cuántica por el agradable que ha sido
colaborar con ellos.
Un agradecimiento especial va a Carlos Fernández y a Carlos González
por haberme ayudado en la multitudes de pequeños problemas de la vida
en esta ciudad, un espectro que cubre desde el tráfico en bicicleta a
la diferencia entre un certificado y un volante.
\vskip 1em \noindent
Un ringraziamento va ovviamente ai miei genitori, che hanno accettato
pazientemente il vedermi lavorare su qualcosa di probabilmente
incomprensibile ai loro occhi, senza mai farmelo pesare.
\vskip 1em \noindent
Un altro ringraziamento ai compagni di Mantova, Enrico, Arianna,
Corrado, Paolo e Cristian, che continuano a volermi bene nonostante le
mie interminabili assenze e che sono capaci ogni volta di farmi
tornare la nostalgia.
\vskip 1em \noindent
Il ringraziamento finale va a Giorgia, che è stata una compagna di
viaggio insostituibile, il cui affetto e sostegno riescono sempre a
superare le distanze (geografiche) che a volte ci separano.

\cleartorecto

\cleartorecto

{
\setcounter{tocdepth}{2}
\tableofcontents
}

\selectlanguage{english} \chapter{Summary}

Since its origins, the field of information theory has had strong ties
to statistical mechanics: the terminology \emph{entropy of information}
was borrowed by Shannon from the thermodynamic entropy, as suggested by
Von Neumann
\cite{avery2003information, PetzEntropy, ScientificAmerican}.
Traditionally information theory studies the storage of information
(coding) and its transmission in noisy channels (communication
capacity). By interpreting the physical interactions as communications
channels, it has been possible to apply the same tools and ideas in
order to understand how the collective behavior of a mechanical system
composed of many (or infinite) parties emerges from the simple and
limited interactions between its individual components. This has lead to
understand the mechanism by which macroscopic properties emerge as
effective behavior from microscopic interactions.

The same relationship has been developed recently between the
corresponding \emph{quantum generalizations} of both theories: quantum
information (which is interested in the storage and manipulation of
information in quantum mechanical systems) and many-body quantum
physics. The ever-growing number of connections between the two fields
goes in both directions, with tools and ideas from quantum information
helping to solve long-standing problems in condensed matter physics, and
new many-body models being developed for the storage and the
transformation of quantum information. At the same time the spectacular
improvements we have seen in the implementation and experimental control
of small quantum systems is fueling the expectation that these
experiments could be scaled up in size. Larger experiments means being
closer to have practical applications, which has driven interest from
top universities and research centers, national funding bodies such as
EPSRC and NSF, to private companies with a strong focus on technological
research as IBM, Microsoft and Google.

It is therefore of great importance to sharpen and deepen our
understanding of these models, especially the ones that more likely
resemble the physical world we live in. This means taking into account
the fact that no experiment can be completely shielded from noise, nor
it can be executed at zero temperature: we have then to consider open
(as opposed to isolated) quantum systems, governed by a dissipative
evolution. To make the problem tractable, a reasonable simplification is
to assume a Markovian evolution, i.e.~an environment which holds no
memory (or loses it fast enough that its effects on the system are
negligible), and for which the future evolution only depends on the
present state of the system and not the way in which it was obtained.
These are evolutions controlled by a Lindblad equation, for which
evolution at a fixed time slice is represented by a completely positive
and trace preserving map, an object called \emph{quantum channel} in the
quantum information terminology.

Only recently has the quantum information community started to be
interested in such models, and has started to consider these type of
evolutions not only an issue to overcome but a resource which we need to
learn how to exploit. This fundamental shift of perspective opened up a
new area in the field, incorporating the idea of engineering artificial
evolutions in which the dissipation works in our favor instead than
against us
\cite{PRL.107.080503, 2010NatPh6.943B, Maurer08062012, verstraete09, Kraus08},
even by protecting the system from the effect of other, uncontrolled,
noisy evolutions. Therefore a range of new interesting problems were
faced: what is the computational power of these models, what conditions
can guarantee the resilience against external noise, how long we have to
wait for obtaining certain states, and so on.

This thesis is part of this effort to improve our knowledge of these
models. We have focused on studying properties of quantum dissipative
evolutions of spin systems on lattices in a particular regime that we
denote \textbf{rapid mixing}. We consider dissipative evolutions with a
unique fixed point, and which compress the whole space of input states
into increasingly small neighborhoods of the fixed point. The time scale
at which this compression takes place, or in other words the time we
have to wait for any input state to become almost indistinguishable from
the fixed point, is called the mixing time of the process. Rapid mixing
is a condition on the scaling of this mixing time with the system size:
if it is logarithmic, then we have rapid mixing.

The main contribution of this thesis is to show that rapid mixing has
profound implications for the corresponding system: it is stable against
external perturbations and its fixed point satisfies an area law for
mutual information. The precise definitions of these properties will be
given later.

It is somehow surprising that these properties can be derived just from
an estimate on the time-scale of the convergence. It is a bit less
surprising when we consider the other important ingredient in this
setting, which are Lieb-Robinson bounds. Well known in the case of
closed systems, but proven to hold also in the case of open systems,
Lieb-Robinson bounds formalize the idea that information in a many-body
model can only spread at a finite velocity, since its propagation has to
be mediated by local interaction terms. Therefore, limiting the
time-scale at which the evolution takes place implies limiting the scale
of the distance at which it can create correlations. This simple
observation lays at the heart of the technical part of this work.

\selectlanguage{spanish} \chapter{Resumen}

Desde su origen, la teoría de la información ha tenido fuertes
conexiones con la mecánica estadística: el mismo término \emph{entropía
de la información} fue elegido por Shannon a partir del término usado en
termodinámica, bajo sugerencia de Von Neumann
\cite{avery2003information, PetzEntropy,
ScientificAmerican}. Tradicionalmente la teoría de la información
estudia el almacenamiento (códigos) y la transmisión a través de canales
con ruido (capacidad de comunicación). Al interpretar las interacciones
físicas como canales de comunicación, ha sido posible aplicar las mismas
técnicas e ideas para entender cómo un sistema mecánico compuesto de
muchas (o infinitas) partes desarrolla un comportamiento colectivo a
partir de las interacciones simples y limitadas entre sus componentes
individuales. Esto ha permitido entender el mecanismo con el cual
propiedades macroscópicas aparecen como efectos de interacciones
microscópicas.

La misma relación se ha desarrollado recientemente entre las
correspondientes \emph{generalizaciones cuánticas} de ambas teorías: la
información cuántica (que estudia el almacenamiento y la manipulación de
la información en sistemas cuánticos) y la física de muchos cuerpos. Las
conexiones entre los dos campos aumentan cada día y van en las dos
direcciones: herramientas e ideas de la información cuántica ayudan a
solucionar problemas abiertos en teoría de la materia condensada, y
nuevos modelos de muchos cuerpos se desarrollan para aplicaciones de la
información cuántica. Al mismo tiempo la implementación y el control
experimental de pequeños sistemas cuánticos ha mejorado de forma
espectacular, aumentando la posibilidad de que estos experimentos se
puedan llevar a cabo a escala más grande. Experimentos más grandes
significa estar cada vez más cerca de aplicaciones prácticas, lo cual ha
orientado hacia el campo el interés de importantes universidades y
centros de investigación, así como agencias nacionales de financiación
como el EPSRC y la NSF, empresas privadas con fuerte inversión en la
investigación y el desarrollo como IBM, Microsoft y Google.

Por lo tanto es muy importante afinar y mejorar nuestro conocimiento de
estos modelos, y en particular los que describen de manera más fidedigna
el mundo físico donde vivimos. Esto significa tener en cuenta que ningún
experimento puede ser perfectamente aislado del ruido, ni puede ser
efectuado a temperatura cero: tenemos que considerar sistemas cuánticos
abiertos en vez de aislados, sujetos a una evolución disipativa. Para
que el problema sea abordable, una simplificación razonable es asumir
que la evolución sea Markoviana, es decir que el sistema ambiente no
tenga memoria de la evolución, o que la pierda lo suficientemente rápido
para que su efecto en el sistema sea despreciable, y tal que la
evolución futura sólo dependa del estado actual del sistema y no de su
historia pasada. Este tipo de sistema está descrito por una ecuación de
Lindblad, donde para cada instante temporal la evolución está dada por
una aplicación completamente positiva y que preserva la traza, un objeto
llamado \emph{canal cuántico} en la terminología de la información
cuántica.

Es reciente el interés de la comunidad de investigadores en información
cuántica por estos modelos, y se ha empezado a ver este tipo de
evoluciones no solamente como un problema que resolver, sino como un
recurso que explotar. Este fundamental cambio de perspectiva ha abierto
una nueva línea de investigación en el campo, incorporando ideas sobre
cómo construir evoluciones artificiales en las cuales la disipación
trabaja en nuestro favor en vez de en nuestra contra
\cite{PRL.107.080503, 2010NatPh6.943B, Maurer08062012, verstraete09,
Kraus08}, incluso protegiendo el sistema de otros ruidos incontrolados.
De esta manera se han planteado nuevos problemas interesantes: ¿cuál es
el poder computacional de estos modelos? ¿Qué condiciones pueden
garantizar la resistencia contra el ruido externo? ¿Cuánto tiempo hay
que esperar para obtener de esta manera cierto tipo de estados?

Esta tesis es parte del esfuerzo para mejorar nuestro conocimiento de
estos modelos. Nos hemos centrado en el estudio de las propiedades de
evoluciones disipativas de sistemas cuánticos de espines en un retículo,
bajo una hipótesis que llamamos \textbf{equilibración rápida}.
Consideramos evoluciones disipativas con un único punto fijo, y que
comprimen todo el espacios de estados iniciales en un entorno
decreciente del punto fijo. La escala temporal a la cuál se tiene esta
compresión, es decir el tiempo que tenemos que esperar para que
cualquier estado inicial se encuentre arbitrariamente cercano al punto
fijo, se llama tiempo de equilibración del proceso. La condición de
equilibración rápida pide que este tiempo sea logarítmico en el tamaño
del sistema.

La contribución principal de esta tesis es demostrar que la
equilibración rápida tiene importantes consecuencias en las propiedades
del sistema: este será estable bajo perturbaciones externas y sus puntos
fijos satisfarán una ley de área para la información mutua. Las
definiciones exactas serán dadas más adelante.

Es en cierta manera sorprendente que estas propiedades se puedan derivar
de una cota en el tiempo de convergencia. Es menos sorprendente una vez
que se considere el otro ingrediente fundamental en este contexto, las
llamadas cotas de Lieb-Robinson. Muy conocidas en el caso de sistemas
cerrados, pero igualmente válidas para sistemas abiertos, las cotas de
Lieb-Robinson formalizan la idea que la información en un sistema de
muchos cuerpos sólo puede moverse con una velocidad finita, dado que su
propagación está mediada por las interacciones locales. Por lo tanto,
controlar la escala temporal con la cuál la evolución converge permite a
su vez controlar la escala espacial a la cuál se pueden crear
correlaciones. Esta simple observación está en la base de la parte más
técnica de este trabajo.

\selectlanguage{english}

\chapter{Introduction}

This dissertation is organized as follows. In
\cref{background-and-current-state-of-the-topic} we will define the main
objects of interest, which are dynamical semigroups of quantum channels.
We will recall the properties of their generators, which are denoted
Lindblad generators. We will discuss why they are a sensible definition
for modeling noisy quantum evolutions, and present the mixing time, a
property which will be crucial in the main assumptions we make in order
to prove our results. We will present the connection between the mixing
time and other important properties of the semigroup, such as the
spectral gap, log-Sobolev inequality, and hypercontractivity. These
connections will show ways to prove rapid mixing. We will also introduce
the notion of mutual information and we will discuss the area law
property of states, together with its connections with efficiency of
simulation and tensor network states. We will also discuss why stability
is a crucial requirement for any mathematical model of a physical
system. In \cref{summary-of-the-results} we define the main assumptions
and present a summary of the results obtained, together with a brief
presentation of the technical tools that have been developed. Finally in
\cref{outlook-and-future-work} we discuss future lines of research which
are being developed at the time being.

The rest of the dissertation is composed of published papers which
collect the results of the work done during the PhD. These chapters
correspond to the following publications.

\begin{enumerate}
\setcounter{enumi}{4}
 \item \cite{Stability-paper} \fullcite{Stability-paper}
 \item \cite{Short-Stability-paper} \fullcite{Short-Stability-paper}
 \item \cite{Area-Law-paper} \fullcite{Area-Law-paper}
\end{enumerate}

The impact of these publications is reflected in the number of citations
they have received, notwithstanding their young age: in particular
\cite{Stability-paper} at the time of writing has already been cited 18
times, while \cite{Short-Stability-paper} received 3 and
\cite{Area-Law-paper} received 1. Moreover, the results obtained have
been presented as contributed talks in the most prestigious and relevant
workshops of the field: the Quantum Information Processing and
Communications 2013 (QIPC2013), the 17th Conference on Quantum
Information Processing (QIP2014), and the Theory of Quantum Computation,
Communication and Cryptography (TQC2015).

\section{Background and current state of the
topic}\label{background-and-current-state-of-the-topic}

\subsection{Notation}\label{notation}

Let us fix the notation that we will use through this dissertation,
although we will present in more detail some of the following objects
later on.

Given a tensor product of two finite dimensional Hilbert spaces
\(\hs_A\otimes \hs_B\), the unique linear map
\(\tr_A: \bounded(\hs_A \otimes \hs_B) \to \bounded(\hs_B)\) such that
\(\tr_A(a\otimes b)=b\tr(a)\) for all \(a \in \bounded(\hs_A)\) and all
\(b \in \bounded(\hs_B)\) will be denoted the \emph{partial trace} over
\(A\). A state over \(\bounded(\hs)\) is given by a linear positive
functional \(\rho : \bounded(\hs) \to \RR\), normalized in such a way
that \(\rho(\identity)=1\). We will denote the trace of an operator
\(X\) by \(\tr X\), and we will denote by \(\norm{\cdot}_p\) the
Schatten \(p\)-norm, i.e. \((\tr \abs{X}^p)^{1/p}\). Where there is no
risk of ambiguity, \(\norm{\cdot}\) will denote the usual operator norm
(i.e.~the Schatten \(\infty\)-norm).

We will consider a cubic lattice \(\Gamma = \ZZ^D\), equipped with the
graph distance metric, and a finite subset \(\Lambda \subset \Gamma\).
We will adhere to the physicists convention of calling any subset of
\(\Gamma\) a lattice, even if it is not a lattice in the graph theoretic
sense. The ball centered at \(x \in \Lambda\) of radius \(r\) will be
denoted by \(b_x(r)\). At each site \(x\) of the lattice we will
associate one elementary quantum system with a finite dimensional
Hilbert space \(\hs_x\). We will use Dirac's notation for vectors:
\(\ket{\phi}\) will denote a vector in \(\hs_x\), \(\bra{\phi}\) its
adjoint, and \(\{ \ket{n}\}_{n=0}^{\dim \hs_x -1 }\) the canonical basis
for \(\hs_x\). Scalar product in \(\hs_x\) will be denoted by
\(\braket{\phi}{\psi}\), and rank-one linear maps by
\(\ketbra{\phi}{\psi}\). For each finite subset
\(\Lambda \subseteq \Gamma\), we associate a Hilbert space given by
\[ \hs_\Lambda = \bigotimes_{x \in \Lambda} \hs_x ,\] and an algebra of
observables defined by
\[ \alg_\Lambda = \bigotimes_{x \in \Lambda} \mathcal B(\hs_x) .\] Since
\(\hs_x\) is finite dimensional, \(\bounded(\hs_x)\) is a matrix
algebra, and we will sometimes denote it by \(\matrixalg_d\), with
\(d = \dim \hs_x\).

If \(\Lambda_1\subset \Lambda_2\), there is a natural inclusion of
\(\alg_{\Lambda_1}\) in \(\alg_{\Lambda_2}\) by identifying it with
\(\alg_{\Lambda_1}\otimes \identity\). The support of an observable
\(O \in \alg_\Lambda\) is the minimal set \(\Lambda^\prime\) such that
\(O = O^\prime\otimes \identity\), for some
\(O^\prime \in \alg_{\Lambda^\prime}\), and will be denoted by
\(\supp O\).

A linear map \(\mcl T: \alg_\Lambda \to \alg_\Lambda\) will be called a
\emph{superoperator} to stress the distinction from operators in
\(\alg_\Lambda\). The support of a superoperator \(\mcl T\) is the
minimal set \(\Lambda^\prime \subseteq \Lambda\) such that
\(\mcl T = \mcl T^\prime \otimes \identity\), where
\(\mcl T^\prime \in \bounded(\alg_{\Lambda^\prime})\). A superoperator
is said to be Hermiticity preserving if it maps Hermitian operators to
Hermitian operators. It is said to be positive if it maps positive
operators (i.e.~operators of the form \(\du O O\)) to positive
operators. \(\mcl T\) is called \emph{completely positive} if
\(\mcl T \otimes \identity : \alg_\Lambda \otimes \matrixalg_n \to \alg_\Lambda \otimes \matrixalg_n\)
is positive for all \(n \ge 1\). Finally, we say that \(\mcl T\) is
trace preserving if \(\trace \mcl T(\rho) =\trace \rho\) for all
\(\rho \in \alg_\Lambda\). The p-Schatten norms on \(\alg_\Lambda\)
induce a corresponding family of norms on \(\bounded(\alg_\Lambda)\): we
denote by \(\norm{\cdot}_{p \to q}\) the operator norm of a
superoperator
\(\mathcal T : (\alg_\Lambda,\norm{\cdot}_p) \to (\alg_\Lambda, \norm{\cdot}_q)\),
i.e.~when we equip its domain with the \(p\)-Schatten norm and the
codomain with the \(q\)-Schatten norm. We will sometimes need the
following norm for superoperators, called diamond norm in the
literature:
\[\norm{\mcl T}_{\diamond} = \sup_n \norm{\mcl T \otimes \identity_n}_{1 \to 1}.\]

\subsection{Dynamical semigroups of quantum
channels}\label{dynamical-semigroups-of-quantum-channels}

\subsubsection{Unitary evolutions and quantum
channels}\label{unitary-evolutions-and-quantum-channels}

Quantum mechanics prescribes that a physical system is represented by
Hilbert space \(\hs\). The measurable properties of the systems are
encoded into a a state \(\rho\), which is a positive operator having
trace one. For simplicity, we will only consider the case in which
\(\hs\) is a \(d\)-dimensional Hilbert space, and therefore
\(\bounded(\hs)\) is a matrix algebra \(\matrixalg_{d}\). In the case of
an isolated system, the physical evolution of the system is described by
a unitary evolution of the state \(\rho\), meaning that the state of the
evolved system is given by \(U \rho U^\dg\), with \(U\) a unitary
operator over \(\hs\). It is immediate to see that this type of
evolution is inherently reversible, meaning that its inverse
\(\rho \to U^\dg \rho U\) is also physically possible. Therefore, in
order to consider dissipative quantum systems, by which we mean quantum
systems which have a non-reversible evolution caused by the interaction
of the system with an environment, we have to replace this unitary
evolution with something more general.

Let us try to be as general as possible, and let us pin down the minimal
assumptions a map \(T: \bounded(\hs) \to \bounded (\hs)\) representing a
physically realizable evolution should satisfy. Let \(\rho\) be the
initial state, and \(T(\rho)\) the evolved state. For sure, \(T\) should
send states to states, and therefore it should be linear \footnote{The
  operational interpretation of an \emph{ensemble} of states
  \(\{(p_i, \dyad{\phi_i})\}\) is that of a probability distribution
  \((p_i)\) over a set of possible states \(\{\ket{\phi_i}\}\).
  Therefore it is reasonable to expect that after the evolution the
  ensemble transformed to \(\{(p_i, T(\dyad{\phi_i}))\}\), or in other
  words that
  \(T(\sum_i p_i \dyad{\phi_i}) = \sum_i p_i T(\dyad{\phi_i})\).},
positive and trace preserving. It is a surprising but important fact
that positivity is not sufficient for such a map to have a consistent
physical interpretation. In fact, imagine we extend our system with an
auxiliary one, with its own Hilbert space \(\mathcal K\) and its state
\(\tau\). Then the state of the joint system is
\(\rho \otimes \tau \in \bounded(\hs \otimes \mathcal K)\). Let us also
assume that the evolution of \(\tau\) is trivial. Is there a map
\(\tilde T\) that extends \(T\) to \(\bounded(\hs \otimes \mathcal K)\),
such that \(\tilde T(\rho \otimes \tau) = T(\rho) \otimes \tau\) for
every possible pair of states \(\rho\) and \(\tau\)? Such a map exists
and it is given by the tensor product \(T\) with the identity map, which
is denoted by \(T \otimes \identity\).

If this were to be a physical evolution, it should again be positive.
But surprisingly, \(T \otimes \identity\) does not need to be positive,
if we only require positivity of \(T\). We have to require the stronger
property of \emph{complete positivity}: we recall that we defined a map
\(T:\matrixalg_d\to\matrixalg_{d^\prime}\) to be completely positive if
\(T\otimes \identity_n\) is a positive map for all \(n\in \NN\), where
\(\identity_n\) is the identity of \(\matrixalg_n\). A completely
positive and trace preserving map is called a \emph{quantum channel},
and they will be our main object of study.

Let us try to justify why quantum channels indeed represent physically
motivated evolutions. Since we want to talk about non-isolated systems,
let us assume we have an environment which is allowed to interact with
our system. Denote the state of the environment by \(\ket\phi\) (which
we can assume, without loss of generality, being pure). The pair
system-environment is a closed system, so it evolves under a unitary
\(U\). At the end of the evolution we discard the environment and look
only at the reduced density matrix of our original system, which we
denote by \(\rho^\prime\). The sequence of operations we have described
takes this form:

\begin{equation}
\label{eq:sequence}
\rho \to \rho \otimes \proj \phi \to U (\rho \otimes \proj{\phi}) U^\dg  \to \tr_E[ U (\rho\otimes \proj{\phi}) U^\dg ] = \rho^\prime
\end{equation}

\noindent
where \(\tr_E\) is the partial trace over the environment. Note that
every step of the process is a completely positive, trace preserving
map: therefore the resulting evolution \(\rho \to \rho^\prime\) is a
quantum channel.

Therefore, if we regard our system as being coupled to an environment,
and that they jointly evolve as a closed system, we are led to the
conclusion that the effective evolution of the system is described by a
quantum channel. Interestingly, this is actually the only thing that can
happen: every quantum channel can be interpreted in this way, as the
restricted action of a unitary evolution acting on a larger system. This
is the content of Stinespring's dilation theorem.

\begin{stinespring}
Let $T: \matrixalg_{d} \to \matrixalg_{d}$ be a completely positive and trace preserving map (a quantum channel). Then there exists a unitary $ U \in \matrixalg_{d^2}$ and a normalized vector $\phi \in \CC^{d}$ such that
    \begin{equation}
        T (\rho) = \tr_E[ U \qty(\rho \otimes \proj{\phi}) U^\dg ]
    \end{equation}
\end{stinespring}

Therefore, evolutions represented by quantum channels are physically
motivated as the restriction of a unitary evolution acting on a larger
system: they are, in a sense, ``effective'' dynamics induced by unitary
evolutions on sub-systems.

\subsubsection{Weak coupling limit}\label{weak-coupling-limit}

Until now we have only considered a \emph{single} application of a
quantum channel, regarding it as a single time-step in a dynamical
evolution, and we have seen that, due to Stinespring's theorem, that is
equivalent to coupling the system with an environment and considering a
unitary evolution of the pair. The clear advantage is, of course, that
it is often easier to reason if we can ignore the internal evolution of
the environment and only consider its effect on the system we are
actually interested in.

But a dynamical system is more than just the application of a single
time-step evolution: it is a composable sequence of such single steps,
or a continuous-time description in which every fixed time slice
\(t\ge0\) gives rise to a physically realizable evolution \(T_t\).
Mathematically we have then a problem: by tracing out the environment,
we have destroyed any possible correlation that the unitary \(U\) might
have created between the system and the environment, both in the form of
quantum entanglement or in the form of classical correlations. We have
only kept a shadow of them in the mixed state obtained, but the loss is
not reversible. The process described in \cref{eq:sequence} is
inherently not composable: the environment has changed because of its
coupled evolution with our system.

While technically true, we might question whether this mathematical
description is relevant for describing physical systems. It turns out
that, in some cases, the effect of the system onto the environment is so
negligible, it can be safely approximated to be irrelevant, and we can
assume the environment to be not evolving. Imagine for example that the
environment is a thermal bath at some temperature: for sure the
interaction with the system will change the equilibrium and the
temperature of the bath, but if the bath is much larger than the system
it is a good approximation to consider a constant temperature,
irrespective of what happens in the system.

Mathematically, this would mean that if
\(U \qty(\rho \otimes \proj{\phi}) U^\dg\) is not too distant from
\(\rho^\prime \otimes \proj{\phi}\), we can approximate the former with
the latter: in the physics literature this is called the \emph{weak
coupling limit} or \emph{Born
approximation}\cite{davies1974, pule1974, accardi1990, breuer2002, Rivas2012}.
At each ``infinitesimal'' time step, the environment is ideally thrown
away, and is replaced by a fresh, identical one, which therefore does
not hold any information about the previous evolution of the system. For
this reason, this type of evolution is also called \emph{Markovian}.

This naturally leads to consider the following dynamical system: the
evolution of the system is described by a
\emph{semigroup}\footnote{Strictly speaking, a representation of the semigroup $(\RR_{+}, +)$}
of quantum channels \(\{T_t: \matrixalg_d \to \matrixalg_d\}_{t\ge 0}\),
such that \(T_0\) is the identity map. The semigroup property
\(T_t T_s = T_{s+t}\) means that it is a homogeneous evolution, and that
it is Markovian. As in the classical theory of dynamical semigroups, if
\(T_t\) is strongly continuous in \(t\) (i.e. \(T_t\) is a
\(C_0\)-semigroup), then it has an infinitesimal generator \(\lind\),
satisfying the following relationships:

\begin{equation}
\label{eq:lindblad-generator}
\lind(x)=  \dv t \eval{T_t(x)}_{t=0} = \lim_{t \downarrow 0} \frac{1}{t}(T_t - \identity)(x).
\end{equation}

Note that for finite dimensional systems, strong continuity implies
uniform continuity, and therefore we can write
\[ \dv t T_t = \lind T_t \qcomma T_t = \exp(t \lind). \]

A generalization of this approximation is to consider an environment
evolving with time, but independently of the system itself (maybe
because of its internal dynamics, or maybe because we have approximated
the effect of the system on the environment in this way). This leads to
consider dynamical cocycles instead of semigroups, i.e.~families
\(\{T_{t,s}\}_{0\le s \le t}\) of quantum channels that satisfy the
property \(T_{t,s} T_{s,k} = T_{t,k}\) for all \(k \le s \le t\). We can
define (time-dependent) generators of cocycles in a similar way as we
did for semigroups: we will not be interested in this non-homogeneous
dynamics here, but we mention it for the sake of completeness.

\subsubsection{Lindbladian generators}\label{lindbladian-generators}

We have seen that the evolution of state \(\rho\) under a semigroup of
quantum channels \(T_t\) is given by the solution of the differential
equation \(\dot{\rho}(t) = \lind \rho(t)\), where
\(\rho(t) = T_t(\rho)\). The (super-)operator \(\lind\) is sometimes
called \emph{Liouvillian}, since this equation is a generalization of
the Liouville-von Neumann equation. It cannot be an arbitrary operator,
since we have imposed some restrictions on the semigroup it generates
(it is a semigroup of quantum channels). Lindblad \cite{lindblad},
Kossakowski, Gorini, and Sudarshan \cite{kossakowski} proved that this
imposes a particular form on the generator \(\lind\), which is called
the Lindlbad-Kossakowski form, and \(\lind\) is usually called
Lindbladian.

\begin{thm}
Let $\lind: \matrixalg_d \to \matrixalg_d$.
The following are equivalent
\begin{enumerate}
\item $\lind$ generates a dynamical semigroup of quantum channels;
\item there exists a completely positive map $\phi:\matrixalg_d \to \matrixalg_d$ and a matrix $\kappa \in \matrixalg_d$ such that
    \begin{equation}
        \lind (\rho) = \phi(\rho) - \kappa \rho - \rho \kappa^\dg; \quad \phi^*(\identity) = \kappa + \kappa^\dg.
    \end{equation}
\item there exists an Hermitian matrix $H$ and a set of matrices $\{ L_j \in \matrixalg_d \}_{j=0,\dots,d^2-1}$ such that
    \begin{equation}\label{eq:lindblad}
        \lind(\rho) = - i \comm{H}{\rho} + \sum_{j=0}^{d^2-1} L_j \rho L_j^\dg - \frac{1}{2} \acomm{L_j^\dg L_j}{\rho};
    \end{equation}
where $\acomm{a}{b} = ab + ba$ is the anticommutator.
\end{enumerate}
\end{thm}

\(H\) is (for obvious reasons) called the Hamiltonian, while the
matrices \(L_j\) are called \emph{Lindblad} or \emph{jump operators}.

By Russo-Dye theorem \cite{Russo1966}, if
\(T:\matrixalg_d \to \matrixalg_d\) is positive and trace preserving,
then it is a contraction under the trace norm: indeed we have that by
duality with respect of the Hilbert-Schmidt product
\[ \norm{T}_{1\to 1} = \norm{T^*}_{\infty \to \infty} = \norm{T^*(\identity)}_\infty = \norm{\identity}_\infty = 1, \]
since the dual of a trace preserving map is a unital map. Therefore, its
eigenvalues lie in the unit disk of the complex plane. Functional
calculus shows that this implies that the spectrum of \(\lind\) is in
the complex semiplane \(\{ z \in \CC \,|\, \Re z \le 0\}\).

Eigenvalues of \(\lind\) lying on the purely imaginary axis correspond
to eigenvalues of \(T_t\) lying on the boundary of the unit complex
disk, and in that context they are also called \emph{peripheral
spectrum}. It can be shown that their associated Jordan blocks are
one-dimensional, and they correspond to periodic states, with stationary
states corresponding to eigenvalue 1\cite{Wolf11}.

For every other eigenvalue \(\lambda\) of \(\lind\), which has a
strictly negative real part, the action of \(T_t\) on the corresponding
generalized eigenspace is the one of a contraction exponential in time:
the subspace is suppressed by a factor of \(\exp(-t \Re \lambda)\).
Therefore, it is the eigenvalue with the largest non-zero real part that
determines the slowest rate of convergence of \(T_t\) to a map
\(T_\infty\) that projects on the space of periodic points (and that
acts unitarily on it). This is the reason for the following definition:

\begin{defn}[Spectral gap]
We denote the \emph{spectral} gap of a Lindbladian $\lind$ the quantity
    \begin{equation} \label{eq:spectral-gap}
    \gap \lind = \min_{\lambda \in \sigma(\lind)\setminus\{0\}} \abs{\Re \lambda}.
    \end{equation}
\end{defn}

Let us assume for a moment that there are no periodic points, or in
other words the peripheral spectrum is trivial: it is composed only of
the eigenvalue 1. In this case \(T_\infty\) is actually a projection.
Then from a simple Jordan decomposition, we can see that the spectral
gap controls the scaling in time of the convergence to the space of
fixed points, and we have that there exists a constant \(c > 0\) such
that

\begin{equation}
\label{eq:gap-convergence}
\norm{T_t(\rho) - T_\infty(\rho)}_1 \le c \exp(-t \gap \lind),
\end{equation}

\noindent
for every initial state \(\rho\).

We will go back to \cref{eq:gap-convergence} later on, when we talk
about families of dynamical systems defined on an increasing sequence of
lattice structures.

\subsubsection{Local generators}\label{local-generators}

Until now, we have only considered finite systems, which can be
considered as one single big physical body subject to some dynamics.
Most applications require instead a description of a \emph{many-body}
model: a system composed of many individual parts, which interact with
each other in a defined and somehow regular way. If we only consider one
single instance of a many-body model, then mathematically speaking there
is no difference in considering the whole system as a single big body,
with some internal degree of freedom evolving according to the mentioned
interactions.

This point of view changes dramatically if we consider instead
increasing sequences of many-body models defined on a graph or lattice
structure. Let us recall the notation we have given for many-body
systems on a lattice. Consider an infinite graph \(\Gamma\) (for
example, \(\ZZ^D\) for some integer \(D\)) equipped with the graph
metric. Associate to every vertex \(x\) in the graph a finite Hilbert
space \(\hs_x\), and let us assume for simplicity that they are all
isomorphic (i.e., they have the same dimension \(d\)). Then for every
finite \(\Lambda \subset \Gamma\) we denote by
\(\hs_\Lambda = \otimes_{x\in \Lambda}\hs_x\), and
\(\alg_\Lambda = \bounded (\hs_\Lambda)\).

In this case, there is a well defined notion of locality: for every pair
of finite subgraphs \(\Lambda_1 \subset \Lambda_2 \subset \Gamma\),
there is a natural embedding of \(\alg_{\Lambda_1}\) into
\(\alg_{\Lambda_2}\), by identifying \(X \in \alg_{\Lambda_1}\) with
\(X \otimes \identity_{\Lambda_2\setminus\Lambda_1} \in \alg_{\Lambda_2}\).
This allowed us to define the notion of \emph{support}: given an
operator \(X \in \alg_{\Lambda}\) we define the support of \(X\),
denoted by \(\supp X\), as the minimal
\(\Lambda^\prime \subset \Lambda\) such that there exists a
\(X^\prime \in \alg_{\Lambda^\prime}\) that satisfy
\(X = X^\prime \otimes \identity\). In a sense, the support of \(X\) is
independent of \(\Lambda\), since
\(\supp X = \supp X \otimes \identity\): therefore considering \(X\) to
be acting on a larger system does not increase its support.

This is the first appearance of a simple but powerful idea that is
behind most of this work: there exist some properties of the objects we
are studying that do not depend on the size of the system, as long as
this is large enough to contain them. If we take an increasing and
absorbing sequence of finite \(\Lambda_n\nearrow \Gamma\), then we can
discuss about properties that are uniform in \(n\).

Physical interactions usually become weaker when the distance between
interacting bodies becomes larger. Therefore, if we can decompose the
generator of the evolution \(\lind\) as a sum of local terms
\(\sum_{Z \subset \Lambda} \lind_Z\), each of which is still of the
Lindblad-Kossakowski form but it is only acting on the subsystem \(Z\) ,
it is reasonable to assume that their norm becomes smaller as the
diameter of their support \(Z\) becomes larger. In this case, we will
say that \(\lind\) is a \textbf{local Lindbladian}.

If we do not specify at which rate \(\norm{\lind_Z}_\diamond\) decays
with \(\diam Z\), this definition can be trivially satisfied by any
Lindbladian. We will postpone for the moment to clarify this point, and
we will do it in \cref{lieb-robinson-assumptions}.

\subsubsection{Mixing time and spectral
gap}\label{mixing-time-and-spectral-gap}

\Cref{eq:gap-convergence} captures the long-time properties of the
dynamical system described by \(T_t\): if \(t\) is larger than
\(\log(\epsilon/c)/\gap \lind\) for some positive \(\epsilon\), then the
set of possible input states will have been compressed inside a
\(\epsilon\)-neighborhood of the space of fixed points. The minimum time
required for this to happen (which might be smaller than the time given
by \cref{eq:gap-convergence}) is called the \textbf{mixing time} of the
dynamical system. We give the formal definition only for systems without
periodic points.

\begin{defn}[mixing time]
We denote the \emph{mixing time} of a dynamical system with no periodic points $T_t:\matrixalg_d\to\matrixalg_d$ the function
$$ \tau(\epsilon) = \min\{ t > 0 : \sup_\rho \norm{T_t(\rho) - T_\infty(\rho)}_1 \le \epsilon \},$$
where the supremum is taken over all states $\rho$.
\end{defn}

Therefore, we can restate what we know about the spectral gap as
follows:

\begin{equation} \label{eq:spectral-gap-bound}
    \tau(\epsilon) \le \frac{\log c -\log \epsilon}{\gap \lind}
\end{equation}

For any single finite-dimensional system, this analysis is usually
satisfactory: more care should be taken if we consider families of
dynamical systems defined on an increasing sequence of lattices
\(\Lambda_n\). In this case we want to control the scaling in \(n\) of
\(\tau(\epsilon)\). First of all, the quantity
\(\lambda_n = \gap \lind_n\) can become smaller, as \(n\) increases, and
therefore the bound on the mixing time will diverge. If instead the
quantity \(\lambda = \inf \lambda_n\) is bounded away from zero, we
informally say that the system is gapped (meaning that it is gapped in
the limit).

Nonetheless, there is another and deeper reason for which the bound
\cref{eq:spectral-gap-bound} will, in general, diverge with \(n\), even
with a strictly positive \(\lambda\): the constant \(c\) will (in
general) also depend on \(n\). In fact, if we obtain this bound via the
Jordan decomposition (which is not the optimal way to do it), it can
grow faster than exponential in \(n\). Some more careful analysis can be
done to improve this dependence: in \cite{Wolf11} it was shown that if
\(\lind\) satisfies a condition called detailed balance, which will be
presented in more detail in \cref{spectral-gap-and-detailed-balance},
with respect to a full-rank state \(\sigma\), then we can take \(c\) to
be equal to \(\norm{\sigma^{-1}}_\infty^{1/2}\), which is equal to
\(\sigma_{\min{}}^{-1/2}\), the minimal eigenvalue of \(\sigma\). This
gives the following result, for which we will also present a different
proof later:

\begin{thm}
  \label{thm:spectral-gap-bound-db}
  If $\lambda$ is the spectral gap of $\lind$, and it satisfies detailed balance with respect to a full-rank state $\sigma$, then
  \begin{equation} \label{eq:spectral-gap-bound-db}
    \tau(\epsilon) \le \frac{\log(\sigma_{\min{}}^{-1/2}) - \log \epsilon}{\lambda}
  \end{equation}
\end{thm}

Note that \(\sigma_{\min{}}^{-1/2}\) has to scale at least exponentially
in the system size, or worse - therefore, we obtain from
\cref{eq:spectral-gap-bound-db} a polynomial time mixing. If we assume
that we know the whole spectrum of \(\lind\) but nothing else, we cannot
improve too much on \cref{thm:spectral-gap-bound-db}: as shown in
\cite{Szehr15}, if we only know the spectrum of \(\lind\) and not some
other property of it, the dependence of \(c\) in the system size cannot
be improved to be slower than exponential.

For some applications, having a polynomial-time mixing is sufficient. In
this work, we require a stronger condition, which because of
\cite{Szehr15} cannot be guaranteed only by some condition on the
spectrum of \(\lind\): that \(\tau(\epsilon)\) scales
\emph{logarithmically} in \(n\) (and in some cases we can have a
relaxation to sub-linear scaling). It is the main contribution of this
thesis to show that under this stronger assumption, some very
interesting properties of the evolution and of its fixed point can be
derived.

We will present such results in \cref{summary-of-the-results}: before
that, let us present one important connection with logarithmic Sobolev
inequalities.

\subsection{Logarithmic Sobolev
inequalities}\label{logarithmic-sobolev-inequalities}

The tools of hypercontractivity and logarithmic Sobolev inequalities
(log-Sobolev for short) have been developed as part of Segal's program
of giving mathematical rigor to Quantum Field Theory \cite{Segal1960}.
Log-Sobolev inequalities have been first introduced by Feissner (at the
time a student of Leonard Gross, who had been in turn student of Segal)
in his PhD thesis \cite{feissner1972gaussian,Feissner1975} in order to
generalize the classical Sobolev inequality to Gaussian measures in
infinite dimensions. Then \cite{Gross1975} used them to study ergodicity
of Markov process in infinite dimensional spaces and were recognized
afterwards to be an effective tool in analyzing finite dimensional
systems too \cite{Diaconis1996}. The application to classical spin
systems has been introduced first by Holley, Stroock and Zegarlinski
\cite{Holley1987,zegarlinski1990A,zegarlinski1990B, Stroock1992D,Stroock1992C}
and thereafter became a standard tool in statistical mechanics. They are
intimately connected to hypercontractivity of semigroups, and have made
an appearance in a wide range of different areas of mathematics.

For a modern review of the classical (commutative) theory of logarithmic
Sobolev inequalities and its connections to Markov semigroups and
concentration of measure, we refer the reader to \cite{Guionnet2002}.
Its quantum generalization has been developed in a series of papers
\cite{qsd1,qsd2,qsd3,hypercontractivitylp}, and the connection between
rapid mixing and log-Sobolev inequalities in the quantum setting is due
to \cite{Quantum-Log-Sobolev}.

Hypercontractivity predates slightly the appearance of log-Sobolev
inequality: the first examples can be traced back to a work of Nelson
(also a student of Segal) \cite{Nelson1973,nelson1966quartic}, when it
was still not called in such a way. For a review of the subject we will
refer to \cite{davies1992hypercontractivity,hyreview}. It is finding its
way into the quantum information community as a tool on its own
\cite{Temme2014,King2014,Cubitt2015,Montanaro2012}.

We will present in the rest of the section a simplified version of the
quantum log-Sobolev theory, and its connection with hypercontractivity
and rapid mixing.

In this dissertation we consider semigroups of trace preserving maps
\(T_t\), therefore describing the evolution of states, but an equivalent
description could be done of the dual (under the Hilbert-Schmidt scalar
product) evolution \(T^*_t\) where observables are evolving, and the
limit state \(\rho_\infty\) is invariant in the sense that
\(\tr(\rho_\infty T^*_t(A))=\tr(\rho_\infty A)\) for all operators
\(A\). In this case the semigroup is unital instead of trace preserving.
This is the approach usually followed in the literature on logarithmic
Sobolev inequalities. We will stick to our notation, therefore denoting
the evolution of observables as \(T_t^*\), but the reader should be
aware of this fact.

\subsubsection{Spectral gap and detailed
balance}\label{spectral-gap-and-detailed-balance}

Before presenting the definition of logarithmic Sobolev inequalities, or
log-Sobolev inequalities for short, let us reformulate
\cref{eq:spectral-gap} in a different but equivalent way, when we assume
to have a full rank state \(\sigma>0\). For such state we can define a
weighted scalar product on \(\bounded(\hs)\) via
\[\innerproduct{A}{B}_{\sigma} = \tr[\sigma^{1/2}A\sigma^{1/2}B] = \innerproduct{\sigma^{1/4}A\sigma^{1/4}}{\sigma^{1/4}B\sigma^{1/4}}_{HS},\]
and the corresponding induced norm
\(\norm{\cdot}_\sigma = \innerproduct{\cdot}_\sigma^{1/2}\). It is easy
to see that
\(\sigma_{\min{}}^{1/2} \norm{A} \le \norm{A}_\sigma \le \norm{A}\),
where \(\sigma_{\min{}}\) is the minimal eigenvalue of \(\sigma\).
Moreover, we can define a generalization of the classical variance, as
\[\variance_\sigma(A) = \norm{A - \innerproduct{A}{\identity}_{\sigma}}^2_{\sigma} = \tr[\sigma^{1/2}A\sigma^{1/2}A] - \tr[\sigma A]^2.\]
Indeed, \(\variance_\sigma(A)\) is positive and invariant under
translations by multiples of the identity. In a similar way, given a
Lindbladian \(\lind\), we can define a non-commutative generalization of
the Dirichlet form:
\[\dirichelet(A,B) = \innerproduct{A}{-\lind^*(B)}_{\sigma} = -\tr[\sigma^{1/2}A\sigma^{1/2}\lind^*(B)];\]
where \(\lind^*\) is the dual of \(\lind\) under the Hilbert-Schmidt
scalar product, i.e.
\(\dirichelet(A,B) = -\tr[\lind(\sigma^{1/2}A\sigma^{1/2})B]\). We will
write \(\dirichelet(A) = \dirichelet(A,A)\). We say that \(\lind\)
satisfies \emph{detailed balance with respect to \(\sigma\)} if
\(\lind(\sigma^{1/2}A\sigma^{1/2})=\sigma^{1/2}\lind^*(A)\sigma^{1/2}\)
for all operators \(A\), and therefore
\[\tr[\sigma^{1/2}A\sigma^{1/2}\lind^*(B)] = \tr[\lind(\sigma^{1/2}A\sigma^{1/2})B] =\tr[\sigma^{1/2}\lind^*(A)\sigma^{1/2}B] .\]
If so, \(\dirichelet\) is a symmetric bilinear form, \(\lind^*\) is
self-adjoint under \(\innerproduct{\cdot}_{\sigma}\), and thus
\(\hat \lind(\cdot) = \sigma^{1/4} \lind^*(\sigma^{-1/4} \cdot \sigma^{-1/4})\sigma^{1/4}\)
is self-adjoint under the Hilbert-Schmidt scalar product. Since
\(\lind^*\) and \(\hat \lind\) are related by a similarity
transformation, \(\lind^*\) has real spectrum, and contractivity of the
generated semigroup implies that it is negative. Note that it also
implies that \(\sigma\) is a steady state for \(\lind\), since for all
\(A\) it holds
\[ \tr[\lind(\sigma) A ] = \tr[ \sigma^{1/2} \identity \sigma^{1/2} \lind^*(A)] = \tr[ \sigma^{1/2} \lind^*(\identity) \sigma^{1/2} A ] = 0, \]
and thus \(\lind(\sigma) = 0\).

If the peripheral spectrum of \(T_t\) is trivial, then kernel of
\(\lind\) is one dimensional, and by the Courant-Fischer-Weyl min-max
principle, the second smaller eigenvalue of \(\lind\), which we have
previously called the spectral gap, is given by:
\[\gap \lind = \min_{A: \innerproduct{A}{\identity}_{\sigma} = 0} \frac{-\innerproduct{\lind^*(A)}{A}_\sigma}{\innerproduct{A}_{\sigma}}
= \min_{A:\variance_\sigma(A)\neq 0} \frac{\dirichelet(A)}{\variance_\sigma(A)}.\]

We have therefore re-expressed \cref{eq:spectral-gap} as a variational
problem: \(\gap \lind\) is the maximal value that can take a constant
\(c\) such that the quadratic functional \(c \variance_\sigma(A)\) lower
bounds the quadratic functional \(\dirichelet(A)\).

\begin{equation}
  \label{eq:spectral-gap-variational}
  c \variance_\sigma(A) \le \dirichelet(A)
\end{equation}

Consider now \(A(t)\) the evolution on an observable \(A\) under
\(\lind^*\), i.e. \(A(t) = T_t^*(A)\). Since \(\sigma\) is invariant
under \(T_t\), it holds that
\(\innerproduct{A(t)}{\identity}_\sigma = \innerproduct{A}{\identity}_\sigma\),
and therefore
\(\lim_{t\to\infty} A(t) = \innerproduct{A}{\identity}_\sigma \identity\).
Therefore \(\variance_\sigma(A(t))\) is equal to
\(\norm{A(t) - A(\infty)}_\sigma^{2}\). We can consider the function
\(t \to \variance_\sigma(A(t))\): its derivative is given by
\(-2\dirichelet[A(t)]\). Therefore \cref{eq:spectral-gap-variational} is
really bounding the derivative of \(\variance_\sigma(\cdot)\) with
respect to the functional itself. This leads to the following bound:
\[ \variance_\sigma(A(t)) \le \variance_\sigma(A) e^{-2ct}.\] Therefore,
the spectral gap controls convergence when measured with
\(\variance_\sigma(\cdot)\). In turn, this implies that:
\[ \norm{A(t) - A(\infty)} \le \sigma_{\min}^{-1/2} \norm{A(t) -A(\infty)}_\sigma \le \sigma_{\min}^{-1/2} \norm{A-A(\infty)} e^{-ct}.\]
By duality this implies the following bound of the form
\cref{eq:gap-convergence}:
\[\sup_{\rho} \norm{T_t(\rho) - \sigma}_1 \le  2 \sigma_{\min{}}^{-1/2} e^{-t\gap \lind}.\]
or equivalently

\begin{thm}
  \label{thm:detailed-balance-gap-bound}
  If $\lambda$ is the spectral gap of $\lind$, then
  \begin{equation} \label{eq:detailed-balance-gap-bound}
    \tau(\epsilon) \le \frac{\log(2\sigma_{\min{}}^{-1/2}) - \log \epsilon}{\lambda}
  \end{equation}
\end{thm}

Notice that \(\sigma^{-1}_{\min{}}\) scales \textbf{at least}
exponentially in the system size (since it has to be at least smaller
that \(1/\dim \hs_\Lambda\)) - but in principle it could be even worse.

We could also have obtained the same bound, but without the
multiplicative constant 2, by using the
fact\cite{Temme2010, Quantum-Log-Sobolev}
\[\norm{\rho - \sigma}_1 \le \variance^{1/2}_\sigma(\sigma^{-1/2}\rho\sigma^{-1/2}),\]
and the fact that
\(\sup_{\rho}\variance_\sigma(\sigma^{-1/2}\rho\sigma^{-1/2})\) is equal
to \(\norm{\sigma^{-1}}_\infty = 1/\sigma_{\min{}}\) (where the sup is
taken over states).

We have seen therefore that the detailed balance condition allows us to
clearly express the relationship between the spectral gap and the mixing
time, obtaining a pretty good prefactor for our bound (definitely better
than what we could have obtained via the Jordan decomposition). The
downside is that we need to assume that the fixed point is unique and
full rank (a condition called \emph{primitivity} of \(\lind\)), and that
\(\sigma_{\min{}}\) can be controlled.

We will show next how in this setting it is natural to define other
conditions on \(\lind\) that allow a better control on the mixing time
than the one obtained via the spectral gap, which in turn will prove to
be sufficient to prove our rapid mixing assumption.

\subsubsection{Entropy and log-Sobolev
inequalities}\label{entropy-and-log-sobolev-inequalities}

The idea of logarithmic Sobolev inequalities and other entropic
inequalities is to generalize what has been done in the previous section
with \(\variance_\sigma(\cdot)\): find a positive functional
\(D(\cdot)\) that bounds the convergence of the semigroup \(T_t\), then
bound the derivative \(\dv{t} D(T_t(\rho)-\sigma)\) in terms of the
function itself, via a comparison with another functional defined in
terms of \(\lind\).

Let us consider the following functional, denoted \textbf{relative
entropy} \[\ent{X}{Y} = \frac{1}{\tr X}\tr[X (\log X - \log Y)].\]
\(\ent{\rho}{\sigma}\) is positive if \(\rho\) and \(\sigma\) are
normalized states, and it is finite if the support of \(\rho\) is
contained in the support of \(\sigma\). It is also monotonically
decreasing under the action of quantum channels \cite{OhyaPetz200405}.
Pinsker inequality \cite{nielsen-chuang} implies that
\(\norm{\rho -\sigma}^2_1 \le 2 \ent{\rho}{\sigma}\). Let us now
differentiate and obtain
\[\dv{t} \ent{\rho(t)}{\sigma} = \tr[ \lind(\rho(t))(\log \rho(t)-\log \sigma)].\]
We can therefore denote by
\(\mcl K(\rho) = - \frac{1}{\tr \rho} \tr[ \lind(\rho)(\log \rho -\log \sigma)]\).
Compare this with the definition of \(\dirichelet\). We can then define
the following log-Sobolev type of inequality:

\begin{equation} \label{eq:log-sobolev-inequality}
   c [ \ent{\rho}{\sigma} - \log \tr \rho] \le \mcl K(\rho)
\end{equation}

\noindent
where the optimal \(c>0\) will be called log-Sobolev constant of
\(\lind\), and denoted by \(\alpha\). As in the case of the spectral gap
inequality, we can then conclude that
\[\sup_{\rho}\norm{\rho(t) - \sigma}_1 \le \sup_{\rho} \sqrt{2 \ent{\rho}{\sigma} } e^{-\alpha t}.\]
We then observe that \(\ent{\rho}{\sigma}\) is upper bounded by
\(\norm{\log \sigma^{-1}}_\infty = -\log(\sigma_{\min{}})\). Therefore,
the convergence bound obtained via the log-Sobolev inequality is
exponentially better than the one obtained through the spectral gap
inequality (see \cref{eq:detailed-balance-gap-bound}):

\begin{thm}
  Let $\alpha$ be the log-Sobolev constant of $\lind$. Then
  \begin{equation}
  \tau(\epsilon) \le \frac{ \log(\log(\sigma^{-1/2}_{\min{}})) - 2\log \epsilon} {2\alpha}.
  \end{equation}
\end{thm}

If \(\sigma^{-1}_{\min{}}\) is only exponential in the system size, then
a system-size uniform log-Sobolev constant implies rapid mixing.
Therefore, log-Sobolev bounds are a way of proving such an assumption
for reversible, detailed balance generators.

This is the approach taken in \cite{MllerHermes2016,1505.04678}. In
\cite{hypercontractivitylp} and \cite{Quantum-Log-Sobolev} a bound
equivalent to \cref{eq:log-sobolev-inequality} was denoted 1-log-Sobolev
inequality, and is obtained by composing
\cref{eq:log-sobolev-inequality} with the map
\(\rho \to \sigma^{-1/2} \rho \sigma^{-1/2}\). If we denote by
\(A = \sigma^{-1/2} \rho \sigma^{-1/2}\) and by \(A(t)\) the evolution
of \(A\) under \(\lind^*\), i.e. \(A(t) = T_t^*(A)\), then detailed
balance implies that
\[A(t) = T_t^*( \sigma^{-1/2} \rho \sigma^{-1/2} ) = \sigma^{-1/2} T_t(\rho) \sigma^{-1/2} = \sigma^{-1/2} \rho(t) \sigma^{-1/2}.\]

Thus \cref{eq:log-sobolev-inequality} can be restated as
follows\footnote{we have removed a unimportant factor of \(1/2\) from
  the original definitions}:
\[c \operatorname{Ent}_1(A) \le \dirichelet_1(A),\] where
\(\operatorname{Ent}_1(A) = \ent{\rho}{\sigma} - \log \tr \rho\) and
\(\dirichelet_1(A)=\mcl K(\rho)\). This version of the bound is clearly
equivalent to \cref{eq:log-sobolev-inequality} if \(\lind\) satisfies
detailed balance. The authors of \cite{Quantum-Log-Sobolev} denote the
optimal constant by \(\alpha_1\).

Unfortunately this is not what is denoted as log-Sobolev inequality in
the classical literature (i.e.~when we define all the above for a
generator of a Markov chain over a probability space, which is the
commutative equivalent of Lindbladian generators over quantum states).
Instead, the classical log-Sobolev inequality is more similar to the
following generalization:
\[c \operatorname{Ent}_2(A) \le \dirichelet(A) ;\] where \(\dirichelet\)
is the Dirichlet form defined earlier,
\(\operatorname{Ent}_2(A) = \operatorname{Ent}_1(I_{1,2}(A))\),
\[I_{1,2}(A) = \sigma^{-1/2} \qty( \sigma^{1/4} A \sigma^{1/4} )^2 \sigma^{-1/2} = \sigma^{-1/4} A \sigma^{1/2} A \sigma^{-1/4}\]
and therefore
\(\operatorname{Ent}_2(A) = \ent{\qty(\sigma^{-1/4} \rho \sigma^{-1/4})^2}{\sigma}\).

This bound is denoted 2-log-Sobolev inequality in
\cite{hypercontractivitylp,Quantum-Log-Sobolev,1505.04678} and the
optimal constant \(\alpha_2\). Unfortunately we do not know if it is
equivalent to \cref{eq:log-sobolev-inequality}: under the additional
hypothesis that \(\dirichelet_1(I_{1,2}(A)) \ge \dirichelet(A)\)
(denoted \emph{\(L_p\)-regularity} in \cite{hypercontractivitylp}), then
one can at least show that \(\alpha_2 \le \alpha_1\), therefore
recovering the classical result. Whether there exist Lindbladians which
satisfy detailed balance but not \(L_p\)-regularity is still an open
problem.

\subsubsection{Hypercontractivity}\label{hypercontractivity}

As we have already seen, as a consequence of the Russo-Dye theorem we
have that a positive and trace preserving map \(T\) is
\emph{contractive} with respect to the 1-norm, since
\(\norm{T}_{1\to 1}=1\) - equivalently, a positive and unital map
\(T^*\) satisfies \(\norm{T^*}_{\infty \to \infty}=1\), i.e.~is
contractive with the \(\infty\)-norm. This applies in particular to
quantum channels. Let us now introduce a non-commutative generalization
of the \(L_p\)-norms\cite{haagerup,qsd1,qsd2,qsd3}: given a full rank
state \(\sigma\), for each \(p\in [1,\infty)\) denote by
\[ \norm{X}_{p,\sigma}^p = \tr\abs{ \sigma^{1/2p} X \sigma^{1/2p} }^p = \norm{ \sigma^{1/2p} X \sigma^{1/2p} }_p^p.\]
It can be checked that \(\norm{\cdot}_{p,\sigma}\) is indeed a norm, and
the usual properties that can be expected from \(L_p\) norms can be
recovered, such as H"older inequality, duality, and interpolation
theorems. In particular, they are increasing in \(p\), thus for all
\(1 \le p \le q \le \infty\) it holds
\(\norm{\cdot}_{1,\sigma} \le \norm{\cdot}_{p,\sigma} \le \norm{\cdot}_{q,\sigma} \le \norm{\cdot}_{\infty}\).
Moreover, \(\lim_{p\to \infty}\norm{X}_{p,\sigma} = \norm{X}_\infty\)
(the usual Schatten \(\infty\)-norm). Note that the norm defined in the
previous section, which we denoted by \(\norm{\cdot}_\sigma\),
corresponds to the \(p=2\) case. The space \(\bounded(\hs)\) equipped
with the \(\norm{\cdot}_{p,\sigma}\) norm will be denoted by
\(L_p(\sigma)\), and the operator norm of
\(T: L_p(\sigma) \to L_q(\sigma)\) will be denoted
\(\norm{T}_{(p,\sigma) \to (q,\sigma)}\).

Let us consider then a quantum channel \(T\) having \(\sigma\) as a
fixed point, and consider its dual \(T^*\) w.r.t. the Hilbert-Schmidt
scalar product. Let us assume again that \(T\) satisfies detailed
balance w.r.t. \(\sigma\). Then we know that \(T^*\) is unital, and thus
\(\norm{T^*}_{\infty \to \infty} = 1\). Moreover, we have that for every
operator \(A\) \[
    \norm{T^*(A)}_{1,\sigma} =
    \norm{ \sigma^{1/2} T^*(A) \sigma^{1/2} }_1 =
    \norm{ T(\sigma^{1/2} A \sigma^{1/2} ) }_1 \le
    \norm{ \sigma^{1/2} A \sigma^{1/2} }_1 = \norm{A}_{1,\sigma},
\] where we have used detailed balance and the fact that
\(\norm{T}_{1\to 1}=1\). This shows that
\(\norm{T^*}_{(1,\sigma) \to (1,\sigma)} = 1\), and therefore by
interpolation it holds that
\(\norm{T^*}_{(p,\sigma) \to (p,\sigma)} = 1\) for all
\(p\in [1,\infty]\).

This leads to define a new property of a linear map
\(T:\bounded(\hs) \to \bounded(\hs)\): we will say that \(T\) is
\textbf{hypercontractive} if there exist some \(p < q\) such that
\(\norm{T}_{(p,\sigma)\to (q,\sigma)}\le 1\). This in particular implies
that \(T\) is contractive with respect to \(\norm{\cdot}_{(p,\sigma)}\).

If we have a dynamical semigroup of quantum channels \(T_t\), then we
can consider \(\norm{T^*_t}_{(p,\sigma)\to (q,\sigma)}\) for some
\(p < q\) as a measure of convergence of the semigroup: indeed for
\(t=0\) we have that \(T_0 = \identity\) and
\(\norm{\identity}_{(p,\sigma) \to (q,\sigma)} =1\) if and only if
\(p\ge q\). On the other hand, if \(\sigma\) is the unique fixed point
of \(T_t\), then \(T^*_\infty(X) = \tr(\sigma X) \identity\) and
therefore
\(\norm{T_\infty^*(A)}_\infty = \tr(\sigma X) \le \norm{X}_{(1,\sigma)}\),
and \(\norm{T_\infty^*}_{(1,\sigma)\to \infty} = 1\).

Let \(1< p < \infty\), and \(q\) its H"older conjugate, i.e.
\(\frac{1}{p}+\frac{1}{q} =1\). Because \(T_t^*\) is self-adjoint in
\(L_2(\sigma)\), then it holds that
\[\norm{T^*_t}_{(p,\sigma)\to(2,\sigma)} = \norm{T^*_t}_{(2,\sigma)\to(q,\sigma)},\]
and therefore if \(1<p\le 2\) then
\[\norm{T^*_{2t}}_{(p,\sigma)\to(q,\sigma)} \le
 \norm{T^*_t}_{(p,\sigma)\to(2,\sigma)}\norm{T^*_t}_{(2,\sigma)\to(q,\sigma)}
 = \norm{T^*_t}_{(2,\sigma)\to(q,\sigma)}^2.\]

In the light of the previous observation, let us focus (as done usually
in the literature), on the behavior of
\(\norm{T_t^*}_{(2,\sigma)\to(q,\sigma)}\). The relationship between
log-Sobolev inequality and hypercontractivity is contained in the
following theorem

\begin{thm}[\cite{hypercontractivitylp}]
  Let $\lind$ satisfy detailed balance and $L_p$-regularity.
  Then the following conditions are equivalent
  \begin{enumerate}
    \item For $q(t) = 1 + e^{2 \alpha t}$,
    $$\norm{T^*_t}_{(2,\sigma)\to(q(t),\sigma)} \le 1.$$
    \item $T^*_t$ satisfies a 2-log-Sobolev inequality with constant $\alpha$.
  \end{enumerate}
\end{thm}

Note that point 1. of the previous theorem implies that, if
\(q(t) = 1 + e^{\alpha t}\), then
\(\norm{T^*_{t}}_{(p(t),\sigma)\to(q(t),\sigma)} \le 1\). For
\(t\to\infty\) we recover that
\(\norm{T^*_{\infty}}_{(1,\sigma)\to(\infty,\sigma)} \le 1\).

\subsection{Area law}\label{area-law}

Another interesting problem in the study of dissipative semigroups is
the description of the fixed point, or steady state, of the evolution.
For some models of noise, the fixed point is the maximally mixed state,
which is the state proportional to the identity. Such a state represents
the situation in which the noise has destroyed all the information on
the physical system, and every measurement will give uniformly
distributed random results. In other cases the noise model is different,
and the steady state will be a thermal state corresponding to some
Hamiltonian, meaning that it will be proportional to \(e^{-\beta H}\)
for some Hermitian \(H\) and some positive \(\beta\) representing the
inverse temperature. Davies maps \cite{davies1976quantum} are such an
example. In other cases the evolution is engineered (or defined) to have
a particular state as a steady state: one starts with a given state,
which is interested in preparing, and from that derives a Lindbladian
generator that produces that state as a fixed point. This is a common
approach of classical Glauber dynamics and Metropolis
sampling\cite{martinelli1999lectures} and of Dissipative State
Preparation \cite{verstraete09,Kraus08}.

One would expect that, if the state satisfies some ``good properties'',
then the resulting evolution would also have nice properties, as for
example would converge quickly. This was proven rigorously in the case
of classical spin systems and Glauber
dynamics\cite{martinelli1993finite,martinelli1999lectures}, where the
``good property'' of the state \(\omega\) is of this type: given two
observables \(A\) and \(B\), supported on separated regions that are
\(d\) distant apart, then the value of \(\omega(A\otimes B)\) becomes
increasingly close to \(\omega(A)\omega(B)\) as \(d\) increases. More
specifically, it is required that the difference between the two goes to
zero exponentially fast in \(d\). This property is called
\emph{exponential decay of correlations}, since the quantity
\(\omega(A\otimes B) - \omega(A)\omega(B)\) measures how correlated the
two region are. Under this assumption, for classical spin systems one
can prove that the corresponding Glauber dynamics is rapid mixing (via
proving a log-Sobolev inequality).

In this thesis we will be interested in the reserve problem: given a
``well behaved'' Lindbladian, what ``good properties'' of the fixed
point can be assumed? We will start by presenting rigorous notions of
correlations in many-body quantum systems.

\subsubsection{Correlation measures}\label{correlation-measures}

Consider a bipartite state \(\rho_{AB}\in \bounded (\hs_{AB})\). If
\(\rho_{AB}\) is of the form \(x \otimes y\) for some states \(x\) in
\(\bounded(\hs_A)\) and \(y\) in \(\bounded(\hs_B)\), we say it is a
product state. In this case measurement over the subsystem \(A\) are
independent from measurements over the subsystem \(B\), and vice versa:
therefore the resulting statistics will be independent and there will be
no correlations between the two subsystems. If \(\rho_{AB}\) is not
product, then there are a number of different measures that quantify
``how far'' it is from being product.

The following notation is borrowed from \cite{Kastoryano2013}. We will
denote by \(\rho_A\) (resp. \(\rho_B\)) the state \(\tr_B \rho_{AB}\)
(resp. \(\tr_A \rho_{AB}\)).

\begin{defn}[Correlation measures]\hfill\\
\label{def:correlations}
  \begin{itemize}
    \item \emph{Covariance correlation}:
      $$C(A : B)
      = \max_{\substack{M \in \alg_A,N \in \alg_B \\ \norm{M} \le 1, \norm{N} \le 1}}
      \abs{ \ev{M\otimes N} - \ev{M}\ev{N} } \\
      = \max_{\substack{M \in \alg_A,N \in \alg_B \\ \norm{M} \le 1, \norm{N} \le 1}}
      \abs{ \trace[ M \otimes N (\rho_{AB} - \rho_A \otimes \rho_B)] } ;$$
    where $\ev{O} = \trace(O \rho_{AB})$ is the expectation value of the observable $O$ acting on $\rho_{AB}$.

  \item \emph{Trace distance correlation}:
      $$T(A : B) = \max_{\substack{F \in \alg_{AB} \\ \norm{F} \le 1}}
      \abs{ \trace[F (\rho_{AB} - \rho_A \otimes \rho_B)] } =
      \norm{\rho_{AB} - \rho_A \otimes \rho_B}_1 .$$

    \item \emph{Mutual information correlation}:
      $$ I(A : B) = S(\rho_A) + S(\rho_B) - S(\rho_{AB}) ;$$
      where $S(\rho) = - \trace (\rho \log_2 \rho)$ is the von Neumann entropy of the state $\rho$.
    \end{itemize}
\end{defn}

\(C(A:B)\) is how correlations are usually measured in the condensed
matter literature. It follows immediately from the definition that
\(C(A:B)\) is upper bounded by \(T(A:B)\) (since \(C(A:B)\) only depends
on what can be measured with product observables, while \(T(A:B)\)
allows for general ones).

The relationship between trace distance and mutual information is given,
in one direction by Pinsker's inequality \cite{nielsen-chuang}, and in
the other by an application of Alicki-Fannes-Audenaert inequalities
\cite{2007JPhA...40.8127A, MR0345574,MR2043448}. We summarize it as
follows:

\begin{thm}\label{thm:fannes-mutual}
  \begin{equation}
   \label{eq:fannes-mutual}
    \frac{1}{4} T(A:B)^2 \le I(A:B) \le 6 T(A:B) \log_2 d_A + 4 h_b( T(A:B));
  \end{equation}
    where $h_b(x) = -x \log_2 x - (1-x) \log_2(1-x)$ denotes the binary entropy function, and $d_A = \dim \hs_A$.
\end{thm}

\subsubsection{Correlations in many-body
systems}\label{correlations-in-many-body-systems}

In our many-body scenario, we can consider the fixed point
\(\rho_\infty\) of \(\lind\) over \(\Lambda\), and for any region
\(A \subset \Lambda\) or any pair of regions \(A, B \subset \Lambda\) we
can consider the reduced density matrices
\(\rho_A = \tr_{\Lambda\setminus A} \rho_\infty\) and
\(\rho_{AB} = \tr_{\Lambda\setminus A \cup B} \rho_\infty\). We can then
ask two types of questions (which we formulate for \(I(A:B)\) but would
be equally interesting for any other measure of correlations):

\begin{itemize}
\item Given $A,B \subset \Lambda$, how does $I(A:B)$ scale with $\dist(A:B)$?
\item Given $A\subset \Lambda$, how does $I(A:A^c)$ scales with the size of $A$?
\end{itemize}

While of similar flavor, in the first case we are only considering
finite regions, while in the second we are considering \(A^c\), which is
growing as \(\Lambda\) gets bigger. Therefore it should not be
surprising that the conditions needed to give an answer to the first are
less restrictive than the ones needed for the second. In the first case,
we speak about \textbf{decay of correlations}: we expect that, if \(A\)
and \(B\) are far apart in the lattice, they become more and more
independent.

The second question is interesting for the following reason. For Haar
random states, \(I(A:A^c)\) is proportional to \(\abs{A}\). Instead,
many physically motivated states show a different behavior, with
\(I(A:A^c)\) scaling as \(\abs{\partial A}\), where \(\partial A\) is
defined as the subset of \(A\) of sites which directly interact with the
complement of \(A\). If the interactions are finite range and \(A\) is a
ball, then \(\abs{A}\) is a polynomial of degree \(D\) while
\(\abs{\partial A}\) has degree \(D-1\). This situation is usually
called \textbf{area law} (the terminology originated in the study of
black hole entropy, where the boundary is indeed a surface).

In the following we will be working with \(T(A:B)\) and \(I(A:B)\), with
the reminder that because of \cref{thm:fannes-mutual} exponential decay
in one of them implies exponential decay in the other.

\subsubsection{Ground states of
Hamiltonian}\label{ground-states-of-hamiltonian}

The problem of studying correlation decay, area laws and their
relationship with dynamics has been extensively treated (although not
completely solved) in the context of groundstates of Hamiltonians. A
Hamiltonian is a Hermitian operator \(H\) acting on the Hilbert space
\(\hs\) which represents the physical system. The action
\(\lind(\rho) = -i[H,\rho]\) generates a group of automorphisms instead
of simply a contraction semigroup, and it can be seen that it is a
special case of \cref{eq:lindblad}. Since any eigenvector of \(H\) is
invariant under the action of \(\lind\), the evolution will have more
than one fixed point: for physical reasons the one corresponding to the
lowest eigenvalue of \(H\) plays a special role, and it is called the
\emph{groundstate} of \(H\). It is a pure state. Given such a state
\(\ket{\phi}_{AB}\), it holds that \(I(A:B) = 2 S(\phi_A)\), where
\(\phi_A = \trace_B \dyad{\phi}_{AB}\) is the reduced density matrix of
\(\dyad{\phi}_{AB}\) over \(A\). Therefore, the mutual information
reduces to (two times) the von Neumann entropy.

The crucial property in this setting is the so-called \textbf{spectral
gap} of \(H\): the difference between the two smallest eigenvalues of
\(H\). Using the standard convention, we will say that a family of
Hamiltonians defined on an increasing and absorbing sequence
\(\Lambda_n \nearrow \Gamma\) is \emph{gapped} if the gap is uniformly
bounded away from zero - in other words, if the gap does not vanishes in
the limit. Otherwise the Hamiltonians will be called \emph{gapless} and
one can be interested in specifying the speed at which the gap closes
(whether polynomially or exponentially fast in \(n\)).

In the seminal work \cite{Hastings2006}, Hastings and Koma proved that
if a family of local Hamiltonians is gapped, then the ground state
satisfies exponential decay of correlations uniformly in \(n\). This
result is interesting because it connects to the condensed matter theory
of quantum phases: a quantum phase is an equivalence class of
Hamiltonian systems, such that two Hamiltonians \(H_1\) and \(H_2\) are
in the same equivalence class if they can be connected by a smooth path
\(H(t)\) of gapped Hamiltonians. Quantum phase transitions are therefore
identified with points in the path where the gap closes. In that
situation it is common for the correlation length to diverge, where by
correlation length we mean a distance \(\xi\) such that
\(C(A:B) \le e^{-\dist(A,B)/\xi}\).

Another property that is expected by condensed matter theorists is that
groundstates of gapped local Hamiltonian satisfy an area law for the
entanglement entropy. The intuitive argument (which unfortunately is not
a formal proof) goes as follow: if we consider a finite region \(A\),
because of exponential decay of correlations, spins which are inside
\(A\) and far away from its boundary are almost independent from the
system outside \(A\). Therefore, correlations and entropy can only come
from spins which are closer to the boundary. Since any given
\(d\)-dimensional spin can only contribute at most by \(\log d\) to the
total entropy, it follows that the total entropy is only scaling as the
size of the boundary.

Whether this argument can be made rigorous is the content of the
\textbf{area law conjecture} (that ground states of local gapped
Hamiltonians satisfy an area law). It is considered a major open problem
in condensed matter physics and has seen active development in recent
years \cite{Eisert2010}. A solution in 1D was obtained by Hastings
\cite{HastingsAreaLaw}, and subsequently a different proof appeared in
\cite{brando2014,brando2013}, where they proved that in 1D exponential
decay of correlations does actually imply an area law. This together
with the result of \cite{Hastings2006} shows that a spectral gap, by
implying exponential decay of correlations, also implies an area law in
1D.

In higher dimensions the problem is still open. Some advances have come
from the computer science community \cite{Arad2012}, with a new proof of
Hastings' and Koma's result, which allowed to greatly improve the
dependence of the correlation length with respect to the spectral gap,
making it fit better with the concrete cases in which we are able to
estimate both, either analytically or numerically. The tools developed
allowed for the construction of the first provable polynomial algorithm
for approximating the groundstate of gapped Hamiltonians in 1D
\cite{1307.5143v1}, as well as other combinatorial tools to study the
structure of ground states. These advances, while very promising, have
not led yet to a proof of an area law for groundstates in dimension
larger or equal than 2.

\subsubsection{Gibbs states and tensor network
states}\label{gibbs-states-and-tensor-network-states}

Gibbs states or thermal states are states proportional to
\(\exp(-\beta H)\), for some Hamiltonian \(H\) and a parameter \(\beta\)
that represents the inverse temperature. They are of interest because
they describe a system in equilibrium at finite temperature \(1/\beta\),
and therefore are naturally suited to fit in the open dynamics scenario:
lots of the dissipative models we have mentioned are attempted
descriptions of a thermalisation process that leads to a Gibbs state.
So, even if they are not the only possible fixed point of dissipative
maps, they are definitely an important class of them. Interestingly,
they all satisfy an area law for the mutual information \cite{Wolf2008}.

Another important class of states (this time pure states) which often
satisfy an area law, are the so called tensor network states
\cite{1603.03039v1} - states whose amplitudes are given by the
contraction of a given network of tensors. To be more specific, in the
large family of tensor network states, the one dimensional Matrix
Product States (MPS) and the two and higher dimensional Projected
Entangled Pair States (PEPS) satisfy an area law by construction. The
interest in these types of states is that they only require a polynomial
(in the number of particles) number of parameters to be described, as
opposed to the exponential dimension of the Hilbert space they live in.
For this reason, they are used extensively in numerics, and they are
believed to give good approximations of groundstates of local gapped
Hamiltonian. While there are examples of states in 2D that satisfy an
area law but are not approximable by a PEPS \cite{1411.2995v2}, it has
been proved that under certain assumptions groundstates of 2D local
Hamiltonians are well approximated by PEPS
\cite{Hastings2007,Molnar2015}.

In 1D, the situation is more clear: groundstates of local gapped
Hamiltonian can be efficiently approximated by MPS. This is not only a
theoretical result, but has also been important in understanding and
developing algorithms that approximate 1D groundstates.

\subsubsection{Area law and correlation
length}\label{area-law-and-correlation-length}

We have mentioned an intuitive -but incomplete- argument that would
connect a finite correlation length with an area law. Let us mention now
an interesting formal connection presented in \cite{Wolf2008}. There the
authors give a different definition of correlation length for the mutual
information: given a finite region \(A \subset \Lambda\), let
\(B_R = \{ x \in \Lambda| \dist(x,A) > R \}\), define \(\xi_\Lambda\) as
the minimal length \(R\) such that

\begin{equation} \label{eq:mutual-information-corr-length}
  I(A: B_R) < \frac{I(A : B_0)}{2} \qcomma \forall A \subset \Lambda.
\end{equation}

(Observe that \(B_0 = \Lambda \setminus A\).) With this definition, then
they can prove that \(I(A:A^C) \le 4 \abs{\partial A} \xi_\Lambda\),
i.e.~an area law.

While the result is sound, one should be careful in considering the
relationship between \cref{eq:mutual-information-corr-length} and the
usual decay of correlations, i.e. \(I(A:B) \le c \exp(-\dist(A,B)/k)\).
It is tempting to argue that, if one has such decay of correlations,
then because \(\dist(A,B_R) = R\) one has that
\(I(A:B) \le c \exp(-R/k)\), then it is sufficient to take
\(\xi_\Lambda\) proportional to \(k\) to satisfy equation
\cref{eq:mutual-information-corr-length}. This argument breaks if the
constant \(c\) in the decay of correlations bound is not independent of
the size of the regions \(A\) and \(B\), which is often the case as we
will see later. If \(A\) is fixed as we change \(\Lambda\), then the
size of \(B_R\) is proportional to the total size of the lattice
\(\Lambda\), and therefore \(\xi_\Lambda\) has to grow with \(\Lambda\).
The resulting bound on \(I(A:A^C)\) would still grow as a polynomial of
lower degree than the geometrical dimension of the lattice, but now
multiplied by a constant which is system-size dependent. This constant
will in most cases make the bound trivial, since it will be larger than
the general worst case bound, which is
\(\log \dim \hs_A = \abs{A} \log d\), where \(d\) is the dimension of
the Hilbert space of a single site.

A similar problem was faced in \cite{Kastoryano2013}, where they
obtained under the assumption of a log-Sobolev inequality a bound of the
form \[ I(A:A^c) \le c \log \log \norm{\sigma^{-1}} \abs{\partial A},\]
where \(\sigma\) is the fixed point of the evolution. Again, the right
hand side of the bound would scale with the correct exponent to talk
about an area law, but the multiplicative constant makes the bound
trivial in most cases.

One of the main results of this thesis is to prove for the first time a
fully satisfactory area law for fixed points of rapidly mixing
evolutions (see \cref{area-law-with-logarithmic-correction}).

\subsection{Stability of quantum
systems}\label{stability-of-quantum-systems}

One of the properties of open quantum systems studied in this
dissertation is stability. Before presenting the results obtained, it is
worth explaining why it is so crucial. The mathematical structure we are
considering is an attempt of describing a physical system composed of
many particles. This might be either a naturally occurring system (for
example, the original motivation of the Ising model was to study
magnetization), or an artificially engineered system created to fulfill
a task (computation, communication, memory, state preparation, etc.).

In the first case, the mathematical model will be of course an
approximation to the real physics: it would be unreasonable to require
that the quantities involved (coupling constants, energy levels,
masses/charges/densities, etc.) can be measured with perfect precision.
The only realistic hope is that they can be known with some level of
precision. Once we plug this information in our mathematical model, we
would like to have a tool that is capable of predicting the results of
future experiments. If they change abruptly for even the smallest change
in the parameters considered, the resulting predictions will rarely
match reality and the model will be deemed to be useless, since it
requires an impossible level of \emph{fine-tuning} to work.

The situation is very similar in the case of artificial and engineered
systems. In this case, the unreasonable assumption is that we have
perfect control over the implementation of the artificial model, meaning
that we can configure its parameters to arbitrary level of precision. No
real system (not even macroscopic and classical) can be perfectly
controlled in this way: the real implementation will be always at best a
very good approximation of the mathematical model. If the resulting
evolution depends heavily on these tiny differences, then we will end up
implementing a different evolution than the one we thought of preparing,
and the result will be different. The only practical models are the ones
for which small errors in the implementation will give rise to small
changes in the resulting system.

In both cases, the theoretical justification of a mathematical model
relies on its \textbf{stability against perturbations}: we can of course
talk about non-stable models, but one should be extremely careful in
considering their physical implications and predictions, since in
practice we will never be able to actually see them in reality. This
argument is only made more stringent when we start considering, apart
from experimental errors, physical sources of noises: no experiment will
be ever completely isolated, no noise will ever be perfectly shielded.

We thus need tools that allow us to justify the soundness of physical
models by proving that they are stable. In the case of local
Hamiltonian, effort has focused in proving stability of the spectral
gap, a parameter which has important consequences on the physical
properties of the resulting models. Starting from the work of
\cite{Bravyi2010,Klich2010} it culminated in \cite{Michalakis2013} where
it was proven that the spectral gap is stable (it does not close) under
some physically reasonable conditions.

It should be stressed here that we are considering a special case of
perturbations here: since we are considering many-body models, where
every particle interacts only with its neighbors, it is natural to
consider a perturbation/error that involves \emph{every single
interacting term}. Therefore, small perturbations will be
microscopically or locally small, but since they will add up as we
consider larger and larger systems, they are actually unbounded
perturbations (but with a local structure). This is why we cannot simply
apply standard perturbation theory but we need to develop specific tools
for this type of perturbations.

Another main result of this thesis is to show that rapidly mixing
systems are indeed stable against perturbations (see
\cref{stability-against-perturbations}).

\section{Summary of the results}\label{summary-of-the-results}

\subsection{Assumptions}\label{assumptions}

In this section, we present and discuss the main assumptions made in
this work. The most characterizing one is for sure rapid mixing, a
condition on the convergence time of the system to its unique fixed
point.

We will be talking about families \(\{\lind^\Lambda\}_\Lambda\) of
Lindbladian generators, where \(\Lambda\) runs over an infinite sequence
of finite subsets of \(\Gamma\). For each of them, we will denote by
\(T_t^\Lambda\) the corresponding evolution, i.e.
\(T_t^\Lambda = \exp(t \lind^\Lambda)\).

\begin{defn}[Unique fixed point]
  Let $\{\lind^\Lambda\}_\Lambda$ be a family of Lindblad generators. We say it has a unique fixed point if, for every $\Lambda$, $\lind^\Lambda$ has a unique fixed point and no periodic point (i.e. it has a trivial peripheral spectrum).
\end{defn}

We will denote by \(T^\Lambda_\infty\) the trace-preserving projector
onto the fixed point of \(T^\Lambda_t = \exp(t\lind^\Lambda)\), given by
\(\lim_{t \to \infty} T_t^\Lambda\).

\subsubsection{Rapid mixing}\label{rapid-mixing}

We have already argued why the spectral gap gives only partial
information about the mixing time of a dissipative evolution, while
log-Sobolev inequalities allow for a stronger control (but require some
strong property of the fixed point). Our approach is more direct, and we
will simply require that the mixing time scales logarithmically with the
system size, leaving aside the question of how to prove such condition.

\begin{defn}[Rapid mixing]
        \label{defn:rapid-mixing}
        Let $\{T_t^\Lambda\}_\Lambda$ be a family of dissipative maps, where $\Lambda$ runs over an infinite sequence of finite subsets of $\Gamma$. We say it satisfies rapid mixing if
        there exist $c, \gamma >0$ and $\delta \ge 1$ such that
        \begin{equation}
                \label{eq:rapid-mixing}
                \sup_{\substack{\rho \ge 0 \\ \trace \rho\, =1 }} \norm{T^\Lambda_t(\rho)- T^\Lambda_{\infty} (\rho)}_1 \le c \abs{\Lambda}^\delta \ e^{-t \gamma} .
        \end{equation}
\end{defn}

As we already mentioned, in some cases it is possible to relax the rapid
mixing assumption: this is covered partially by
\cite[sec. 4.5]{Stability-paper}.

Just like proving the existence of a spectral gap for a Hamiltonian
system, proving rapid mixing for a dissipative model is a hard task.
Apart from ``easy'' examples, such as non-interacting models and
dissipative state preparation for graph states \cite{Kastoryano12}, the
other important class of models satisfying such assumption are the
reversible Lindbladians satisfying a Log-Sobolev inequality
\cite{Quantum-Log-Sobolev}, which includes classical models such as
Glauber dynamics for the Ising model in the appropriate range of
parameters \cite{martinelli1999lectures}.

\subsubsection{Uniform families.}\label{uniform-families.}

As we have explained in the previous sections, we are interested in
studying the \emph{scaling} of some properties of a family of
Lindbladian generators \(\lind_n\), defined on an increasing and
absorbing sequence of finite lattices \(\Lambda_n\) converging to an
infinite graph \(\Gamma\) (in our case, \(\Gamma\) will be \(\ZZ^D\),
but the same reasoning goes through if we consider any other graph in
which balls grow polynomially in the diameter). But at the same time,
since we are interested in describing physical models, we would like
that different \(\lind_n\) represent the ``same'' physical system on a
different scale, in such a way that the scaling actually tells us
something about the physics we are trying to understand.

What does it mean for operators defined on different lattices to
represent the ``same physical system''? Of course the question is
ill-defined, so no definitive answer can be given, but we can try to
make some assumptions about a \emph{rule} or \emph{recipe} to obtain,
from the same ingredients, all the \(\lind_n\) at different scales.

One possible way would be to assume that the local terms of each
\(\lind_n\) are just the translation of a single local generator
\(l_0\): that is, there exists some finite \(r>0\) and a \(l_0\) acting
on \(\alg_{b_0(r)}\) such that for every \(n\) we have
\[ \lind_n = \sum_{x : b_x(r) \subset \Lambda_n} l_x \] where \(l_x\) is
the translated of \(l_0\) by the vector \(x\). This situation is usually
referred to as \textbf{translational invariant}, since in the limit the
interactions are invariant under translations of \(\Gamma\) (of course,
it does not makes sense to talk about invariance under translation for
finite lattices).

It should be noted that this is an excessively restrictive assumption:
not only because we might want to study systems in which the
interactions depend of the position in the lattice, but also because
near the boundary of \(\Lambda_n\) the system becomes
``under-determined'': since there is no room to fit the support of
\(l_0\) there, there are fewer and fewer interactions involving sites
near the boundary. Sometimes this case is denoted \textbf{open boundary
conditions}. Since we are interested in systems with a unique fixed
point, this assumption can be especially problematic, given that
under-determination around the boundary might cause multiple fixed
points to appear - therefore, we would be requiring two incompatible
conditions on \(\lind_n\).

To overcome this limitations, we have proposed a definition of what we
called \textbf{uniform families}, which we believe is a general enough
way of describing ``meaningful'' sequences of Lindbladian generators.
Let us denote by

\begin{equation}
\partial_d \Lambda = \{ x \in \Lambda \,|\, \dist(x,\Gamma \setminus \Lambda) \le d \}.
\end{equation}

By convention, we will write \(\partial \Lambda\) for
\(\partial_1 \Lambda\).

\begin{defn}\label{defn:boundary-condition}
  Let $\Lambda \subset \Gamma$, a \emph{boundary condition} for $\Lambda$ is given by a Lindbladian $\mcl B^{\partial \Lambda} = \sum_{d\ge 1} B^{\partial \Lambda}_d$, where $\supp B^{\partial \Lambda}_d \subset \partial_d \Lambda $.
\end{defn}

The definition of boundary condition involves a different notion of
locality than the one we used for defining local generators: the decay
in norm is only required as interactions get \emph{inside} the bulk of
the system, but they are allowed to be strong between spins that are as
distant as we want, as long as they have the same distance from the
boundary. For example, if \(\Lambda\) is a square, this definition
allows to couple opposite spins in the boundary: this situation is
denoted \emph{periodic boundary condition}, since one can imagine of
wrapping up \(\Lambda\) on a torus, and therefore making opposite spins
in the boundary become nearest neighbors. This and other exotic ways of
coupling spins in the boundary can be all described by the definition
given above.

\begin{defn}
  \label{def:uniform-family}
  A \emph{uniform family} of Lindbladians is given by the following:
  \begin{enumerate}[(i)]
  \item \textit{bulk interaction}: a Lindbladian $\mcl M_Z$ for every finite set $Z \subset \ZZ^D$;
  \item \textit{boundary conditions}: a family of \textit{boundary conditions} $\{ \mcl B^{\partial \Lambda} \}_{\Lambda}$, for every finite $\Lambda \subset \ZZ^D$.
  \end{enumerate}
\end{defn}

Given a uniform family of Lindbladians as we have just defined, for each
finite \(\Lambda \subset \Gamma\) we can define two Lindbladian
generators acting on it:

\begin{align}
\mathcal L^\Lambda =  \sum_{Z \subset \Lambda} \mathcal M_Z &\quad \text{open boundary evolution} ; \\
\mathcal L^{\bar \Lambda} = \mathcal L^\Lambda + \mathcal B^{\partial \Lambda} &\quad \text{closed boundary evolution}.
\end{align}

When speaking about \(\lind^{\bar \Lambda}\), we will refer to the terms
\(\mathcal M_Z\) as ``bulk'' interactions and to
\(\mathcal B_d^{\partial \Lambda}\) as ``boundary'' interactions.

Some comments on this definition are needed: the definition of a uniform
does not involve a particular sequence of increasing lattice
\(\Lambda_n\), but instead allows to define one (actually two)
Lindbladian for every finite \(\Lambda\) for which a boundary condition
is given. If we take two finite
\(\Lambda_1 \subset \Lambda_2 \subset \Gamma\), and we look at the
interactions involving the particles in the ``bulk'' of \(\Lambda_1\),
meaning the sites that are far away from \(\Gamma \setminus \Lambda_1\),
then it is easy to see that \(\lind^{\overline \Lambda_1}\) and
\(\lind^{\overline \Lambda_2}\) have the same short-range interactions,
and they only differ over the long-range ones: either because of the
effect of the terms \(\mathcal M_Z\) with \(Z\) extending outside
\(\Lambda_1\), or because of the difference between
\(\mathcal B^{\partial \Lambda_1}\) and
\(\mathcal B^{\partial \Lambda_2}\). Speaking informally, the
microscopic details of the interactions are the same apart from some
long-range terms. In the following section we will assume that the
strength of the interactions, i.e.~the norm of the corresponding
operators, decays in their range: therefore, for uniform families, we
have that the difference between the bulk interactions of
\(\lind^{\overline \Lambda_1}\) and \(\lind^{\overline \Lambda_2}\) will
be small. This is the fundamental property and defining characteristic
of uniform families of Lindbladian: up to small errors, the microscopic
details of the interactions in the bulk do no depend on how large the
system is taken (as long as it is large enough).

\subsubsection{Lieb-Robinson
Assumptions}\label{lieb-robinson-assumptions}

Up to now, our notion of local Lindbladian is incomplete: if we do not
specify at what rate the norms of the interactions decay, then we can
always trivially decompose a Lindbladian as a sum of local terms that
are all zero but the one with support on the full space. If instead we
impose that the norms decay as a function of the diameter of the support
we obtain a highly non trivial condition. Since the decay rate is
related to a property we will define later called Lieb-Robinson
velocity, we denote these conditions \textbf{Lieb-Robinson assumptions},
and we give them only for the uniform families defined earlier.

\begin{defn}[Lieb-Robinson Assumptions]
\label{def:lr-assumptions}
There exists an \emph{increasing} function $\nu(r)$ satisfying $\nu(x+y)\le \nu(x) \nu(y)$, and positive constants $v$, $b$, and $c$, such that the following conditions hold:
\begin{equation}
  \label{eq:assumption-a1}
  \tag{A-1}
  \sup_{x\in \Gamma} \sum_{Z \ni x} \norm{\mcl M_Z}_\diamond \abs{Z} \nu(\diam Z) \le v < \infty,
\end{equation}
\begin{equation}
  \label{eq:assumption-a2}
  \tag{A-2}
  \sup_{x\in \Gamma} \sup_{r} \nu(r) \sum_{d = r}^N \norm{ B_d^{\partial b_x(N)}}_\diamond  \le c N^b.
\end{equation}
\end{defn}

Note that if \(\norm{\mcl M_Z}_\diamond\) decays exponentially in
\(\diam Z\) (or is zero for all \(Z\) with a large enough diameter, a
situation denoted \textbf{finite range interactions}), then we can take
\(\nu(r) = \exp(\mu r)\) for some positive \(\mu\). If instead it decays
polynomially, we are forced to consider slower functions, such as
\(\nu(r) = (1+r)^\mu\). In particular, if
\(\norm{\mcl M_Z}_\diamond \sim (\diam Z)^{-\alpha}\), then
\cref{eq:assumption-a2} is satisfied for \(\mu < \alpha - (2D+1)\),
where \(D\) is the geometrical dimension of \(\Gamma\) (which means that
\(\alpha\) has to be larger than \(2D+1\) for the condition to hold).

The motivation for such assumptions is that systems which satisfy them
exhibit a finite speed of propagation: the support of a local observable
spreads in time, usually linearly, up to an exponentially small tail.
This implies that regions which are spatially separated, if they are
uncorrelated at time zero, remain almost uncorrelated for a finite time,
which depends (most of the time linearly) on their distance. More
details will be given in \cref{lieb-robinson-bounds}

\subsubsection{Frustration freeness}\label{frustration-freeness}

Another condition we will need to impose in some cases is frustration
freeness, inspired by a similar condition for Hamiltonians. It is worth
noticing that in this case we do not require any specific behavior from
the boundary terms, but only from the bulk ones.

\begin{defn}
We say that a uniform family $\mcl L = \{ \mcl M, \mcl B\}$ satisfies \emph{frustration freeness} (or is \emph{frustration free}) if for all $\Lambda$ and all
fixed points $\rho_\infty$ of $\lind^{\bar \Lambda}$
\begin{equation}
  \mcl M_Z (\rho_\infty) = 0 \quad \forall Z \subset \Lambda.
\end{equation}
\end{defn}

Many interesting and natural examples of Lindbladians have this
property: Davies generators and other types of Gibbs samplers for
commuting Hamiltonians \cite{arxiv1409.3435}, as well as dissipative
state engineering maps of PEPS.

\subsection{Technical tools}\label{technical-tools}

Before presenting the main results, let us present some of the technical
tools used to obtain them. They have been developed specifically in
order to prove such results, but might have interest and applications in
other contexts. We will start by presenting Lieb-Robinson bounds, a key
tool in almost every many-body work, and then continue with some derived
or inspired results.

In this section, we will assume that the generator \(\lind\) satisfies
assumptions \eqref{eq:assumption-a1} and \eqref{eq:assumption-a2}.

\subsubsection{Lieb-Robinson bounds}\label{lieb-robinson-bounds}

In many-body systems interactions are assumed to be local or
quasi-local: the spin present at each site of the lattice can only
interact directly with its neighbors (finite range interactions), or if
it can interact with distant spins the strength of the interactions has
to decay quickly with the distance. Therefore, the interaction between
distant spins is not direct, but it is mediated by the intermediate
spins which have to ``relay'' between them. It is natural to expect then
that interaction between distant spins is not instantaneous, but it will
show a delay, which will get larger as more and more spins have to be
involved in the relaying process, or in other words as the distance
increases. This is not at all a relativistic effect, since there is no
finite speed of light in our models: a better metaphor would be the
speed of sound, since it is given by the medium in which the
``information'' propagates.

This intuitive argument is formalized by Lieb-Robinson bounds, and the
resulting velocity of propagation is called Lieb-Robinson velocity. The
first formal proof was obtained in the setting of Hamiltonian systems
and groups of automorphisms \cite{lieb1972, robinson1968}, and for this
reason the propagation speed is also called group velocity. It was later
on generalized to dissipative evolutions \cite{Nachtergaele12,Poulin10}.

A consequence of Lieb-Robinson bounds and the existence of such a finite
velocity is that, if we consider a finite region \(A\) and we modify the
generator of the evolution on sites that are distant from \(A\), the
resulting modified evolution will be almost indistinguishable from the
original one on \(A\), at least for short times: for times shorter than
the one needed for information to propagate from the modified sites to
\(A\), spins in \(A\) have no way to ``know'' about the modification,
and therefore will evolve in the same way as if the change had not been
made. Since the velocity is given by assumption
\eqref{eq:assumption-a1}, it will be uniform for all system sizes. This
fact allowed in \cite{Nachtergaele12} to prove existence of the
thermodynamic limit of the finite-systems evolutions.

The effect of perturbing the dynamics in a distant region is given by
the following lemma, which is derived from the usual Lieb-Robinson
bounds.

\begin{lemma}[{\cite[Lemma 5.4]{Stability-paper}}]
  \label{lemma:lieb-robinson-localization}
  Let $\lind_1$ and $\lind_2$ be two local Lindbladians,
  and suppose $\lind_2$ satisfies assumption \eqref{eq:assumption-a1} with parameters $v$ and $\nu(r)$.
  Consider an operator $O_X$ supported on $X \subset \Lambda$, and denote by $O_i(t)$ its evolution under $\lind^*_i,\, i=1,2$.
  Suppose that $\lind_1 - \lind_2 = \sum_{r\ge 0} M_r$,
  where $M_r$ is a superoperator supported on $Y_r$ which vanishes on $\identity$, and $\dist(X, Y_r) \ge r$.
  Then the following holds:
  \begin{equation}
    \norm{O_1(t) - O_2(t)} \le  \norm{O_X} \abs{X} \frac{e^{vt} -vt -1}{v} \sum_{r=0}^\infty  \norm{M_r}_{\diamond} \nu^{-1}(r) .
  \end{equation}
\end{lemma}

\subsubsection{Open and closed
evolution}\label{open-and-closed-evolution}

The definition of uniform family allowed us to define two evolution
families, one with boundary condition and one without. The definition of
boundary condition we have given is justified by the following result:
if assumptions \cref{eq:assumption-a1} and \cref{eq:assumption-a2} are
satisfied, then the effect of the boundary conditions spreads from the
boundary towards the bulk of the system with the same finite speed of
propagation that characterizes Lieb-Robinson bounds. Therefore, for
short times and observables far away from the boundary, the two
evolutions will be indistinguishable. This is proven by applying
\cref{lemma:lieb-robinson-localization}.

\begin{lemma}[{\cite[Lemma 5.6]{Stability-paper}}]
\label{lemma:localizing-boundary}
Let $O_A$ be an observable supported on $A\subset \Lambda$,
and let $O_A(t)$ (resp. $\bar O_A(t)$) its evolution under $\lind^{\Lambda *}$ (rep., $\lind^{\bar \Lambda *}$).
Let $r = \dist(A,\Gamma\setminus\Lambda)$.
There exist positive constants $c$, $v$, and $\beta$ such that:
\begin{equation}
\norm{ O_A(t) - \bar O_A(t)} \le c \norm{O_A}\abs{A} \frac{e^{vt} - 1 -vt}{v} \nu^{-\beta}(r).
\end{equation}
\end{lemma}

\subsubsection{Frustration-free
localization}\label{frustration-free-localization}

All the bounds presented in this section are valid for any observable
localized in a certain region. The dual statement involving the
evolution of states would then say something about the evolution of any
state which at time zero decomposes as a product with respect to that
given region. While working on the problem of determining an area law
for the mutual information, we were faced with a similar but different
problem: what happens for some specific states, which are not product
but satisfy some other good property? More specifically, imagine that we
prepare our system on \(\Lambda_n = b_0(n)\) in the state
\(\rho^\infty_n\) which is the fixed point of the (closed boundary)
evolution on \(\Lambda_n\), but then we extend the system with an
arbitrary state \(\tau\) in \(\Lambda_{n+1}\setminus \Lambda_n\) and
look at the evolution of the state \(\rho^\infty_n\otimes \tau\) under
the generator defined on \(\Lambda_{n+1}\).

What can we say in this case? Standard Lieb-Robinson bounds will not
give any useful insight, since the regions we are considering
(\(\Lambda_n\) and \(\Lambda_{n+1}\setminus \Lambda_{n}\)) are at zero
distance, and we cannot assume that \(\rho^\infty_n\) is product (or
close to product) in any other bipartition of the system. On the other
hand, if we assume frustration freeness, then since most of the terms in
the generator of the evolution will be zero, the only non-zero ones
corresponding to terms near the boundary of \(\Lambda_n\). We will
therefore expect that the evolution will be approximately trivial in the
bulk of \(\Lambda_n\), and it will begin invading it from the boundary
at the Lieb-Robinson speed.

This intuitive idea is made formal in the following lemma. It should be
noted that, as far as we know, it is not a direct consequence of the
standard Lieb-Robinson bounds. Instead, in order to prove it, we
replicated the ideas and techniques that are present in the proof of the
Lieb-Robinson bound and have adapted them to this specific situation.

\begin{lemma}[{\cite[Lemma 12]{Area-Law-paper}}]
        \label{lemma:localization}
        Let $ \mcl L = (\mcl M, \mcl B)$ be an uniform family of Lindbladians, satisfying frustration freeness.
        Let $A \subset \Gamma$ be a finite region, and fix a positive natural number $m$. Let $B = A(m+1)$,
        $R = A(m+1)\setminus A(m)$ and
        $\rho_\infty^m$ a fixed point of $T_t^{\bar A(m)}$ and $\tau$ an arbitrary state on $R$.
        \begin{equation}
                \label{eq:localizing}
                  \norm{\left( T_t^{\bar B} - T_t^{ \bar B\setminus A} \right)(\rho_\infty^m \otimes \tau)}_1 \le
                  \poly(m) \nu^{-1}(m) \left[ e^{vt} -1 + t \right] ;
        \end{equation}
        where $T_t^{\bar B \setminus A}$ denotes the evolution
        generated by
        \[ \mcl L^{\bar B \setminus A} = \sum_{Z \subset {B\setminus A}} \mcl M_Z + \sum_{d \le m+1} \mcl B^{\partial B}_d .\]
\end{lemma}

\subsection{Main results}\label{main-results}

Let us now present the main results obtained in
\cite{Stability-paper, Short-Stability-paper,Area-Law-paper}.

\subsubsection{Local rapid mixing}\label{local-rapid-mixing}

\begin{defn}[Local rapid mixing]
  \label{def:local-mixing}
Take $A \subset \Lambda$, and define the \emph{contraction of $T_t$ relative to $A$} as
\begin{equation}
    \eta^A(T_t) := \sup_{\substack{\rho \ge 0 \\ \trace \rho\, =1 }}
    \norm{\trace_{A^c} \qty[ T_t(\rho) - T_{\infty}(\rho) ]}_1
    = \sup_{\substack{O_A \in \mathcal A_A \\ \norm{O_A}=1}}
    \norm{ T^*_t(O_A) - T^*_{\infty}(O_A) }.
\end{equation}
We say that $\mathcal L$ satisfies \emph{local rapid mixing} if, for each $A \subset \Lambda$, we have that
  \begin{equation}
    \label{eq:local-rapid-mixing}
    \eta^{A}(T_t) \le k(\abs{A}) e^{-\gamma t},
  \end{equation}
where $k(r)$ grows polynomially in $r$, $\gamma >0$ and all the constants appearing above are independent of the system size.
\end{defn}

We can also define a \textbf{local mixing time} as the inverse of
\(\eta^A\):

\begin{equation}
    \tau^A(\epsilon) =\min \{ t>0 : \sup_\rho \norm{\trace_{A^c} [ T_t(\rho) - T_{\infty}(\rho) ]} \le \epsilon \}.
\end{equation}

Then local rapid mixing implies that \(\tau^A\) depends on \(\abs{A}\)
and \(\epsilon\), but not on \(\Lambda\): this implies that (up to small
errors) local observables converge to their limit in a time scale which
depends only on the observable support and \textbf{not} on the system
size. Together with Lieb-Robinson bounds, it implies also that local
observables are able to interact only with a finite region around them
independent of the system size.

It seems surprising that just based on an estimate on the mixing time of
a uniform family we could derive such stronger property, but this is
what it is proved in \cite[Proposition 6.6]{Stability-paper}.

\begin{thm}
For uniform families satisfying Lieb-Robinson assumptions, rapid mixing implies local rapid mixing.
\end{thm}

\subsubsection{Local
indistinguishability}\label{local-indistinguishability}

The implication of local rapid mixing for observables can be in a sense
``dualized'' to a corresponding property of the family of fixed points.
The reasoning goes as follows: the limit
\(O(\infty) = \lim_{t\to \infty} O(t)\) of the evolution of an
observable is its expectation value against the limit state, i.e.
\(O(\infty) = \trace(\rho_\infty O)\identity\), since:
\[ \trace \rho O(\infty) = \lim_{t \to \infty} \trace \rho T_t^*(O) = \lim_{t \to \infty} \trace T_t(\rho) O = \trace \rho_\infty O.\]
We have that on the one hand, Lieb-Robinson bounds imply that if \(O\)
is a local observable, then for short times \(O(t)\) does not depend on
the interactions that are far away from its support; on the other hand,
\(O(t)\) converges to \(O(\infty)\) independently of the system size.
Therefore, \(O(\infty)\) can in a sense only depend on interactions that
are close to its support at time zero, and not on the ones that are too
far away. Since \(O(\infty)\) is equal to \(\tr \rho_\infty O\), this
means that the observable \(O\) cannot distinguish between different
fixed point of evolutions with increasing system size, as the only
change between them is in the interactions far away from the support of
\(O\). This is formalized in the following lemma
\cite[Lemma 6.2]{Stability-paper}

\begin{lemma}
    \label{lemma:rapidmixing-localization}
    Let $\mcl L = \{ \mcl M, \mcl B\}$ be a uniform family of dissipative evolutions that
    satisfies rapid mixing, and suppose each $T_t^{\bar \Lambda}$ has a unique fixed point and no other periodic points.
    Fix a $\Lambda$ and let $\rho_\infty$ be the unique fixed point of $T^{\bar \Lambda}_t$.
    Given $A \subset \Lambda$, for each $s \ge 0$ denote by $\rho^s_\infty$ the unique fixed point of $T_t^{\bar A(s)}$.

     Then we have:
    \begin{equation}
      \norm{ \trace_{A^c} ( \rho_\infty - \rho_\infty^s ) }_1 \le \abs{A}^\delta \Delta_0(s),
    \end{equation}
    where $\Delta_0(s) = c (\abs{A(s)}/\abs{A})^{\delta v/(v+\gamma)} \nu^{-\beta \gamma /(v+\gamma)}$, and $c$ is a positive constant while $\beta$ and $v$ are given in lemma \ref{lemma:localizing-boundary} and $\delta$ and $\gamma$ in definition \ref{defn:rapid-mixing}.
\end{lemma}

This property is coherent with the idea that uniform families represent
some model in which the microscopic dynamic is well defined and
independent of the system size.

\subsubsection{Stability against
perturbations}\label{stability-against-perturbations}

We have presented, in \cref{stability-of-quantum-systems}, the
importance of stability against perturbation for theoretically
justifying the models we consider. Let us give a formal definition of
stability. Since there are definitely different notions that might be
useful in different context, we will take a very conservative approach,
and impose as few assumptions as possible on the perturbations, while
requiring the strongest possible notion of stability. There are of
course a number of possible relaxations of this result.

Given a uniform family of Lindbladians \(\lind\), defined by its bulk
terms \(\mcl M_Z\) and its boundary conditions \(\mcl B_d\), we will
consider a perturbation of both the bulk terms
\(\mcl M_Z^\prime = \mcl M_Z + E_Z\) and of the boundary conditions
\(\mcl B^\prime_d = \mcl B_d + E_d\). The perturbation should be small
compared to the norm of the original Lindbladians, so we will assume
that for all \(E_Z\) and \(E_d\) it holds that
\(\norm{E_Z}_\diamond \le \epsilon \norm{M_Z}_\diamond\) and
\(\norm{E_d}_\diamond \le \epsilon \norm{B_d}_\diamond\), where
\(\epsilon >0\).

Notice that such perturbation are small microscopically, but since they
act on every local term, their sum
\(E^{\bar \Lambda} = \sum_{Z \subset \Lambda} E_Z + \sum_d E_d\) has
norm which diverges with the system size, so that it is an unbounded
perturbation once we forget about its local structure. This implies that
we cannot simply apply standard perturbation theory results.

We still need some condition on the perturbation for it to be
``physically realistic''. Requiring \(\mcl M_Z^\prime\) and
\(\mcl B^\prime_d\) to be Lindbladians would suffice, but we can relax
this restriction and only impose the following conditions (which are
always satisfied if the perturbed generators are Lindbladians):

\begin{itemize}
\item $\du E_Z(\identity) = \du E_d(\identity) = 0$;
\item $S_t = \exp[t(\lind^{\bar \Lambda} + E^{\bar \Lambda})]$ is a contraction for each $t\ge 0$.
\end{itemize}

With this perturbation model, under the rapid mixing assumption, we are
able to prove the following stability result
\cite[Theorem 6.7]{Stability-paper}.

\begin{thm}
  \label{thm:stability}
  Let $\mathcal L$ be a uniform family of local Lindbladians with a unique fixed point, satisfying rapid mixing.
    Let $S_t$ be defined as above.
    For an observable $O_A$ supported on $A  \subset \Lambda$, we have for all $t \ge 0$:
  \begin{equation}\label{eq:stability}
    \norm{T_t^{*}(O_A) - S_t^{*}(O_A)} \le c(\abs{A})\, \norm{O_A} \left( \epsilon + \abs{\Lambda} \nu^{-\eta}(d_A) \right),
  \end{equation}
  where $d_A = \dist(A, \Lambda^c)$;
  $\eta$ is positive and independent of $\Lambda$;
  $c(|A|)$ is independent of $\Lambda$ and $t$, and is bounded by a polynomial in $\abs{A}$.
\end{thm}

Let us discuss the r.h.s. of \cref{eq:stability}. The multiplicative
term \(\norm{O_A}\) is an expected normalization constant, that takes
into account the fact that the l.h.s. is 1-homogeneous. The constant
\(c(\abs{A})\) only depends on the support of \(A\) and not on
\(\Lambda\): therefore, if \(A\) is fixed once for all and does not
scale with \(\Lambda\), this is simply a constant prefactor. We will
discuss later on what happens if this is not the case.

The local norm of the perturbation is \(\epsilon\), so it is expected
that it would appear on the r.h.s. of the bound. More surprisingly it
only appears as a linear factor. Let us stress again that if the
perturbation is a sum of local terms acting on single sites, then
\(\epsilon\) is the (diamond) norm of one single term, not of the whole
perturbation, and in particular is independent of the system size.

The other term \(\abs{\Lambda} \nu^{-\eta}(d_A)\) is a boundary
correction term, which takes into account the effect of the perturbed
boundary on the observable. As it is expected, it is decaying with the
distance of \(A\) to the boundary, and it vanishes as \(\Lambda\) goes
to the infinite lattice \(\Gamma\). Therefore, for large enough
\(\Lambda\), is will be smaller than \(\epsilon\) and can be considered
negligible. In some intermediate cases it might not be optimal: for
example, one would not expect such a term to appear in the case of
translational invariant interactions and periodic boundary conditions,
even if the observable is localized near the boundary. After all, in
this setting, the location of the boundary is only a mathematical
necessity, and it does not correspond to any difference in the physical
interactions between the spins. In fact, we can ``shift'' the system to
put the boundary as distant as possible from \(A\), and therefore in
such cases one should always consider \(d_A\) to be the half-diameter of
the complement of \(A\) in \(\Lambda\). With this observation, the
decaying term \(\nu^{-\eta}(d_A)\) will quickly cancel the contribution
of the term \(\abs{\Lambda}\) and become negligible compared to
\(\epsilon\).

It is worth noting that if we consider non-local observables
(i.e.~observables whose support grows with \(\Lambda\)), then the factor
\(c(\abs{A})\) cannot be assumed to be smaller than linear, and in
particular this means that the bound cannot be improved to be
non-divergent in \(\Lambda\). In fact, it is easy to construct simple
examples of non-interacting spins such that there are observables
supported on the full lattice for which the l.h.s. of
\cref{eq:stability} grows linearly with the system size. This means in
particular that the factor \(c(\abs{A})\) has to be at least linear. See
\cite[Example 4.8]{Stability-paper} for such a construction.

\subsubsection{Area law with logarithmic
correction}\label{area-law-with-logarithmic-correction}

Regarding the problem of correlations decay for the fixed point of the
evolution, we have the following result
\cite[Theorem 14]{Area-Law-paper}:

\begin{thm}
Let $\mathcal L$ be a uniform family of local Lindbladians with a unique fixed point, satisfying rapid mixing. Then the fixed point of every $\lind^{\bar \Lambda}$ satisfies:
\begin{equation}
\label{eq:correlation-decay}
T(A : B) \le 3 (\abs{A} + \abs{B})^\delta \Delta_0\left(\frac{d_{AB}}{2}\right),
\end{equation}
where $\Delta_0$ is the fast-decaying function given in \cref{lemma:rapidmixing-localization}.
\end{thm}

Because of \cref{thm:fannes-mutual}, \(I(A:B)\) will show a similar
decay.

Let us now consider the question of whether \(\rho_\infty\) satisfies an
area law or not. We have not been able to give a definite answer, but we
have obtained a relaxed version: a scaling of \(I(A:A^c)\) as
\(\abs{\partial A} \log \abs{A}\), which in terms of the radius of \(A\)
goes as \(r^{D-1}\log r\). In order to obtain such a result, we will
need some extra assumptions
\cite[Proposition 16,Theorem 17]{Area-Law-paper}.

\begin{thm}
\label{thm:area-law}
Let $\lind$ be a uniform family of local Lindbladians with a unique fixed point, satisfying rapid mixing.
Moreover, let us assume that $\lind$ satisfies either of the following conditions:
\begin{itemize}
\item $\rho_\infty$ is pure for every $\Lambda$;
\item $\lind$ satisfy frustration-freeness;
\end{itemize}
Then the following holds for fixed point of every $\lind^{\bar \Lambda}$, for some constant $c$ independent of $\Lambda$:
\begin{equation}
I(A:A^c) \le c \abs{\partial A} \log \abs{A}
\end{equation}
\end{thm}

Interestingly, the two alternative conditions required in
\cref{thm:area-law} are independent from each other. Therefore, we
suspect that their need is an artifact of the proof, and that the result
holds without either assumption.

\section{Outlook and future work}\label{outlook-and-future-work}

We have seen that a condition on the scaling of the mixing time of a
quantum dynamical system can have deep impact on the properties it
exhibits: both dynamical (like the stability property) and static (as
the area law and the indistinguishability of its fixed points).

This brings us to reconsider the mixing time as a fundamental and
characterizing property of these models, in the same spirit that the
spectral gap plays in the context of Hamiltonian systems, as it was
already proposed in \cite{PhysRevB.90.045101}. In this dissertation we
have identified a class of systems which, because of the mentioned
properties, clearly plays a special role in a possible classification of
dissipative evolutions. Whether this class is to be considered the
``good'' case or the ``trivial'' case is still to be determined: it will
be important to see which properties can be recovered and which cannot
hold when we lift the rapid mixing condition for something less
restrictive, and how large and diverse is the class of rapid mixing
models. This is an interesting line of future research, and the
following observations can be considered as the first steps in its
development.

\subsection{Polynomial mixing time}\label{polynomial-mixing-time}

If we consider Lindbladians as ``dissipative machines'' or preparation
procedures to obtain useful quantum states, then from an algorithmic
point of view polynomial time mixing is perfectly acceptable, and
constitutes a more natural assumption than rapid mixing. In the same
spirit, for computing purposes, it is sometimes useful to consider
gapless Hamiltonians whose spectral gap only closes \emph{polynomially}
in the system size. In both cases, we have a situation that is clearly
unfavorable in the thermodynamic limit, but that for finite systems can
still be somehow treated in a reasonable amount of time and with a
reasonable amount of resources.

In this situation, the connection with condensed matter theory (both
formal results as wells as the ``philosophy'') becomes less useful: it
would be akin to study the efficiency of different sorting algorithms by
looking at how they perform on a infinite set of elements. Instead, a
different set of tools would be needed

Describing what is the computational power and the properties of such
class is an extremely interesting property and would be a way of
clarifying whether dissipative state preparation can be scaled up in
realistic experiments. In this setting stability, not necessarily in the
same form as we have proven here, should be a crucial characteristic to
consider.

\subsection{Preparation of topological
models}\label{preparation-of-topological-models}

The area law result could be interpreted as a negative property of rapid
mixing models: they are not able to generate states which violate area
laws. On the other hand is it also known that some topological states
cannot be obtained in sub-linear time \cite{PhysRevB.90.045101}, at
least with realistic assumptions about the Lindbladian generators, even
if they satisfy an area law. Since topological states are regarded as
the key ingredient for a robust quantum memory, it would be interesting
to study whether:

\begin{enumerate}
\def\labelenumi{\roman{enumi}.}
\tightlist
\item
  they can be prepared dissipatively in a stable way;
\item
  there exist ``good'' dissipative processes that preserve the state,
  once the system has been prepared by other means.
\end{enumerate}

The difference between the two lies in the fact that in the second case
we are not asking for the process to robustly or rapidly prepare the
topological model when started in any arbitrary state, but the only
requirement is that these good properties hold when the process is
started exactly in the state we want to preserve. If the system is
robust, then any state which is close enough to the desired state would
converge back to it, and any small perturbation of the generators will
only slightly move this special fixed point. In principle, we would not
be requiring any good properties outside these: there might be other
fixed point, either stable or unstable.

In this line of research, we should mention the following works: the
proposal of an ``encoder'' \cite{Dengis2014}, which not only prepares a
state in the 2D Toric Code groundstate, but it also allows to encode
logical information, in the sense that a specific pair of qubits in the
system will get mapped into the virtual qubits of the Toric Code. The
process requires linear time, and is highly not translation invariant.
What is the effect of noise on this encoding procedure is still an open
problem.

In \cite{arxiv1409.3435}, they study two different families of
Lindbladian models arising from a commuting quantum Hamiltonian, both of
which prepare the Gibbs state of the corresponding Hamiltonian. They
show that for high enough temperature, both processes are gapped and
therefore they have polynomial time mixing. In the corresponding
classical case it is possible to show the stronger property of a
log-Sobolev inequality and rapid mixing, so one could hope of
strengthening the result. Notice that this would not contradict the
result on the preparation of topological models: since here we are
preparing the Gibbs state at some finite temperature, and we know that
for the 2D Toric Code this shows no topological properties, there is in
principle nothing forbidding a rapid mixing process to generate the
Gibbs state of the 2D Toric Code.

It is known that if we take the temperature to zero, the Gibbs state
converges to the ground state - under some assumption on the density of
states at different energy levels, to get a good approximation it is
sufficient that the temperature scales as the inverse of the logarithm
of the number of particles \cite{Hastings2007}. Therefore, it would be
interesting to study the behavior of the models considered in
\cite{arxiv1409.3435} as the temperature goes to zero, for specific
commuting Hamiltonian such as the Toric Code. It is not clear whether
the resulting map will have fixed points outside of the groundstate
space or not, and whether the gap will close or stay open. In
\cite{1602.01108v1} it was considered a similar problem of preparing a
Gibbs state for a Hamiltonian with a critical temperature, and it was
shown that in the regime of phase coexistence there cannot be a unique
fixed point. This does not solve the case of the Toric Code, since there
is no critical temperature, but it can be an interesting comparison.

\subsection{Proving rapid mixing}\label{proving-rapid-mixing}

In this dissertation we have shown how rapid mixing implies properties
of the fixed point of the evolution, and we have discussed log-Sobolev
inequalities as a way of proving such condition in the case of detailed
balance Lindbladians. It would be extremely useful to have tools and
conditions that might allow to prove the rapid mixing hypothesis, either
via log-Sobolev inequality or not. The work done in
\cite{arxiv1409.3435} can be considered as going in this direction:
describing properties of the fixed point that imply conditions on the
mixing time. This is inspired by the results obtained for classical
models (see \cite{martinelli1993finite,martinelli1999lectures} for a
review), for which it was shown that a condition on the correlation
decay of the fixed point of Glauber dynamics implies a log-Sobolev
inequality and in turn rapid mixing. Moreover, this type of correlation
decay is usually present for Gibbs states at high temperature.

In \cite{arxiv1409.3435} a similar approach was taken, and a specific
definition of correlation decay has been used to show that the
corresponding dynamics has a spectral gap. It is also proven that this
condition holds for high temperature, but also in the 1D case. It would
be interesting to see if one could strengthen the result and prove a
log-Sobolev inequality instead, which would be expected given the
classical result.

The results presented in \cref{hypercontractivity} show that when we
generalize log-Sobolev inequalities from the classical to the quantum
setting, we do not obtain a single inequality but a family (indexed by
\(p \in [1,\infty)\)) of inequalities. Classically they are all
equivalent, but in the quantum case this is not known, and therefore
different authors have resorted to assume \(L_p\)-regularity: the
assumption that the \(L_2\) constant lower bounds all the others. To
recover the connection with hypercontractivity one indeed needs the
whole family of log-Sobolev inequalities to hold, but as we showed in
\cref{entropy-and-log-sobolev-inequalities} one only needs the case
\(L_1\) in order to prove rapid mixing.

It is not known how generic the \(L_p\)-regularity condition is: the
only general result being that some important class, such as Davies
generators \cite{Quantum-Log-Sobolev} and unital processes
\cite{hypercontractivitylp}, satisfy it. In \cite{Quantum-Log-Sobolev}
it is conjectured that every detailed balanced Lindbladian satisfies
\(L_p\)-regularity, and every primitive Lindbladian satisfies a weak
version of it. It this conjecture fails to hold, then one might want to
only look at strategies to prove the \(L_1\) inequality, since that is
the one which provides the rapid mixing bound. Showing which conditions
have to be imposed on the fixed point in order to prove the log-Sobolev
inequalities will help us understand whether the class of rapid mixing
systems is ``small'' or ``large''.

\subsection{Other work in different research
lines}\label{other-work-in-different-research-lines}

Another line of research which was developed during the PhD, other than
the study of open dissipative dynamics, is the study of whether
properties of the thermodynamic limit of a sequence of Hamiltonians can
be inferred from the study of an increasing sequence of finite systems.
More specifically, we were interested in the possible pitfalls of the
common approach taken to study the large-system limit of Hamiltonian
models, which consists in analyzing a sequence of finite cases, either
experimentally or numerically, to then extrapolate some properties of
the limit. In a number of important cases
\cite{march1992electron, LandauLifshitzVolume5,domb1983phase,Tagliacozzo2008,Pirvu2012}
this approach has been successful and has led to important insight on
the properties of the physical models in the large-system limit. On the
other hand, there is a general negative result: the problem of deciding
whether a sequence of translation invariant local Hamiltonians is gapped
or gapless in the limit is an undecidable problem
\cite{Spectralgapundecidability}. This result shows that unpredictable
behavior can be shown by this type of models, which lead us to be
interested in exploring the possibilities of constructing such exotic
examples.

The resulting work has been presented in a (yet to be published) paper
\cite{size-driven}, in which we present two families of models that show
a surprising property: for any finite region smaller than a fixed
threshold, the ground state and low-excited states are classical states
(product states in the computational basis); above the threshold they
show instead topological properties, which are characteristic of quantum
models. By increasing the local dimension of the spins, the threshold
can be made arbitrarily large, and already for local dimension \(d=10\)
becomes bigger than the estimated number of particles in the universe.
We denoted this phenomenon \textbf{size-driven phase transition}, as it
can be seen as an abrupt change from a classical model to a quantum
model driven by the change of the system size parameter.

The two constructions are based on different ideas, and have different
thresholds scaling. They are both based on \emph{tilings models}: a
tiling is a covering of a region of the plane with unit squares with
colored edges, such that colors on neighboring squares matches. It has
been shown that the problem of deciding, given a finite set of tiles,
whether they can tile the whole plane or not is undecidable
\cite{berger1966undecidability,robinson1971undecidability}. This result
was a building block of the undecidability result of
\cite{Spectralgapundecidability}. We modified their construction by
using \emph{plaquette} and \emph{star} interactions (instead of just
having plaquettes), and with this we produced two families of models.
The first one is based on the idea of constructing periodic patterns
with very large periods (compared to the number of colors used in the
tiling), in such a way that a specific pattern only appears once every
period. By penalizing that pattern, we can induce a energy frustration
for every lattice size larger than the period: this allows us to
implement the transition between the classical and quantum models.

The other construction is based on the idea, already present in the
results of undecidability, of encoding the history of a Turing Machine
into the groundstate of the Hamiltonian. In this way, by giving an
energy penalty if the machine halts, we have the same phenomenon as
before of an energy frustration when the system size becomes large
enough for the machine to halt. Since determining whether (and when) a
Turing machine will halt is an undecidable problem, this was one of the
key ingredient into showing undecidability of the spectral gap. We were
able to highly optimize the cost of the encoding, in the sense of the
minimal local Hilbert space dimension needed to write the history state
of the Turing machine into the spin model. With this optimized encoding,
we considered the so called \emph{Busy Beavers} Turing machines: a
machine with a very small dimension, but whose halting time is
surprising large - it actually grows faster than any computable
function. This gives us models which have a relatively small local
dimension, but for which there is a frustration only for extremely large
system sizes. Again, this frustration allows us to implement the phase
transition.

\selectlanguage{spanish} \chapter{Introducción}

Esta tesis está organizada como sigue. En el
\cref{objetos-de-estudio-y-resultados-previos} definimos los principales
objetos de interés, que son los semigrupos dinámicos de canales
cuánticos. Recordaremos las propiedades de sus generadores, llamados
Lindbladianos. Se explicará por qué son un buen modelo para evoluciones
cuánticas con ruido, y se definirá el tiempo de equilibración, una
propiedad central en las hipótesis necesarias para probar los resultados
principales. Se presentará la relación entre tiempo de equilibración y
otras importantes propiedades del semigrupo, como el gap espectral, la
desigualdad de log-Sobolev e hipercontractividad. Estas conexiones nos
permitirán encontrar técnicas para probar la condición de equilibración
rápida. Introduciremos también la noción de información mutua y se
discutirá la ley de área para estados, así como sus conexiones con la
dificultad de simulación y los estados de redes tensoriales. Finalmente,
se discutirá por qué la estabilidad es una condición fundamental para
cualquier modelo matemático de un sistema físico. En el
\cref{resumen-de-los-resultados} se definirán las hipótesis principales
y una síntesis de los resultados obtenidos, junto con una breve
presentación de las herramientas técnicas desarrolladas para probarlos.
En el \cref{perspectivas-y-trabajos-futuros} se discutirán las lineas
futuras de investigación en las cuales estamos trabajando actualmente.

El resto de la tesis está compuesta por las publicaciones que recogen
los resultados obtenidos a lo largo del Doctorado. Los capítulos
corresponden a las publicaciones siguientes.

\begin{enumerate}
\setcounter{enumi}{4}
\item \cite{Stability-paper} \fullcite{Stability-paper}
\item \cite{Short-Stability-paper} \fullcite{Short-Stability-paper}
\item \cite{Area-Law-paper} \fullcite{Area-Law-paper}
\end{enumerate}

El impacto de estas publicaciones está reflejado en el número de citas
recibidas, a pesar de su reciente publicación: en particular,
\cite{Stability-paper} en el momento de la publicación de esta tesis ha
recibido ya 18 citas, mientras que \cite{Short-Stability-paper} ha
recibido 3 y \cite{Area-Law-paper} una. Además, los resultados obtenidos
han sido presentados como ponencia oral en las conferencias más
prestigiosas del área: en Quantum Information Processing and
Communications 2013 (QIPC2013), en el 17th Conference on Quantum
Information Processing (QIP2014), y en Theory of Quantum Computation,
Communication and Cryptography (TQC2015).

\section{Objetos de estudio y resultados
previos}\label{objetos-de-estudio-y-resultados-previos}

\subsection{Notación}\label{notaciuxf3n}

Fijemos primero la notación que usaremos a lo largo de toda la tesis,
aunque se presentarán estos objetos con más detalle más adelante.

Dado un producto tensorial de dos espacios de Hilbert de dimensión
finita \(\hs_A\otimes \hs_B\), la única función lineal
\(\tr_A: \bounded(\hs_A \otimes \hs_B) \to \bounded(\hs_B)\) tal que
\(\tr_A(a\otimes b)=b\tr(a)\) para todo \(a \in \bounded(\hs_A)\) y todo
\(b \in \bounded(\hs_B)\) se llamará la \emph{traza parcial} sobre
\(A\). Un estado sobre \(\bounded(\hs)\) está dado por un funcional
lineal positivo \(\rho : \bounded(\hs) \to \RR\), normalizado de manera
que \(\rho(\identity)=1\). Indicaremos la traza de un operador \(X\) con
\(\tr X\), y la norma \(p\) de Schatten como \(\norm{\cdot}_p\), es
decir \((\tr \abs{X}^p)^{1/p}\). Si no hay riesgo de ambigüedad,
\(\norm{\cdot}\) denotará la norma de operador usual (la norma de
Schatten \(\infty\)).

Consideraremos un retículo cúbico \(\Gamma = \ZZ^D\), con la distancia
de grafo, y un subconjunto finito \(\Lambda \subset \Gamma\). Como es
común en física, llamaremos a todo subconjunto de \(\Gamma\) un
retículo, aunque no lo sea en el sentido de teoría de grafos. La bola
centrada en \(x \in \Lambda\) de radio \(r\) se denotará con \(b_x(r)\).
A cada vértice \(x\) del retículo le vamos a asociar un sistema cuántico
elemental con un espacio de Hilbert de dimensión finita \(\hs_x\).
Usaremos la notación de Dirac para vectores: \(\ket{\phi}\) será un
vector en \(\hs_x\), \(\bra{\phi}\) su adjunto, y
\(\{ \ket{n}\}_{n=0}^{\dim \hs_x -1 }\) la base canónica de \(\hs_x\).
El producto escalar en \(\hs_x\) se denotará como
\(\braket{\phi}{\psi}\), y las funciones de rango uno como
\(\ketbra{\phi}{\psi}\). A cada subconjunto finito
\(\Lambda \subseteq \Gamma\) le asociamos un espacio de Hilbert dado por
\[ \hs_\Lambda =
\bigotimes_{x \in \Lambda} \hs_x ,\] y un álgebra de observables
definida por \[ \alg_\Lambda = \bigotimes_{x \in \Lambda} \mathcal
B(\hs_x) .\]

Dado que \(\hs_x\) tiene dimensión finita, \(\bounded(\hs_x)\) es una
álgebra de matrices, y la indicaremos a veces como \(\matrixalg_d\),
donde \(d = \dim \hs_x\).

Si \(\Lambda_1\subset \Lambda_2\), hay una inclusión natural de
\(\alg_{\Lambda_1}\) en \(\alg_{\Lambda_2}\), identificándola con
\(\alg_{\Lambda_1}\otimes \identity\). El soporte de un observable
\(O \in \alg_\Lambda\) es el mínimo conjunto \(\Lambda^\prime\) tal que
\(O = O^\prime\otimes \identity\), para algún
\(O^\prime \in \alg_{\Lambda^\prime}\), y será indicado por \(\supp O\).

Una función lineal \(\mcl T: \alg_\Lambda \to \alg_\Lambda\) se llamará
\emph{superoperador} para subrayar la distinción con los operadores en
\(\alg_\Lambda\). El soporte de un superoperador \(\mcl T\) es el
conjunto mínimo \(\Lambda^\prime \subseteq \Lambda\) tal que
\(\mcl T = \mcl T^\prime \otimes \identity\), donde
\(\mcl T^\prime \in \bounded(\alg_{\Lambda^\prime})\). Diremos que un
superoperador preserva la hermiticidad si manda operadores hermíticos a
operadores hermíticos. Diremos que es positivo si manda operadores
positivos (es decir, operadores de la forma \(\du O O\)) a positivos.
Diremos que \(\mcl T\) es \emph{completamente positivo} si
\(\mcl T \otimes \identity : \alg_\Lambda \otimes \matrixalg_n \to \alg_\Lambda \otimes \matrixalg_n\)
es positivo para todo \(n \ge 1\). Diremos que \(\mcl T\) preserva la
traza si \(\trace \mcl T(\rho) =\trace \rho\) para todo
\(\rho \in \alg_\Lambda\). Las normas \(p\) de Schatten en
\(\alg_\Lambda\) inducen una familia correspondiente de normas sobre
\(\bounded(\alg_\Lambda)\): denotamos con \(\norm{\cdot}_{p \to q}\) la
norma de operador de un superoperador
\(\mathcal T : (\alg_\Lambda,\norm{\cdot}_p) \to (\alg_\Lambda, \norm{\cdot}_q)\),
es decir, cuando el dominio está equipado con la norma \(p\) y la imagen
con la norma \(q\). A veces nos hará falta la siguiente norma, llamada
norma diamante: \[\norm{\mcl T}_{\diamond} = \sup_n \norm{\mcl T
\otimes \identity_n}_{1 \to 1}.\]

\subsection{Semigrupos dinámicos de canales
cuánticos}\label{semigrupos-dinuxe1micos-de-canales-cuuxe1nticos}

\subsubsection{Evolución unitaria y canales
cuánticos}\label{evoluciuxf3n-unitaria-y-canales-cuuxe1nticos}

La mecánica cuántica nos dice que un sistema físico viene representado
por un espacio de Hilbert \(\hs\). Las propiedades medibles del sistema
se codifican en un estado \(\rho\), que es un operador positivo con
traza uno. Para simplificar sólo consideraremos espacios de Hilbert de
dimensión finita \(d\). Por lo tanto \(\bounded(\hs)\) es una álgebra de
matrices \(\matrixalg_{d}\). En el caso de sistemas aislados, la
evolución física del sistema se describe con una evolución unitaria del
estado \(\rho\), donde el estado del sistema después de la evolución
está dado por \(U \rho U^\dg\), con \(U\) un operador unitario de
\(\hs\). Es inmediato ver que este tipo de evolución es necesariamente
reversible, dado que su inversa \(\rho \to U^\dg \rho U\) es también
físicamente posible. Por lo tanto, para incluir a sistemas disipativos,
donde la evolución no es reversible a causa de una interacción con un
espacio ambiente, tenemos que reemplazar la evolución unitaria por algo
más general.

Vamos a intentar fijar unos requisitos lo más generales posibles que una
aplicación \(T: \bounded(\hs) \to \bounded (\hs)\) tiene que satisfacer
para representar a una evolución físicamente realizable. Sea \(\rho\) el
estado inicial, y \(T(\rho)\) el estado evolucionado. \(T\) debe mandar
estados a estados, y por lo tanto tiene que ser lineal \footnote{La
  interpretación de un \emph{ensamble} de estados
  \(\{(p_i, \dyad{\phi_i})\}\) es la de una distribución de probabilidad
  \((p_i)\) sobre un conjunto de estados posibles \(\{\ket{\phi_i}\}\).
  Por lo tanto es razonable esperar que después de la evolución el
  ensamble se haya transformado en \(\{(p_i, T(\dyad{\phi_i}))\}\), o
  que en otras palabras
  \(T(\sum_i p_i \dyad{\phi_i}) = \sum_i p_i T(\dyad{\phi_i})\).},
positivo y que preserve la traza. Es un hecho sorprendente pero muy
importante que la positividad no sea suficiente para dar a \(T\) una
interpretación física consistente. Imaginemos extender nuestro sistema
con uno auxiliar, con su propio espacio de Hilbert \(\mathcal K\) y
estado \(\tau\). Entonces el estado del conjunto de los dos sistemas es
\(\rho \otimes \tau \in \bounded(\hs \otimes \mathcal K)\). Asumimos
también que la evolución de \(\tau\) sea trivial. ¿Existe una aplicación
\(\tilde T\) que extienda \(T\) en \(\bounded(\hs \otimes \mathcal K)\),
de manera que \(\tilde T(\rho \otimes \tau) = T(\rho) \otimes \tau\)
para todos los estados \(\rho\) y \(\tau\)? Esa función existe y está
dada por el producto tensorial de \(T\) con la función identidad, y se
denota por \(T \otimes \identity\).

Si esta fuese una evolución física, debería ser otra vez positiva. Pero
sorprendentemente hay aplicaciones \(T\) positivas tales que
\(T \otimes \identity\) no es positiva. Por esta razón necesitamos pedir
la condición más fuerte de que \(T\) sea completamente positiva:
recordamos que una aplicación \(T:\matrixalg_d\to\matrixalg_{d^\prime}\)
es completamente positiva si \(T\otimes \identity_n\) es positiva para
todo \(n\in \NN\), donde \(\identity_n\) es la identidad de
\(\matrixalg_n\). Una aplicación completamente positiva y que preserva
la traza se suele llamar un \emph{canal cuántico}, y será nuestro objeto
principal de estudio.

Vamos ahora a justificar por qué los canales cuánticos realmente
representan una evolución motivada físicamente. Dado que nuestro interés
está en sistemas no aislados, imaginemos tener un espacio ambiente que
pueda interactuar con nuestro sistema original. Indicamos el estado del
ambiente con \(\ket\phi\) (y podemos asumir, sin pérdida de generalidad,
que sea un estado puro). El conjunto sistema-ambiente es un sistema
aislado, y por lo tanto evoluciona con una unitaria \(U\). Al final de
la evolución, descartamos el ambiente y sólo miramos a la matriz de
densidad reducida de nuestro sistema original, que llamaremos
\(\rho^\prime\). La secuencia de operaciones que hemos descrito tiene la
forma siguiente:

\begin{equation}
\label{es:eq:sequence} \rho \to \rho \otimes \proj \phi \to U (\rho
\otimes \proj{\phi}) U^\dg \to
\tr_E[ U (\rho\otimes \proj{\phi}) U^\dg ] = \rho^\prime
\end{equation}

\noindent donde \(\tr_E\) es la traza parcial sobre el ambiente. Nótese
que cada paso es una aplicación completamente positiva y que preserva la
traza : por lo tanto la evolución resultante \(\rho \to \rho^\prime\) es
un canal cuántico.

Por lo tanto, si consideramos nuestro sistema acoplado con un espacio
ambiente, y que los dos evolucionan juntos como un sistema aislado,
llegamos a la conclusión de que la evolución del sistema original está
dada por un canal cuántico. Esto es realmente lo único que puede pasar:
todo canal cuántico se puede interpretar de esta manera, como la
restricción a un subsistema de una evolución unitaria en un sistema más
grande. Este es el contenido del teorema de dilatación de Stinespring.

\begin{stinespring} Sea $T: \matrixalg_{d} \to \matrixalg_{d}$ una
aplicación completamente positiva y que preserva la traza (un canal
cuántico). Entonces existen una unitaria $U \in \matrixalg_{d^2}$ y un
vector normalizado $\phi \in \CC^{d}$ tales que \begin{equation} T
(\rho) = \tr_E[ U \qty(\rho \otimes \proj{\phi}) U^\dg ]
\end{equation} \end{stinespring}

Por lo tanto, la evolución que representa un canal cuántico se justifica
físicamente como la restricción de una evolución unitaria que actúa en
un sistema más grande: eso son, en cierto sentido, las evoluciones
efectivas inducidas en subsistemas por evoluciones unitarias.

\subsubsection{Límite de acoplamiento
débil}\label{luxedmite-de-acoplamiento-duxe9bil}

Hasta ahora hemos considerado \emph{una sola} aplicación de un canal
cuántico, interpretándola como un paso temporal individual de una
evolución dinámica, y hemos visto que, por el teorema de Stinespring, es
equivalente a acoplar el sistema con un ambiente y considerar una
evolución unitaria conjunta. La ventaja es, claramente, que a menudo es
más simple razonar ignorando la evolución interna del ambiente y
considerar únicamente su efecto en el sistema que nos interesa.

Pero un sistema dinámico es más que una sola aplicación de un paso
temporal: es una composición secuencial de tales pasos, o una
descripción a tiempo continuo en la cual cada instante temporal
\(t \ge 0\) produce una evolución físicamente realizable \(T_t\).
Matemáticamente tenemos un problema: al trazar el ambiente, hemos
perdido toda correlación que la unitaria \(U\) podría haber creado entre
el sistema y su ambiente, tanto en la forma de entrelazamiento cuántico
como en la forma de correlaciones clásicas. Sólo hemos conservado una
sombra de ellas en el estado mixto resultante, pero la pérdida es
irreversible. El proceso descrito por la \cref{es:eq:sequence} no se
puede componer de manera natural: el ambiente ha cambiado porque ha
evolucionado junto con nuestro sistema.

Aunque técnicamente correcta, quizás nuestra descripción matemática no
es del todo relevante para describir sistemas reales. Resulta que en
algunos casos, el efecto del sistema en el ambiente es despreciable, y
se puede aproximar asumiendo que el ambiente no evoluciona. Imaginemos
por ejemplo que el ambiente sea un baño térmico de cierta temperatura:
seguramente la interacción con el sistema cambiará el equilibrio y la
temperatura del ambiente, pero si este es mucho más grande que el
sistema no será una mala aproximación asumir que la temperatura es
constante a lo largo de la evolución.

Matemáticamente, esto significa que si
\(U \qty(\rho \otimes \proj{\phi}) U^\dg\) no es demasiado distinto de
\(\rho^\prime \otimes \proj{\phi}\), podemos sustituir uno por el otro:
en la literatura física esto se llama \emph{límite de acoplamiento
débil} o \emph{aproximación de Born}
\cite{davies1974, pule1974, accardi1990, breuer2002,
Rivas2012}. En cada paso ``infinitesimal'', el ambiente se elimina y se
reemplaza por uno nuevo e idéntico al original, que por lo tanto no
contiene ninguna información sobre la evolución previa del sistema. Por
esta razón, este tipo de evolución se llama también \emph{Markoviana}.

Esto nos permite considerar el siguiente sistema dinámico: la evolución
del sistema está descrita por un \emph{semigrupo} \footnote{O de
manera más exacta, de una representación del semigrupo $(\RR_{+}, +)$.}
de canales cuánticos
\(\{T_t: \matrixalg_d \to \matrixalg_d\}_{t\ge 0}\), tales que \(T_0\)
es la aplicación identidad. La propiedad de semigrupo
\(T_t T_s = T_{s+t}\) implica que la evolución es homogénea y
Markoviana. Como en la teoría clásica de semigrupos dinámicos, si
\(T_t\) es fuertemente continuo en \(t\) (\(T_t\) es un
\(C_0\)-semigrupo), entonces tiene un generador infinitesimal \(\lind\),
que verifica las relaciones siguientes:

\begin{equation}
\label{es:eq:lindblad-generator} \lind(x)= \dv t \eval{T_t(x)}_{t=0} =
\lim_{t \downarrow 0} \frac{1}{t}(T_t - \identity)(x).  \end{equation}

Nótese que para sistemas de dimensión finita, la continuidad fuerte
implica la continuidad uniforme, y que por lo tanto podemos escribir
\[ \dv t T_t = \lind T_t \qcomma T_t = \exp(t \lind). \]

Una generalización de esta aproximación es considerar un ambiente que
evoluciona en el tiempo, pero de manera independiente del sistema (por
su propia dinámica interna, o quizás porque hemos aproximado el efecto
del sistema sobre el ambiente de esta manera). Esto nos lleva a
considerar co-ciclos en vez de semigrupos, es decir familias
\(\{T_{t,s}\}_{0\le s \le t}\) de canales cuánticos que satisfacen la
propriedad \(T_{t,s} T_{s,k} = T_{t,k}\) para todo \(k \le s \le t\).
Podemos definir generadores de co-ciclos (que serán dependientes del
tiempo) de manera similar a como se hizo para semigrupos: dejaremos
fuera de este trabajo estas evoluciones no homogéneas, pero las
mencionamos por completitud.

\subsubsection{Generadores
Lindbladianos}\label{generadores-lindbladianos}

Hemos visto que la evolución del estado \(\rho\) bajo un semigrupo de
canales cuánticos \(T_t\) está dada por la solución de la ecuación
diferencial \(\dot{\rho}(t) = \lind \rho(t)\), donde
\(\rho(t) = T_t(\rho)\). El superoperador \(\lind\) se llama también
\emph{Liouvilliano}, dado que esta ecuación es una generalización de la
ecuación de Liouville-von Neumann. No puede ser cualquier operador, dado
que hemos impuesto algunas restricciones sobre el semigrupo que genera
(es un semigrupo de canales cuánticos). Lindblad \cite{lindblad},
Kossakowski, Gorini, y Sudarshan \cite{kossakowski} probaron que tales
generadores tienen una forma particular, llamada forma de
Lindblad-Kossakowski, y \(\lind\) se suele llamar Lindbladiano.

\begin{thm} Sea $\lind: \matrixalg_d \to \matrixalg_d$.  Los
siguientes hechos son equivalentes
\begin{enumerate} \item $\lind$ es
generador de un semigrupo dinámico de canales cuánticos;
\item existen una aplicación completamente positiva $\phi:\matrixalg_d \to
\matrixalg_d$ y una matriz $\kappa \in \matrixalg_d$ tal que
\begin{equation}
\lind (\rho) = \phi(\rho) - \kappa \rho - \rho
\kappa^\dg; \quad \phi^*(\identity) = \kappa + \kappa^\dg.
\end{equation}
\item existen una matriz hermítica $H$ y un conjunto de
matrices $\{ L_j \in \matrixalg_d \}_{j=0,\dots,d^2-1}$ tales que
\begin{equation}\label{es:eq:lindblad}
\lind(\rho) = - i \comm{H}{\rho} + \sum_{j=0}^{d^2-1} L_j \rho L_j^\dg
- \frac{1}{2} \acomm{L_j^\dg L_j}{\rho}; \end{equation}
donde $\acomm{a}{b} = ab + ba$ es el anticonmutador.
\end{enumerate}
\end{thm}

A \(H\) se le llama (por razones obvias) el hamiltoniano, mientras que a
las matrices \(L_j\) se les llama \emph{operadores de salto} o \emph{de
Lindblad}.

Por el teorema de Russo-Dye \cite{Russo1966}, si
\(T:\matrixalg_d \to \matrixalg_d\) es una aplicación positiva y que
preserva la traza, entonces es una contracción con respecto a la norma
traza: de hecho, por dualidad con respecto al producto escalar de
Hilbert-Schmidt, \[
\norm{T}_{1\to 1} = \norm{T^*}_{\infty \to \infty} =
\norm{T^*(\identity)}_\infty = \norm{\identity}_\infty = 1, \] dado que
el dual de una aplicación que preserva la traza es una aplicación que
preserva la identidad. Por lo tanto, los autovalores de un canal
cuántico están en el disco unidad complejo. Usando el cálculo funcional,
se puede ver que esto implica que los autovalores de \(\lind\) están
contenidos en el semiplano \(\{ z \in \CC \,|\, \Re z \le 0\}\).

Los autovalores de \(\lind\) que se quedan en el eje imaginario
corresponden a autovalores de \(T_t\) en la frontera del disco unidad, y
por lo tanto forman el llamado \emph{espectro periférico}. Se puede
probar que los bloques de Jordan asociados tienen dimensión 1, y
corresponden a estados periódicos de la evolución, mientras que los
estados estacionarios corresponden al autovalor 1.

Para cualquier otro autovalor \(\lambda\) de \(\lind\), al tener parte
real estrictamente negativa, la acción de \(T_t\) en el autoespacio
generalizado correspondiente es la de una contracción exponencial en el
tiempo: el subespacio es aniquilado por un factor
\(\exp(-t \Re \lambda)\). Por lo tanto, es el autovalor con la parte
real no-nula más grande el que determina la razón de convergencia más
lenta de \(T_t\) hacia una aplicación \(T_\infty\) que proyecta en el
espacio generado por los estados periódicos (y que sobre ese espacio
actúa como una unitaria). Esto justifica la siguiente definición:

\begin{defn}[Gap espectral] Definimos el \emph{gap espectral} de un
Lindbladiano $\lind$ como \begin{equation} \label{es:eq:spectral-gap}
\gap \lind = \min_{\lambda \in \sigma(\lind)\setminus\{0\}} \abs{\Re
\lambda}.  \end{equation} \end{defn}

Asumimos por un momento que no haya estados periódicos, o dicho de otra
manera que el espectro periférico esté trivialmente compuesto sólo del
autovalor 1. En este caso \(T_\infty\) es realmente una proyección.
Desde la descomposición de Jordan podemos ver que el gap espectral
controla la razón de convergencia en el tiempo al espacio de puntos
fijos, y que existe una constante \(c>0\) tal que

\begin{equation} \label{es:eq:gap-convergence} \norm{T_t(\rho) -
T_\infty(\rho)}_1 \le c \exp(-t \gap \lind), \end{equation}

\noindent
para todo estado inicial \(\rho\).

Volveremos a la \cref{es:eq:gap-convergence} más adelante, cuando
hablemos de familias de sistemas dinámicos definidos en una sucesión
creciente de retículos.

\subsubsection{Generadores locales}\label{generadores-locales}

Hasta ahora, sólo hemos considerado sistemas finitos, que pueden ser
considerados como un único cuerpo físico con su dinámica. La mayoría de
aplicaciones requieren en su lugar una descripción de un modelo \emph{a
muchos cuerpos}: un sistema compuesto de muchas piezas individuales, que
interactúan entre sí de una manera definida y con cierta regularidad. Si
tenemos en cuenta una única instancia de un modelo de muchos cuerpos,
matemáticamente hablando es lo mismo que considerar todo el sistema como
un solo cuerpo grande, con unos cuantos grados de libertad interna
evolucionando de acuerdo a las interacciones mencionadas.

Este punto de vista cambia si consideramos una sucesión de modelos de
muchos cuerpos definidos en una estructura de grafo o de retículo.
Recordamos la notación mencionada anteriormente. \(\Gamma\) será un
grafo infinito con la métrica de grafo, por ejemplo \(\ZZ^D\) para algún
entero \(D\). Asociamos a cada vértice \(x\) en el grafo un espacio de
Hilbert complejo de dimensión finita \(\hs_x\), y asumimos por
simplicidad que sean todos isomorfos (es decir, todos tienen la misma
dimensión \(d\)). Para cada subconjunto finito
\(\Lambda \subset \Gamma\) indicamos con
\(\hs_\Lambda = \otimes_{x\in \Lambda}\hs_x\), y
\(\alg_\Lambda = \bounded (\hs_\Lambda)\).

En este caso hay una noción bien definida de localidad: para cada par de
subgrafos finitos \(\Lambda_1 \subset \Lambda_2 \subset \Gamma\), existe
una inclusión natural de \(\alg_{\Lambda_1}\) en \(\alg_{\Lambda_2}\),
dada por la identificación de \(X \in \alg_{\Lambda_1}\) con
\(X \otimes \identity_{\Lambda_2\setminus\Lambda_1} \in \alg_{\Lambda_2}\).
Esto nos ha permitido definir la noción de \emph{soporte}: dado un
operador \(X \in \alg_{\Lambda}\) definimos el soporte de \(X\), y lo
indicamos como \(\supp X\), como el mínimo
\(\Lambda^\prime \subset \Lambda\) tal que existe un
\(X^\prime \in \alg_{\Lambda^\prime}\) de manera que
\(X = X^\prime \otimes \identity\). En cierto sentido, el soporte de
\(X\) es independiente de \(\Lambda\), dado que
\(\supp X = \supp X \otimes \identity\): considerar \(X\) actuando en un
conjunto más grande no incrementa su soporte.

Esta es la primera aparición de una idea simple pero potente que está en
la base de este trabajo: existen propiedades de los objetos que
estudiamos que no dependen del tamaño del sistema, con ser este de
tamaño suficiente para contenerlos. Si consideramos una sucesión
creciente y absorbente de \(\Lambda_n\nearrow \Gamma\) finitos, entonces
podemos estudiar propiedades que son uniformes en \(n\).

Las interacciones físicas suelen ser más débiles cuando la distancia
entre los cuerpos que interactúan se hacen más largas. Por lo tanto si
podemos descomponer el generador de la evolución \(\lind\) como una suma
de términos locales \(\sum_{Z\subset \Lambda} \lind_Z\), cada uno de los
cuales es nuevamente de la forma de Lindblad-Kossakowski pero solo actúa
en un subsistema \(Z\), es razonable pedir que sus normas sean mas
pequeñas al crecer el diámetro de su soporte \(Z\). En este caso diremos
que \(\lind\) es un \textbf{Lindbladiano local}.

Si no especificamos a qué velocidad \(\norm{\lind_Z}_\diamond\) decrece
con respecto a \(\diam Z\), cualquier Lindbladiano satisface esta
condición de manera trivial. Vamos a detallar más el decaimiento que
necesitaremos en el \cref{hipuxf3tesis-de-lieb-robinson}.

\subsubsection{Tiempo de equilibración y gap
espectral}\label{tiempo-de-equilibraciuxf3n-y-gap-espectral}

La \cref{es:eq:gap-convergence} describe las propiedades del sistema
dinámico descrito por \(T_t\) para tiempos largos: si \(t\) es más
grande que \(\log(\epsilon/c)/\gap \lind\) para algún \(\epsilon\)
positivo, entonces el conjunto de estados iniciales posibles ha sido
comprimido en un \(\epsilon\)-entorno del espacio de puntos fijos. El
tiempo mínimo para que esto pase (que podría ser más pequeño que el
tiempo dado por la \cref{es:eq:gap-convergence}) se llamará
\textbf{tiempo de equilibración} del sistema dinámico. Daremos la
definición formal sólo para sistemas sin puntos periódicos.

\begin{defn}[Tiempo de equilibración] Llamamos \emph{tiempo de
equilibración} de un sistema dinámico
$T_t:\matrixalg_d\to\matrixalg_d$ sin puntos periódicos a la
función $$ \tau(\epsilon) = \min\{ t > 0 : \sup_\rho \norm{T_t(\rho) -
T_\infty(\rho)}_1 \le \epsilon \},$$ donde el supremo se toma sobre
todos los estados $\rho$.  \end{defn}

Por lo tanto, podemos reformular lo que sabemos sobre el gap espectral
como sigue:

\begin{equation} \label{es:eq:spectral-gap-bound}
\tau(\epsilon) \le \frac{\log c -\log \epsilon}{\gap \lind}
\end{equation}

Para todo sistema de dimensión finita, este análisis suele ser
suficiente: más cuidado será necesario al considerar familias de
sistemas dinámicos definidas en secuencias crecientes de retículos
\(\Lambda_n\). En este caso queremos controlar el crecimiento de
\(\tau(\epsilon)\) con respecto a \(n\). Por emepzar, la cantidad
\(\lambda_n = \gap \lind_n\) puede decrecer al crecer \(n\), haciendo
diverger la cota al tiempo de equilibración. Si por el contrario la
cantidad \(\lambda = \inf \lambda_n\) está acotada lejos de cero,
diremos informalmente que el sistema tiene gap (en el sentido que tiene
gap en el límite).

Sin embargo, puede haber una razón más profunda para que la cota de la
\cref{es:eq:spectral-gap-bound} diverja en general en \(n\), aunque se
tenga un \(\lambda\) estrictamente positivo: la constante \(c\)
dependerá en general de \(n\) también. De hecho si obtenemos esa cota a
través de la descomposición de Jordan (aunque esta no sea necesariamente
la manera óptima para obtenerla), puede crecer mas rápido que una
exponencial en \(n\). Un análisis más cuidadoso puede mejorar esta
dependencia: en \cite{Wolf11} se demuestra que si \(\lind\) satisface
una condición llamada reversibilidad con respecto a un estado de rango
máximo \(\sigma\), que presentaremos con más detalle en el
\cref{gap-espectral-y-reversibilidad}, entonces podemos elegir \(c\)
igual a \(\norm{\sigma^{-1}}_\infty^{1/2}\), que a su vez es igual a
\(\sigma_{\min{}}^{-1/2}\), el mínimo autovalor de \(\sigma\). Esto nos
permite probar el siguiente resultado, del cual daremos después una
demostración alternativa:

\begin{thm}
\label{es:thm:spectral-gap-bound-db} Si $\lambda$ es el gap espectral
de $\lind$, que es reversible con respecto a un estado de rango
máximo $\sigma$, entonces \begin{equation}
\label{es:eq:spectral-gap-bound-db} \tau(\epsilon) \le
\frac{\log(\sigma_{\min{}}^{-1/2}) - \log \epsilon}{\lambda}
\end{equation} \end{thm}

Nótese que \(\sigma_{\min{}}^{-1/2}\) tiene que escalar por lo menos de
manera exponencial con el tamaño del sistema, o peor - por lo tanto de
la \cref{es:eq:spectral-gap-bound-db} podemos obtener como mucho una
cota polinomial al tiempo de equilibración. Si conocemos el espectro de
\(\lind\) y nada más, no es posible mejorar mucho la cota del
\cref{es:thm:spectral-gap-bound-db}: como se muestra en \cite{Szehr15},
si no tenemos más información que el espectro de \(\lind\), la
dependecia de \(c\) en el tamaño del sistema no puede ser mejorada en
una cota más lenta que una exponencial.

Para algunas aplicaciones, tener un tiempo de equilibración polinomial
es suficiente. En este trabajo, pediremos una condición más fuerte, que
gracias a \cite{Szehr15} sabemos que no puede ser garantizada solamente
a partir de la información sobre el espectro de \(\lind\): que
\(\tau(\epsilon)\) tenga un crecimiento \emph{logarítmico} en \(n\) (en
algunos casos podemos relajar esta hipótesis a un crecimiento
sub-lineal). Es la contribución principal de esta tesis probar que a
partir de esta condición podemos probar algunas propiedades muy
interesantes de la evolución y de su punto fijo.

Presentaremos tales resultados en el \cref{resumen-de-los-resultados}:
antes vamos a presentar una conexión importante con la desigualdad
logarítmica de Sobolev.

\subsection{Desigualdad logarítmica de
Sobolev}\label{desigualdad-logaruxedtmica-de-sobolev}

Las herramientas de hipercontractividad y desigualdad logarítmica de
Sobolev (en adelante, log-Sobolev) fueron introducidas como parte del
programa de Segal de dar rigor matemático a la Teoría Cuántica de Campos
\cite{Segal1960}. La desigualdad de log-Sobolev fue introducida por
primera vez por Feissner (en esa época, un estudiante de Leonard Gross,
que a su vez había sido estudiante de Segal) en su Tesis Doctoral
\cite{feissner1972gaussian,Feissner1975} con el fin de generalizar la
desigualdad clásica de Sobolev a medidas Gaussianas en dimensión
infinita. Luego Gross \cite{Gross1975} la usó para estudiar la
ergodicidad de procesos de Markov en dimensión infinita, y finalmente
fue reconocida como una herramienta útil para estudiar también procesos
en dimensión finita \cite{Diaconis1996}. La aplicación a sistemas de
espín clásicos fue introducida por primera vez por Holley, Stroock, y
Zegarlinski \cite{Holley1987,zegarlinski1990A,zegarlinski1990B,
Stroock1992D,Stroock1992C} y luego se convirtió en una herramienta
estándar en mecánica estadística. Está fuertemente relacionada con la
contractividad de semigrupos, y ha jugado un papel importante en muchas
áreas de las matemáticas.

Para una reseña moderna de la teoría clásica (conmutativa) de la
desigualdad de log-Sobolev y su conexión con semigrupos de Markov y
concentración de la medida, véase \cite{Guionnet2002}. Su generalización
cuántica ha sido desarrollada en una serie de artículos
\cite{qsd1,qsd2,qsd3,hypercontractivitylp}, y la conexión entre
equilibración rápida y desigualdad de log-Sobolev en el contexto de
partículas cuánticas ha sido introducida en \cite{Quantum-Log-Sobolev}.

La hipercontractividad es ligeramente anterior a la desigualdad de
log-Sobolev: el primer ejemplo de su uso se puede encontrar en un
trabajo de Nelson (también estudiante de
Segal)\cite{Nelson1973,nelson1966quartic}, aunque todavía no había sido
llamada así. Para una reseña del argumento véase
\cite{davies1992hypercontractivity,hyreview}. Está siendo reconocida en
la comunidad de información cuántica como una herramienta muy potente
\cite{Temme2014,King2014,Cubitt2015,Montanaro2012}.

Presentaremos ahora una versión simplificada de la teoría de
log-Sobolev, y su conexión con hipercontractividad y equilibración
rapida.

En esta tesis consideramos semigrupos de aplicaciones que preservan la
traza \(T_t\), y que por lo tanto describen la evolución de estados,
pero una descripción equivalente se obtiene de manera dual (con respecto
al producto escalar de Hilbert-Schmidt) al considerar la aplicación
\(T^*_t\), que describe la evolución de los observables, y el estado
límite \(\rho_\infty\) es invariante en el sentido que
\(\tr(\rho_\infty T^*_t(A))=\tr(\rho_\infty A)\) para todo operador
\(A\). En este caso el semigrupo preserva la identidad en vez de la
traza. Este enfoque es el que usualmente es elegido en la literatura
sobre desigualdades de log-Sobolev. Mantendremos nuestra notación, e
indicaremos la evolución de observables con \(T_t^*\), pero el lector
debería ser consciente de la diferencia.

\subsubsection{Gap espectral y
reversibilidad}\label{gap-espectral-y-reversibilidad}

Antes de presentar la definición de desigualdad de log-Sobolev, vamos a
reformular la \cref{es:eq:spectral-gap} de una manera distinta pero
equivalente, en el caso de considerar un estado de rango máximo
\(\sigma>0\). Dado un estado de este tipo, podemos definir un producto
escalar pesado sobre \(\bounded(\hs)\) como
\[\innerproduct{A}{B}_{\sigma} = \tr[\sigma^{1/2}A\sigma^{1/2}B] =
\innerproduct{\sigma^{1/4}A\sigma^{1/4}}{\sigma^{1/4}B\sigma^{1/4}}_{HS},\]
y la norma inducida correspondiente
\(\norm{\cdot}_\sigma = \innerproduct{\cdot}_\sigma^{1/2}\). Es fácil
ver que
\(\sigma_{\min{}}^{1/2} \norm{A} \le \norm{A}_\sigma \le \norm{A}\),
donde \(\sigma_{\min{}}\) es el autovalor mínimo de \(\sigma\). Además,
podemos definir una generalización de la varianza clásica, como
\[\variance_\sigma(A) = \norm{A -
\innerproduct{A}{\identity}_{\sigma}}^2_{\sigma} =
\tr[\sigma^{1/2}A\sigma^{1/2}A] - \tr[\sigma A]^2.\] En efecto,
\(\variance_\sigma(A)\) es positiva e invariante bajo traslaciones por
múltiplos de la identidad. De manera similar, dado un Lindbladiano
\(\lind\), podemos definir una generalización no-conmutativa de la forma
de Dirichlet: \[\dirichelet(A,B) =
\innerproduct{A}{-\lind^*(B)}_{\sigma} =
-\tr[\sigma^{1/2}A\sigma^{1/2}\lind^*(B)];\] donde \(\lind^*\) es el
dual de \(\lind\) bajo el producto escalar de Hilbert-Schmidt, es decir
\(\dirichelet(A,B) = -\tr[\lind(\sigma^{1/2}A\sigma^{1/2})B]\).
Escribiremos \(\dirichelet(A) = \dirichelet(A,A)\). Diremos que
\(\lind\) es \emph{reversible con respecto a \(\sigma\)} si
\(\lind(\sigma^{1/2}A\sigma^{1/2})=\sigma^{1/2}\lind^*(A)\sigma^{1/2}\)
para todo operador \(A\), y por lo tanto
\[\tr[\sigma^{1/2}A\sigma^{1/2}\lind^*(B)] =
\tr[\lind(\sigma^{1/2}A\sigma^{1/2})B]
=\tr[\sigma^{1/2}\lind^*(A)\sigma^{1/2}B] .\] En este caso,
\(\dirichelet\) es una forma bilinear simétrica, \(\lind^*\) es
autoadjunto con respecto a \(\innerproduct{\cdot}_{\sigma}\), y por lo
tanto
\(\hat \lind(\cdot) = \sigma^{1/4} \lind^*(\sigma^{-1/4} \cdot \sigma^{-1/4})\sigma^{1/4}\)
es autoadjunto con respecto al producto escalar de Hilbert-Schmidt. Dado
que \(\lind^*\) y \(\hat \lind\) están relacionados por una semejanza,
\(\lind^*\) tiene espectro real, y la contractividad del semigrupo
generado implica que es negativo. Nótese que también implica que
\(\sigma\) es un estado invariante por \(\lind\), dado que para todo
\(A\) se cumple que \[\tr[\lind(\sigma) A ] =
\tr[ \sigma^{1/2} \identity \sigma^{1/2} \lind^*(A)] =
\tr[ \sigma^{1/2} \lind^*(\identity) \sigma^{1/2} A ] = 0, \] y por lo
tanto \(\lind(\sigma) = 0\).

Si el espectro periférico de \(T_t\) es trivial, entonces el núcleo de
\(\lind\) tiene dimensión uno, y por el principio min-max de
Courant-Fischer-Weyl el segundo autovalor más pequeño de \(\lind\), que
antes hemos llamado el gap espectral, está dado por \[\gap \lind =
\min_{A: \innerproduct{A}{\identity}_{\sigma} = 0}
\frac{-\innerproduct{\lind^*(A)}{A}_\sigma}{\innerproduct{A}_{\sigma}}
= \min_{A:\variance_\sigma(A)\neq 0}
\frac{\dirichelet(A)}{\variance_\sigma(A)}.\]

Hemos reexpresado entonces la \cref{es:eq:spectral-gap} como un problema
variacional: \(\gap \lind\) es el valor máximo que puede tomar una
constante \(c\) tal que el funcional cuadrático
\(c \variance_\sigma(A)\) esté acotado superiormente por el funcional
cuadrático \(\dirichelet(A)\).

\begin{equation}
\label{es:eq:spectral-gap-variational} c \variance_\sigma(A) \le
\dirichelet(A) \end{equation}

Considérese ahora la evolución \(A(t)\) de un observable \(A\) bajo
\(\lind\), es decir \(A(t) = T_t^*(A)\). Dado que \(\sigma\) es
invariante por \(T_t\), se tiene que
\(\innerproduct{A(t)}{\identity}_\sigma = \innerproduct{A}{\identity}_\sigma\),
y por lo tanto
\(\lim_{t\to\infty} A(t) = \innerproduct{A}{\identity}_\sigma \identity\).
Esto implica que \(\variance_\sigma(A(t))\) es igual a
\(\norm{A(t) - A(\infty)}_\sigma^{2}\). Consideremos la función
\(t \to \variance_\sigma(A(t))\): su derivada está dada por
\(-2\dirichelet[A(t)]\). Por lo tanto la
\cref{es:eq:spectral-gap-variational} está realmente acotando la
derivada de \(\variance_\sigma(\cdot)\) por la función misma. Esto lleva
a la cota siguiente: \[ \variance_\sigma(A(t)) \le \variance_\sigma(A)
e^{-2ct}.\] Por lo tanto, el gap espectral controla la convergencia
cuando esta es medida por \(\variance_\sigma(\cdot)\). A su vez, esto
implica que \[ \norm{A(t) - A(\infty)} \le \sigma_{\min}^{-1/2}
\norm{A(t) -A(\infty)}_\sigma \le \sigma_{\min}^{-1/2}
\norm{A-A(\infty)} e^{-ct}.\] Por dualidad esto implica la siguiente
cota de la forma de la \cref{es:eq:gap-convergence}: \[\sup_{\rho}
\norm{T_t(\rho) - \sigma}_1 \le 2 \sigma_{\min{}}^{-1/2} e^{-t\gap
\lind}.\] o de manera equivalente

\begin{thm}
\label{es:thm:detailed-balance-gap-bound} Si $\lambda$ es el gap
espectral de $\lind$, entonces \begin{equation}
\label{es:eq:detailed-balance-gap-bound} \tau(\epsilon) \le
\frac{\log(2\sigma_{\min{}}^{-1/2}) - \log \epsilon}{\lambda}
\end{equation} \end{thm}

Nótese que \(\sigma^{-1}_{\min{}}\) escala \textbf{por lo menos}
exponencialmente con el tamaño del sistema (dado que tiene que ser por
lo menos más pequeño que \(1/\dim \hs_\Lambda\)), pero en principio
podría ser peor.

Podríamos haber obtenido la misma cota, pero sin el factor
multiplicativo de 2, usando el hecho que \cite{Temme2010,
Quantum-Log-Sobolev} \[\norm{\rho - \sigma}_1 \le
\variance^{1/2}_\sigma(\sigma^{-1/2}\rho\sigma^{-1/2}),\] y que
\(\sup_{\rho}\variance_\sigma(\sigma^{-1/2}\rho\sigma^{-1/2})\) es igual
a \(\norm{\sigma^{-1}}_\infty = 1/\sigma_{\min{}}\) (donde el supremo
está tomado sobre estados).

Hemos visto que la condición de reversibilidad nos permite expresar
exactamente la relación entre el gap espectral y el tiempo de
equilibración, obteniendo un prefactor bastante bueno a la cota (mucho
mejor de lo que podríamos haber obtenido a través de la descomposición
de Jordan). La desventaja es que hemos tenido que suponer que el punto
fijo sea único y de rango máximo (en este caso decimos que \(\lind\) es
\emph{primitivo}), y que tengamos algún control sobre
\(\sigma_{\min{}}\).

En seguida vamos a mostrar cómo en este contexto es natural definir
otras condiciones sobre \(\lind\) que nos permitirán tener un control
mejor sobre el tiempo de equilibración que el que obtuvimos de la cota
sobre el gap espectral. A su vez estas nuevas cotas serán suficientes
para probar la condición de equilibración rápida.

\subsubsection{Entropía y desigualdad de
log-Sobolev}\label{entropuxeda-y-desigualdad-de-log-sobolev}

La idea de la desigualdad de log-Sobolev y de otras desigualdades
entrópicas es de generalizar lo que hemos hecho en la sección anterior
con \(\variance_\sigma(\cdot)\): encontrar un funcional positivo
\(D(\cdot)\) que acote la convergencia de \(T_t\), luego acotar la
derivada \(\dv{t} D(T_t(\rho)-\sigma)\) en términos de la función misma,
comparándola con otro funcional definido en términos de \(\lind\).

Consideremos el funcional siguiente, que llamaremos \textbf{entropía
relativa} \[\ent{X}{Y} = \frac{1}{\tr X}\tr[X (\log X - \log Y)].\]
\(\ent{\rho}{\sigma}\) es positivo si \(\rho\) y \(\sigma\) son estados
normalizados, y es finito si el soporte de \(\rho\) está contenido en el
soporte de \(\sigma\). Decrece de manera monótona bajo la acción de
canales cuánticos \cite{OhyaPetz200405}. La desigualdad de Pisker
\cite{nielsen-chuang} implica que
\(\norm{\rho -\sigma}^2_1 \le 2 \ent{\rho}{\sigma}\). Al derivar
obtenemos que \[\dv{t}
\ent{\rho(t)}{\sigma} =
\tr[ \lind(\rho(t))(\log \rho(t)-\log \sigma)].\] Podemos por lo tanto
indicar con
\(\mcl K(\rho) = - \frac{1}{\tr \rho} \tr[ \lind(\rho)(\log \rho -\log \sigma)]\).
Compárese esta definición con la de \(\dirichelet\). Podemos entones
definir la siguiente desigualdad de tipo log-Sobolev:

\begin{equation}
\label{es:eq:log-sobolev-inequality} c
[ \ent{\rho}{\sigma} - \log \tr \rho] \le \mcl K(\rho)
\end{equation}

\noindent donde la constante óptima \(c>0\) será llamada constante de
log-Sobolev de \(\lind\), y la indicaremos con \(\alpha\). Como en el
caso de la desigualdad del gap espectral, podemos concluir que
\[\sup_{\rho}\norm{\rho(t) - \sigma}_1 \le \sup_{\rho}
\sqrt{2 \ent{\rho}{\sigma} } e^{-\alpha t}.\] Observamos que
\(\ent{\rho}{\sigma}\) está acotado por
\(\norm{\log \sigma^{-1}}_\infty = -\log(\sigma_{\min{}})\). Por lo
tanto, se obtiene una cota en la convergencia del semigrupo
exponencialmente mejor que la que obtuvimos con el gap espectral (ver la
\cref{es:eq:detailed-balance-gap-bound}):

\begin{thm} Sea $\alpha$ la constante de log-Sobolev de
  $\lind$. Entonces \begin{equation} \tau(\epsilon) \le \frac{
  \log(\log(\sigma^{-1/2}_{\min{}})) - 2\log \epsilon} {2\alpha}.
  \end{equation} \end{thm}

Si \(\sigma^{-1}_{\min{}}\) es exponencial en el tamaño del sistema y la
constante de log-Sobolev es uniforme en él, entonces el sistema tiene
equilibración rápida. Por lo tanto, la desigualdad de log-Sobolev es una
manera de probar esta hipótesis para sistemas reversibles.

Este es el enfoque elegido en \cite{MllerHermes2016,1505.04678}. En
\cite{hypercontractivitylp} y \cite{Quantum-Log-Sobolev} una cota
equivalente a la \cref{es:eq:log-sobolev-inequality} se denota como
desigualdad 1-log-Sobolev, y se obtiene al componer la
\cref{es:eq:log-sobolev-inequality} con la aplicación
\(\rho \to \sigma^{-1/2} \rho \sigma^{-1/2}\). Si indicamos con
\(A = \sigma^{-1/2} \rho \sigma^{-1/2}\) y con \(A(t)\) la evolución de
\(A\) bajo \(\lind^*\), es decir \(A(t) = T_t^*(A)\), entonces la
condición de reversibilidad implica que
\[A(t) = T_t^*( \sigma^{-1/2} \rho \sigma^{-1/2} ) =
\sigma^{-1/2} T_t(\rho) \sigma^{-1/2} = \sigma^{-1/2} \rho(t)
\sigma^{-1/2}.\]

Por lo tanto la \cref{es:eq:log-sobolev-inequality} puede reexpresarse
como sigue\footnote{Hemos eliminado un factor \(1/2\) de la definición
  original.} \[c \operatorname{Ent}_1(A) \le \dirichelet_1(A),\] donde
\(\operatorname{Ent}_1(A) = \ent{\rho}{\sigma} - \log \tr \rho\) y
\(\dirichelet_1(A)=\mcl K(\rho)\). Esta versión de la cota es claramente
equivalente a la \cref{es:eq:log-sobolev-inequality} si \(\lind\) es
reversible. Los autores de \cite{Quantum-Log-Sobolev} llaman a la
constante óptima \(\alpha_1\).

Desafortunadamente esto no es lo que se suele llamar desigualdad de
log-Sobolev en la literatura clásica (es decir, cuando todo lo anterior
se define para generadores de cadenas de Markov sobre un espacio de
probabilidad, que es el equivalente conmutativo de Lindbladianos sobre
estados cuánticos). En lugar de ello, la desigualdad clásica es más bien
parecida a la siguiente generalización: \[c
\operatorname{Ent}_2(A) \le \dirichelet(A) ;\] donde \(\dirichelet\) es
la forma de Dirichlet definida anteriormente,
\(\operatorname{Ent}_2(A) = \operatorname{Ent}_1(I_{1,2}(A))\),
\[I_{1,2}(A) = \sigma^{-1/2} \qty( \sigma^{1/4} A \sigma^{1/4} )^2
\sigma^{-1/2} = \sigma^{-1/4} A \sigma^{1/2} A \sigma^{-1/4}\] y por lo
tanto
\(\operatorname{Ent}_2(A) = \ent{\qty(\sigma^{-1/4} \rho \sigma^{-1/4})^2}{\sigma}\).

A esta cota se le llama desigualdad 2-log-Sobolev en
\cite{hypercontractivitylp,Quantum-Log-Sobolev,1505.04678} y a su
constante óptima \(\alpha_2\). Lamentablemente no sabemos si es
equivalente a la \cref{es:eq:log-sobolev-inequality}: bajo la hipótesis
adicional de que \(\dirichelet_1(I_{1,2}(A)) \ge \dirichelet(A)\)
(llamada \emph{\(L_p\)-regularidad} en \cite{hypercontractivitylp}), al
menos se puede probar que \(\alpha_2 \le \alpha_1\), recuperando el
resultado clásico. El hecho de que existan Lindbladianos que sean
reversibles pero que no sean \(L_p\)-regulares es todavía un problema
abierto.

\subsubsection{Hipercontractividad}\label{hipercontractividad}

Como hemos visto, a consecuencia del teorema de Russo-Dye, una
aplicación positiva y que preserva la traza \(T\) es \emph{contractiva}
con respecto a la norma de la traza, dado que \(\norm{T}_{1 \to 1} =1\)
- o de manera equivalente, una aplicación positiva y que preserva la
unidad \(T^*\) verifica que \(\norm{T^*}_{\infty \to \infty}\)=1, es
decir es contractiva con respecto a la norma \(\infty\). Esto se aplica
en particular a los canales cuánticos. Vamos ahora a introducir una
versión no conmutativa de las normas \(L_p\)
\cite{haagerup,qsd1,qsd2,qsd3}: dado un estado de rango máximo
\(\sigma\), para cada \(p\in [1,\infty)\) definimos
\[\norm{X}_{p,\sigma}^p = \tr\abs{ \sigma^{1/2p} X \sigma^{1/2p} }^p = \norm{ \sigma^{1/2p} X \sigma^{1/2p} }_p^p.\]
Se puede comprobar que \(\norm{\cdot}_{p,\sigma}\) es de hecho una
norma, y que se recuperan la propiedades usuales de los espacios
\(L_p\), como desigualdad de Hölder, dualidad, y teoremas de
interpolación. En particular, estas normas son crecientes en \(p\), y
por lo tanto para todo \(1 \le p \le q \le \infty\) se cumple que
\(\norm{\cdot}_{1,\sigma} \le \norm{\cdot}_{p,\sigma} \le \norm{\cdot}_{q,\sigma} \le \norm{\cdot}_{\infty}\).
Además, \(\lim_{p\to \infty}\norm{X}_{p,\sigma} = \norm{X}_\infty\) (la
norma \(\infty\) de Schatten usual). Nótese que la norma definida en la
sección anterior, que denotamos por \(\norm{\cdot}_\sigma\), corresponde
al caso \(p=2\). El espacio \(\bounded(\hs)\) con la norma
\(\norm{\cdot}_{p,\sigma}\) se denotará con \(L_p(\sigma)\), y la norma
de operador de una aplicación \(T: L_p(\sigma) \to L_q(\sigma)\) se
indicará con \(\norm{T}_{(p,\sigma) \to (q,\sigma)}\).

Consideremos entonces un canal cuántico \(T\) que tenga a \(\sigma\)
como su punto fijo, y su dual \(T^*\) con respecto al producto escalar
de Hilbert-Schmidt. Asumimos que \(T\) sea reversible con respecto a
\(\sigma\). Entonces sabemos que \(T^*\) preserva la unidad, y por lo
tanto \(\norm{T^*}_{\infty \to \infty} = 1\). Además, tenemos que para
cada operador \(A\)
\[ \norm{T^*(A)}_{1,\sigma} = \norm{ \sigma^{1/2} T^*(A) \sigma^{1/2} }_1 = \norm{ T(\sigma^{1/2} A \sigma^{1/2} ) }_1 \le \norm{ \sigma^{1/2} A \sigma^{1/2} }_1 = \norm{A}_{1,\sigma}, \]
donde hemos usado reversibilidad y el hecho de que
\(\norm{T}_{1\to 1}=1\). Podemos concluir entonces que
\(\norm{T^*}_{(1,\sigma) \to (1,\sigma)} = 1\), y por interpolación que
\(\norm{T^*}_{(p,\sigma) \to (p,\sigma)} = 1\) para todo
\(p\in [1,\infty]\).

Esto nos lleva a definir una nueva propiedad de una aplicación lineal
\(T:\bounded(\hs) \to \bounded(\hs)\): diremos que \(T\) es
\textbf{hipercontractiva} si existen \(p < q\) tales que
\(\norm{T}_{(p,\sigma)\to (q,\sigma)}\le 1\). En particular esto implica
que \(T\) sea contractiva con respecto a la norma
\(\norm{\cdot}_{(p,\sigma)}\).

Si consideramos un semigrupo dinámico de canales cuánticos \(T_t\),
entonces podemos considerar \(\norm{T^*_t}_{(p,\sigma)\to (q,\sigma)}\)
para algun \(p < q\) como una medida de convergencia del semigrupo: de
hecho para \(t=0\) tenemos que \(T_0 = \identity\) y por lo tanto
\(\norm{\identity}_{(p,\sigma) \to (q,\sigma)} =1\) si y sólo si
\(p\ge q\). Por otra parte, si \(\sigma\) es el único punto fijo de
\(T_t\), entonces \(T^*_\infty(X) = \tr(\sigma X) \identity\) y por lo
tanto
\(\norm{T_\infty^*(A)}_\infty = \tr(\sigma X) \le \norm{X}_{(1,\sigma)}\),
y \(\norm{T_\infty^*}_{(1,\sigma)\to \infty} = 1\).

Sea \(1< p < \infty\), y \(q\) su conjugado de Hölder, es decir
\(\frac{1}{p}+\frac{1}{q} =1\). Dado que \(T_t^*\) es auto-adjunto en
\(L_2(\sigma)\), se cumple que \[\norm{T^*_t}_{(p,\sigma)\to(2,\sigma)}
= \norm{T^*_t}_{(2,\sigma)\to(q,\sigma)},\] y por lo tanto si
\(1<p\le 2\) entonces \[\norm{T^*_{2t}}_{(p,\sigma)\to(q,\sigma)} \le
\norm{T^*_t}_{(p,\sigma)\to(2,\sigma)}\norm{T^*_t}_{(2,\sigma)\to(q,\sigma)}
= \norm{T^*_t}_{(2,\sigma)\to(q,\sigma)}^2.\]

Dada la observación anterior, nos centramos (como es común en la
literatura) en el comportamiento de
\(\norm{T_t^*}_{(2,\sigma)\to(q,\sigma)}\). La relación entre
desigualdad de log-Sobolev e hipercontractividad está contenida en el
teorema siguiente:

\begin{thm}[\cite{hypercontractivitylp}] Sea
$\lind$ un Lindbladiano reversible y que sea $L_p$-regular. Entonces
las condiciones siguientes son equivalentes \begin{enumerate} \item
Para $q(t) = 1 + e^{2 \alpha t}$,
$$\norm{T^*_t}_{(2,\sigma)\to(q(t),\sigma)} \le 1.$$ \item $T^*_t$
satisface una desigualdad 2-log-Sobolev con constante óptima $\alpha$.
\end{enumerate} \end{thm}

Obsérvese que el punto 1. del teorema anterior implica que, si
\(q(t) = 1 + e^{\alpha t}\), entonces
\(\norm{T^*_{t}}_{(p(t),\sigma)\to(q(t),\sigma)} \le 1\). Pasando al
límite para \(t\to\infty\) recuperamos que
\(\norm{T^*_{\infty}}_{(1,\sigma)\to(\infty,\sigma)} \le 1\).

\subsection{Ley de área}\label{ley-de-uxe1rea}

Otro problema interesante en el estudio de semigrupos disipativos es la
descripción del punto fijo, o estado invariante, de la evolución. Para
algunos modelos de ruido el punto fijo es el estado máximamente mixto,
proporcional a la identidad. Este estado representa la situación en la
cual el ruido ha destruido toda información sobre el sistema físico, y
cualquier medida producirá resultados uniformemente distribuidos. En
otros casos el modelo de ruido es distinto, y el estado invariante será
un estado térmico correspondiente a algún hamiltoniano, proporcional a
\(e^{-\beta H}\) para algún operador hermítico \(H\) y un \(\beta\)
positivo que representa el inverso de la temperatura. Este es el caso
por ejemplo de las aplicaciones de Davies \cite{davies1976quantum}. En
otros casos todavía la evolución es artificial y construida para tener
un estado particular como punto fijo: se considera con un estado que
queremos preparar, y de ahí derivamos un generador Lindbladiano que
produce ese estado como punto fijo. Este es el enfoque que tienen la
dinámica de Glauber clásica y el muestreo de Metropolis
\cite{martinelli1999lectures} y la Preparación Disipativa de Estados
\cite{verstraete09,Kraus08}.

Uno se esperaría que, si el estado satisficiera alguna propiedad
``buena'', la evolución resultante también tendría algunas propiedades
buenas, por ejemplo de convergencia rápida. Esto ha sido probado de
manera rigurosa en el caso de espines clásicos y dinámica de Glauber
\cite{martinelli1993finite,martinelli1999lectures}, donde la propiedad
``buena'' del estado \(\omega\) es de este tipo: dados dos observables
\(A\) y \(B\), soportados en regiones que distan \(d\), el valor de
\(\omega(A\otimes B)\) se acerca al de \(\omega(A)\omega(B)\) al crecer
\(d\). Mas precisamente, se requiere que la diferencia entre los dos
decaiga a cero exponencialmente rápido en \(d\). Esta propiedad se suele
llamar \emph{decaimiento exponencial de correlaciones}, dado que la
cantidad \(\omega(A\otimes B) - \omega(A)\omega(B)\) mide cómo de
correladas son las dos regiones. Bajo esta hipótesis, para sistemas de
espines clásicos se puede probar que la dinámica de Glauber
correspondiente tiene equilibración rápida (a través de una desigualdad
de log-Sobolev).

En esta tesis hemos afrontado el problema inverso: dado un Lindbladiano
con ``buenas'' propiedades, ¿qué propiedades del punto fijo podemos
derivar? Para empezar, presentaremos de manera rigurosa la noción de
correlaciones en sistemas de muchos cuerpos.

\subsubsection{Medidas de correlaciones}\label{medidas-de-correlaciones}

Considérese un estado bipartito \(\rho_{AB}\in \bounded (\hs_{AB})\). Si
\(\rho_{AB}\) es de la forma \(x \otimes y\) para algún estado \(x\) en
\(\bounded(\hs_A)\) y un estado \(y\) en \(\bounded(\hs_B)\), diremos
que es un estado producto. En este caso cada medida sobre el subsistema
\(A\) será independiente de las medidas en el subsistema \(B\), y
viceversa: por lo tanto las estadísticas obtenidas serán independientes
y no habrá correlaciones entre los dos subsistemas. Si \(\rho_{AB}\) no
es un producto, hay varias maneras distintas de cuantificar ``cuánto de
lejos'' está de ser un producto.

La notación siguiente está tomada de \cite{Kastoryano2013}. Denotaremos
con \(\rho_A\) (resp. \(\rho_B\)) el estado \(\tr_B \rho_{AB}\) (resp.
\(\tr_A \rho_{AB}\)).

\begin{defn}[Medidas de correlaciones]\hfill\\
\label{es:def:correlations} \begin{itemize} \item \emph{Correlaciones en
covarianza}: $$C(A : B) = \max_{\substack{M \in \alg_A,N \in \alg_B \\
\norm{M} \le 1, \norm{N} \le 1}} \abs{ \ev{M\otimes N} - \ev{M}\ev{N}
} \\ = \max_{\substack{M \in \alg_A,N \in \alg_B \\ \norm{M} \le 1,
\norm{N} \le 1}} \abs{
\trace[ M \otimes N (\rho_{AB} - \rho_A \otimes \rho_B)] } ;$$ donde
$\ev{O} = \trace(O \rho_{AB})$ es el valor esperado del observable $O$
con respecto a $\rho_{AB}$.

  \item \emph{Correlaciones en traza}: $$T(A : B) = \max_{\substack{F
      \in \alg_{AB} \\ \norm{F} \le 1}} \abs{
      \trace[F (\rho_{AB} - \rho_A \otimes \rho_B)] } =
      \norm{\rho_{AB} - \rho_A \otimes \rho_B}_1 .$$

    \item \emph{Correlaciones en información mutua}: $$ I(A : B) =
      S(\rho_A) + S(\rho_B) - S(\rho_{AB}) ;$$ donde $S(\rho) = -
      \trace (\rho \log_2 \rho)$ es la entropía de von Neumann del
      estado $\rho$.  \end{itemize} \end{defn}

En la teoría de la materia condensada, las correlaciones suelen medirse
con \(C(A:B)\). De la definición se sigue de manera inmediata que
\(C(A:B)\) está acotada superiormente por \(T(A:B)\) (dado que
\(C(A:B)\) sólo depende de medidas con observables productos, mientras
que \(T(A:B)\) permite operadores más generales).

La relación entre la distancia en traza y la información mutua está
dada, en una dirección, por la desigualdad de Pinsker
\cite{nielsen-chuang}, y en la otra por una aplicación de las
desigualdades de Alicki-Fannes-Audenaert \cite{2007JPhA...40.8127A,
MR0345574,MR2043448}. Sintetizamos las dos como sigue:

\begin{thm}\label{es:thm:fannes-mutual} \begin{equation}
\label{es:eq:fannes-mutual} \frac{1}{4} T(A:B)^2 \le I(A:B) \le 6
T(A:B) \log_2 d_A + 4 h_b( T(A:B)); \end{equation} donde $h_b(x) = -x
\log_2 x - (1-x) \log_2(1-x)$ es la función de entropía binaria, y
$d_A = \dim \hs_A$.  \end{thm}

\subsubsection{Correlaciones en sistemas de muchos
cuerpos}\label{correlaciones-en-sistemas-de-muchos-cuerpos}

En el caso de sistemas de muchos cuerpos, consideremos el punto fijo
\(\rho_\infty\) de \(\lind\) en \(\Lambda\), y para cada región
\(A \subset \Lambda\) o cada par de regiones \(A, B \subset \Lambda\)
consideremos la matriz reducida
\(\rho_A = \tr_{\Lambda\setminus A} \rho_\infty\) y
\(\rho_{AB} = \tr_{\Lambda\setminus A \cup B} \rho_\infty\). Podemos
entonces preguntar dos tipos de cuestiones (donde usaremos \(I(A:B)\)
pero hubiesen sido igualmente interesantes para cualquier otra medida de
correlaciones):

\begin{itemize} \item Dados $A,B \subset \Lambda$,
¿cómo escala $I(A:B)$ con respecto a $\dist(A:B)$?  \item Dado
$A\subset \Lambda$, ¿cómo escala $I(A:A^c)$ con respecto al tamaño de
$A$?  \end{itemize}

Aunque sean preguntas similares, en el primer caso sólo consideramos
regiones finitas, mientras que en el segundo consideramos \(A^c\), que
crece al crecer \(\Lambda\). Entonces no debería sorprender que la
condiciones necesarias para dar una respuesta a la primera pregunta sean
menos restrictivas que por la segunda. En el primer caso hablamos de
\textbf{decaimiento de correlaciones}: esperamos que al considerar
regiones \(A\) y \(B\) más alejadas, estas se vuelvan más
independientes.

La segunda pregunta es interesante por lo siguiente. Para estados
aleatorio elegidos con la medida de Haar, \(I(A:A^c)\) es proporcional a
\(\abs{A}\). Por el otro lado, muchos estados de interés físico tienen
un comportamiento muy distinto, y \(I(A:A^c)\) escala como
\(\abs{\partial A}\), donde \(\partial A\) se define como el subconjunto
de \(A\) de vértices que interactúan directamente con el complementario
de \(A\). Si las interacciones son finitas y \(A\) es una bola, entonces
\(\abs{A}\) es un polinomio de grado \(D\) mientras que
\(\abs{\partial A}\) es de grado \(D-1\). Esta situación se denomina
\textbf{ley de área} (con una terminología tomada del estudio de la
entropía de agujeros negros, donde la frontera es efectivamente una
superficie).

En lo que sigue trabajaremos con \(T(A:B)\) y con \(I(A:B)\), pero
recordamos que a causa del \cref{es:thm:fannes-mutual} el decaimiento
exponencial de una cantidad implica el decaimiento exponencial de la
otra.

\subsubsection{Estados fundamentales de
hamiltonianos}\label{estados-fundamentales-de-hamiltonianos}

El problema de estudiar decaimientos de correlaciones, leyes de área y
sus relaciones con la dinámica ha sido afrontado de manera extensa en el
contexto de estados fundamentales de hamiltonianos, aunque estas
relaciones no han sido totalmente determinadas aún. Un hamiltoniano es
un operador hermítico \(H\) sobre un espacio de Hilbert \(\hs\) que
representa un sistema físico. El generador \(\lind(\rho) = -i[H,\rho]\)
genera un grupo de automorfismos en vez de simplemente un semigrupo de
contracciones, y se puede ver como caso especial de la
\cref{es:eq:lindblad}. Dado que todo autovector de \(H\) es invariante
bajo la acción de \(\lind\) la evolución tendrá más de un punto fijo:
pero hay motivaciones físicas que justifican que los que corresponden al
autovalor mínimo de \(H\) tengan un papel especial, y son llamados
\textbf{estados fundamentales} de \(H\). Son estados puros. Dado un
estado puro \(\ket{\phi}_{AB}\), se verifica que
\(I(A:B) = 2 S(\phi_A)\), donde \(\phi_A = \trace_B \dyad{\phi}_{AB}\)
es la matriz de densidad reducida de \(\dyad{\phi}_{AB}\) sobre \(A\).
Por lo tanto, la información mutua se reduce a (dos veces) la entropía
de von Neumann.

La propiedad crucial en este contexto es el llamado \textbf{gap
espectral} de \(H\): la diferencia entre los dos autovalores más
pequeños de \(H\). Según convención, diremos que una familia de
hamiltonianos definidos en una secuencia creciente y absorbente
\(\Lambda_n \nearrow \Gamma\) tiene un \textbf{gap} si el gap es
uniformemente acotado por encima de cero - en otras palabras, si el gap
no se cierra en el límite. En caso contrario se dirá que el hamiltoniano
no tiene gap, y se puede estudiar con que velocidad el gap se cierra (si
polinomialmente o exponencialmente rápido en \(n\)).

En su artículo fundacional \cite{Hastings2006}, Hastings y Koma probaron
que si una familia de hamiltonianos locales tiene gap, entonces el
estado fundamental verifica un decaimiento exponencial de correlaciones
uniforme en \(n\). Este resultado es interesante porque conecta con la
teoría de fases de la materia condensada: una fase cuántica es una clase
de equivalencia de hamiltonianos, tal que dos hamiltonianos \(H_1\) y
\(H_2\) son equivalentes si se pueden conectar con un camino regular
\(H(t)\) de hamiltonianos con gap uniforme en \(t\). Las transiciones de
fases corresponden por lo tanto a los puntos del camino donde el gap se
cierra. En esa situación es común que la longitud de correlación
diverja, donde por longitud de correlación intendemos la distancia
\(\xi\) tal que \(C(A:B) \le e^{-\dist(A,B)/\xi}\).

Otra propiedad prevista por la teoría de materia condensada es que el
estado fundamental de hamiltonianos con gap verifique una ley de área
para la entropía de entrelazamiento. El argumento intuitivo (pero que no
constituye una prueba rigurosa), es el siguiente: si consideramos una
región finita \(A\), por el decaimiento exponencial de correlaciones los
espines que están dentro de \(A\) y lejos de la frontera serán casi
independientes de los que están fuera de \(A\). Por lo tanto, las
correlaciones y la entropía sólo pueden ser dadas por los espines que
están cerca de la frontera. Dado que cada espín de dimensión \(d\) sólo
puede contribuir con un factor \(\log d\) a la entropía total, se sigue
que esta escalará con el tamaño de la frontera.

Que este argumento se pueda hacer riguroso es el contenido de la
\textbf{conjetura de la ley de área} (que los estados fundamentales de
hamiltonianos con gap verifiquen una ley de área). Es un importante
problema abierto en la teoría de materia condensada y en los últimos
años ha visto importantes avances \cite{Eisert2010}. Una solución fue
obtenida para el caso de dimensión 1 por Hastings
\cite{HastingsAreaLaw}, y luego una demostración alternativa fue
presentada en \cite{brando2014,brando2013}, donde se probó que en 1D el
decaimiento exponencial de correlaciones implica una ley de área. Junto
con el resultado de Hastings y Koma \cite{Hastings2006}, esto implica
que un gap espectral, al implicar un decaimiento de correlaciones,
también implica una ley de área en 1D.

En dimensión más alta el problema sigue abierto. Algunos avances han
llegado de la comunidad de ciencias de la computación \cite{Arad2012},
con una nueva prueba del resultado de Hastings y Koma, que ha permitido
mejorar la dependencia de la longitud de correlación con el gap
espectral, produciendo unas cotas más fieles en los casos concretos para
los cuales somos capaces de calcular (analíticamente o numéricamente)
las dos cantidades. Las herramientas desarrolladas han permitido también
la construcción del primer algoritmo para aproximar estados
fundamentales de hamiltonianos con gap en 1D para el que es posible
probar una complejidad polinomial \cite{1307.5143v1}, así como otras
herramientas combinatorias para estudiar la estructura de los estados
fundamentales. Estos avances, aunque muy prometedores, todavía no han
llevado a una demostración de una ley de área para estados en dimensión
mayor o igual a 2.

\subsubsection{Estados de Gibbs y estados de redes de
tensoriales}\label{estados-de-gibbs-y-estados-de-redes-de-tensoriales}

Los estados de Gibbs o estados térmicos son estados proporcionales a
\(\exp(-\beta H)\), por algún hamiltoniano \(H\) y un parámetro
\(\beta\) que representa al inverso de la temperatura. Son interesantes
porque describen un sistema en equilibrio a temperatura finita
\(1/\beta\), y porque encajan de manera natural en el caso de dinámica
abierta: muchos de los modelos disipativos que hemos mencionado son
intentos de describir un proceso de termalización que lleva a un estado
de Gibbs. Por lo tanto, aunque no son los únicos puntos fijos posibles
de los sistemas disipativos, constituyen ciertamente una clase
importante. Cabe destacar que verifican una ley de área para la
información mutua \cite{Wolf2008}.

Otra clase importante de estados (esta vez puros) que a menudo
satisfacen una ley de área son los llamados estados de redes tensoriales
\cite{1603.03039v1} - estados cuyas amplitudes vienen dadas por la
contracción de una red de tensores. Para ser más específicos, en la gran
familia de estados de redes tensoriales, en 1D los \textbf{Matrix
Product States (MPS)} y en dimensión mayor o igual a 2 los
\textbf{Projected Entangled Pair States (PEPS)} satisfacen una ley de
área por construcción. El interés en ese tipo de estados es que
solamente requieren una cantidad polinomial de parámetros (en el número
de partículas) para describirlos, al contrario de la dimensión
exponencial del espacio de Hilbert en el cual viven. Por esta razón, son
ampliamente usados en el calculo numérico, y se cree que son buenas
aproximaciones de los estados fundamentales de hamiltonianos con gap.
Aunque haya ejemplos de estados en 2D que verifiquen una ley de área
pero que no sean aproximables por un PEPS \cite{1411.2995v2}, se probó
que bajo ciertas hipótesis en 2D el estado fundamental de un
hamiltoniano local se puede aproximar por un PEPS
\cite{Hastings2007,Molnar2015}.

En 1D, la situación es bastante más clara: los estados fundamentales de
hamiltonianos locales con gap se pueden aproximar de manera eficiente
con MPS. Esto no solo es un resultado teórico, sino que ha sido muy
importante a la hora de entender y desarrollar algoritmos que aproximen
estados fundamentales en 1D.

\subsubsection{Ley de área y longitud de
correlación}\label{ley-de-uxe1rea-y-longitud-de-correlaciuxf3n}

Hemos mencionado una demostración intuitiva -pero incompleta- que
conectaría una longitud de correlación finita con una ley de área. Hay
que mencionar al respecto un resultado riguroso presentado en
\cite{Wolf2008}. En ese trabajo los autores dan una noción distinta de
longitud de correlación para la información mutua: dada una región
finita \(A \subset \Lambda\), sea
\(B_R = \{ x \in \Lambda| \dist(x,A) > R \}\), y definimos
\(\xi_\Lambda\) como la longitud mínima \(R\) tal que

\begin{equation}
\label{es:eq:mutual-information-corr-length} I(A: B_R) < \frac{I(A :
B_0)}{2} \qcomma \forall A \subset \Lambda.  \end{equation}

(Nótese que \(B_0 = \Lambda \setminus A\).) Con esta definición, pueden
probar que \(I(A:A^C) \le 4 \abs{\partial A} \xi_\Lambda\), es decir una
ley de área.

Aunque el resultado es correcto, hay que tener cuidado en considerar la
relación entre la \cref{es:eq:mutual-information-corr-length} y el
decaimiento de correlaciones usual, es decir
\(I(A:B) \le c \exp(-\dist(A,B)/k)\). Se podría razonar que, en el caso
de decaimiento exponencial de correlaciones, al cumplirse que si
\(\dist(A,B_R) = R\) entonces \(I(A:B) \le c \exp(-R/k)\), entonces es
suficiente elegir \(\xi_\Lambda\) proporcional a \(k\) para verificar la
\cref{es:eq:mutual-information-corr-length}. Este argumento no funciona
de la forma esperada si la constante \(c\) que aparece en el decaimiento
de correlaciones no es independiente del tamaño de las regiones \(A\) y
\(B\), lo que suele ser el caso como veremos más adelante. Si \(A\) está
fijado y hacemos crecer \(\Lambda\), entonces el tamaño de \(B_R\) es
proporcional al tamaño total del sistema, con lo cual también
\(\xi_\Lambda\) crecerá con \(\Lambda\). La cota obtenida de esta manera
será todavía polinómica de grado menor que la dimensión geométrica del
retículo, pero será multiplicada por una constante que depende del
tamaño del sistema. Esta constante hará que en muchos casos la cota sea
trivial, dado que será más grande que la cota general dada por
\(\log \dim \hs_A = \abs{A} \log d\), donde \(d\) es la dimensión del
espacio de Hilbert de una única partícula.

Un problema similar se encontró en \cite{Kastoryano2013}, donde se
demostró bajo la hipótesis de una desigualdad de log-Sobolev una cota de
la forma \[ I(A:A^c) \le c \log \log \norm{\sigma^{-1}}
\abs{\partial A},\] donde \(\sigma\) es el punto fijo de la evolución.
Otra vez, el término de la derecha escala con el exponente adecuado a
una ley de área, pero la constante multiplicativa hace que esta cota sea
peor que la trivial en la mayor parte de los casos.

Uno de los resultados principales de esta tesis es demostrar por primera
vez una ley de área satisfactoria para puntos fijos de evoluciones con
equilibración rapida (véase el
\cref{ley-de-uxe1rea-con-correcciuxf3n-logaruxedtmica}).

\subsection{Estabilidad de sistemas
cuánticos}\label{estabilidad-de-sistemas-cuuxe1nticos}

Una de las propiedades de los sistemas cuánticos abiertos estudiada en
esta tesis es la estabilidad. Antes de presentar el resultado obtenido,
merece la pena explicar por qué es tan crucial. La estructura matemática
que estamos considerando es un intento de describir un sistema físico
compuesto de muchas partículas. Este podría ser o bien un sistema que
ocurre en la naturaleza (por ejemplo, la motivación original del modelo
de Ising fue la de estudiar la magnetización), o bien un sistema
artificial creado para cumplir con un trabajo (computación,
comunicación, memorias, preparación de estados, etc.)

En el primer caso, el modelo matemático será obviamente una aproximación
de la física real: sería poco razonable pedir que las cantidades
involucradas (constantes de interacción, niveles de energías,
masas/cargas/densidades, etc.) se puedan medir con precisión infinita.
La única expectativa realista es que se puedan medir con cierto nivel de
precisión. Una vez que introduzcamos esta información en nuestro modelo
matemático, nos gustaría tener una herramienta que sea capaz de predecir
los resultados de nuestros experimentos. Si estas predicciones cambian
de manera brusca al más mínimo cambio en los parámetros considerados,
entonces las predicciones que nos proporciona raramente tendrán algo que
ver con la realidad y el modelo resultará ser bastante inútil, dado que
requeriría un \emph{afinamiento} imposible para funcionar.

La situación es muy similar en el caso de sistemas artificiales. En este
caso, la hipótesis irrazonable es que se pueda tener control perfecto
sobre la implementación del modelo artificial, en el sentido de que
podemos configurar sus parámetros a cualquier nivel de precisión. Ningún
sistema real (ni siquiera macroscópico y clásico) puede ser controlado
de manera perfecta: la implementación en realidad siempre será como
mucho una aproximación fiel del modelo matemático. Si las evoluciones
resultantes dependen de manera crucial de estas pequeñas diferencias,
entonces acabaremos implementando una evolución muy distinta de la que
pensábamos, y con resultados diferentes. Por lo tanto los únicos modelos
prácticos son aquellos para los que pequeños errores en la
implementación darán a lugar a pequeños cambios en el sistema final.

En los dos casos, la justificación teórica de un modelo matemático
requiere que éste sea \textbf{estable contra perturbaciones}: podemos
obviamente hablar de modelos que no sean estables, pero hay que tener
mucho cuidado al considerar su implementación física y sus predicciones,
dado que en la práctica no seremos nunca capaces de verlas en la
realidad. Este razonamiento sólo se hace más fuerte al considerar,
además de errores experimentales, fuentes de ruidos físico: ningún
experimento estará perfectamente aislado, ningún ruido será
completamente eliminado.

Por lo tanto necesitamos herramientas que justifiquen la robustez de los
modelos físicos gracias a su estabilidad. En el caso de hamiltonianos
locales, el enfoque ha sido probar estabilidad del gap espectral, un
parámetro que tiene importantes consecuencias en las propiedades físicas
de los modelos correspondientes. Al partir del trabajo de
\cite{Bravyi2010,Klich2010} se llegó al de \cite{Michalakis2013}, en el
cual se probó que el gap espectral es estable (es decir que no se
cierra) bajo algunas condiciones físicamente razonables.

Hay que subrayar que estamos considerando un tipo especial de
perturbaciones: dado que trabajamos con modelos de muchos cuerpos, donde
cada partícula solo interactúa con sus vecinas, es natural considerar
que las perturbaciones y los errores involucren a \emph{cada término de
interacción local}. Por lo tanto, las perturbaciones serán pequeñas a
nivel microscópico y localmente despreciables, pero se sumarán al
considerar sistemas más grandes, y serán realmente perturbaciones no
acotadas (pero con mucha estructura local). Por esta razón no podemos
simplemente aplicar la teoría de perturbaciones estándar, sino que
tenemos que desarrollar técnicas específicas para este tipo de
perturbaciones.

Otro resultado importante de esta tesis es probar que los sistemas con
equilibración rápida son estables bajo perturbaciones (véase el
\cref{estabilidad-bajo-perturbaciones}).

\section{Resumen de los resultados}\label{resumen-de-los-resultados}

\subsection{Hipótesis}\label{hipuxf3tesis}

En este apartado presentaremos y discutiremos las hipótesis principales
hechas en este trabajo. La más importante de ellas es la de
equilibración rápida, una condición sobre el tiempo de convergencia del
sistema hacia su punto fijo.

Hablaremos de familias de generadores Lindbladianos
\(\{\lind^\Lambda\}_\Lambda\) donde \(\Lambda\) es una sucesión
creciente de subconjuntos finitos de \(\Gamma\). Para cada uno de ellos,
denotaremos con \(T_t^\Lambda\) la evolución correspondiente, es decir
\(T_t^\Lambda = \exp(t \lind^\Lambda)\).

\begin{defn}[Punto fijo único]
  Sea $\{\lind^\Lambda\}_\Lambda$ una familia de generadores Lindbladianos. Diremos que
  tiene un único punto fijo si, para todo $\Lambda$, $\lind^\Lambda$
  tiene un único punto fijo y ningún punto periódico (es decir, tiene un espectro periférico trivial).
\end{defn}

Indicaremos con \(T^\Lambda_\infty\) el proyector sobre el punto fijo de
\(T^\Lambda_t = \exp(t\lind^\Lambda)\) que preserva la traza, dado por
\(\lim_{t \to \infty} T_t^\Lambda\).

\subsubsection{Equilibración rápida}\label{equilibraciuxf3n-ruxe1pida}

Ya hemos descrito por qué el gap espectral sólo nos proporciona
información parcial sobre el tiempo de equilibración de un sistema
disipativo, mientras que la desigualdad de log-Sobolev permite tener un
control mayor (pero también requiere alguna propiedad más fuerte del
punto fijo). Nuestro enfoque será más directo, y pediremos simplemente
que el tiempo de equilibración escale de manera logarítmica con el
tamaño del sistema, dejando al lado el problema de cómo probar tal
condición.

\begin{defn}[Equilibración rápida]
        \label{es:defn:rapid-mixing}
        Sea $\{T_t^\Lambda\}_\Lambda$ una familia de aplicaciónes disipativas, donde $\Lambda$
        varía en una secuencia infinita de subconjuntos de $\Gamma$.
        Diremos que tiene equilibración rápida si existen
        $c, \gamma >0$ y $\delta \ge 1$ tales que
        \begin{equation}
                \label{es:eq:rapid-mixing}
                \sup_{\substack{\rho \ge 0 \\ \trace \rho\, =1 }} \norm{T^\Lambda_t(\rho)- T^\Lambda_{\infty} (\rho)}_1 \le c \abs{\Lambda}^\delta \ e^{-t \gamma} .
        \end{equation}
\end{defn}

Como hemos mencionado anteriormente, en algunos casos es posible relajar
la hipótesis de equilibración rápida: este caso está tratado en parte en
\cite[sec. 4.5]{Stability-paper}.

Probar que un modelo disipativo tiene equilibración rápida es difícil,
de la misma manera que lo es probar la existencia del gap espectral para
un sistema hamiltoniano. Aparte de casos ``simples'', como modelos sin
interacciones y preparación de \emph{graph states} \cite{Kastoryano12},
la otra clase importante de modelos que verifican esta propiedad son
Lindbladianos reversibles que satisfacen una desigualdad de log-Sobolev
\cite{Quantum-Log-Sobolev}, lo que incluye modelos clásicos como la
dinámica de Glauber para el modelo de Ising en el rango apropiado de
parámetros \cite{martinelli1999lectures}.

\subsubsection{Familias uniformes}\label{familias-uniformes}

Como ya hemos explicado, estamos interesados en estudiar el
comportamiento de algunas propiedades de los generadores Lindbladianos
\(\lind_n\), definidos en una sucesión creciente y absorbente de
retículos finitos \(\Lambda_n\) que convergen a un grafo infinito
\(\Gamma\) (en nuestro caso, \(\Gamma\) será \(\ZZ^D\), pero el mismo
razonamiento hubiese funcionado con cualquier otro grafo en el cual las
bolas crecen de manera polinómica con su diámetro). Pero al mismo
tiempo, dado que estamos interesados en modelos físicos, queremos que
distintos \(\lind_n\) representen ``el mismo'' sistema físico en escalas
distintas, de manera que estudiar esta sucesión realmente nos diga algo
sobre la física que estamos modelizando.

¿Qué significa que operadores definidos en retículos distintos
representen ``el mismo'' sistema físico? Obviamente la pregunta tiene
muchas respuestas posibles, pero intentaremos hacer algunas hipótesis
sobre una \emph{regla} o \emph{receta} para obtener, a partir de los
mismos ingredientes, todos los \(\lind_n\) en las distintas escalas.

Una hipótesis posible podría ser que todos los términos locales de cada
\(\lind_n\) sean simplemente el trasladado de un único generador local
\(l_0\): es decir, existe un \(r>0\) finito y un \(l_0\) que actua en
\(\alg_{b_0(r)}\) de manera que para todo \(n\) se cumpla
\[ \lind_n = \sum_{x : b_x(r) \subset \Lambda_n} l_x \] donde \(l_x\) es
el trasladado de \(l_0\) por el vector \(x\). A esta situación se le
suele denominar \textbf{invarianza translacional}, dado que en el límite
las interacciones son invariantes bajo translaciones de \(\Gamma\) (no
tiene sentido obviamente hablar de invarianza bajo traslaciones para
retículos finitos).

Se debería notar que esta es una restricción excesiva: no solamente
porque nos gustaría poder estudiar sistemas donde las interacciones
dependen de la posición en el retículo, sino porque también cerca de la
frontera de \(\Lambda_n\) el sistema deviene ``indeterminado'': dado que
no hay espacio para contener el soporte de \(l_0\) ahí, habrá cada vez
menos interacciones que involucren los vértices cercanos a la frontera.
A veces a esta situación se le llama \textbf{condiciones de contorno
abiertas}. Dado que estamos interesados en sistemas con un único punto
fijo, esta hipótesis puede ser especialmente problemática, puesto que la
indeterminación cerca de la frontera puede crear múltiples puntos fijos
- y estaremos pidiendo dos condiciones incompatibles entre sí.

Para superar esta limitación, hemos propuesto una definición que hemos
llamado \textbf{familias uniformes}, y creemos que puede ser una
definición lo suficientemente general para describir sucesiones
``interesantes'' de generadores Lindbladianos. Denotaremos con

\begin{equation} \partial_d \Lambda = \{ x \in \Lambda \,|\,
\dist(x,\Gamma \setminus \Lambda) \le d \}.  \end{equation}

Por convención, escribiremos \(\partial \Lambda\) en lugar
\(\partial_1 \Lambda\).

\begin{defn}\label{es:defn:boundary-condition}
  Sea $\Lambda \subset \Gamma$. Una \emph{condición de contorno} para $\Lambda$ viene dada
  por un Lindbladiano $\mcl B^{\partial \Lambda} = \sum_{d\ge 1} B^{\partial \Lambda}_d$,
  donde $\supp B^{\partial \Lambda}_d \subset \partial_d \Lambda $.
\end{defn}

En la definición de condición de contorno se usa una noción de localidad
distinta de la que hemos usado para definir los generadores locales: el
decaimiento en la norma es sólo necesario cuando las interacciones
entran \emph{dentro} del centro del sistema, mientras que se permite que
sean fuertes entre espines muy distantes siempre que estén a la misma
distancia de la frontera. Por ejemplo, si \(\Lambda\) es un cuadrado,
esta condición permite acoplar espines opuestos en la frontera, una
condición conocida como \emph{condiciones de contorno periódicas}, dado
que podemos imaginar haber envuelto \(\Lambda\) en un toro, de manera
que vértices opuestos en la frontera se vuelvan vecinos. Estas y otras
condiciones más exóticas pueden ser descritas por la definición dada
arriba.

\begin{defn}
  \label{es:def:uniform-family}
  Una \emph{familia uniforme} de Lindbladianos está dada por:
  \begin{enumerate}[(i)]
  \item \textit{interacciones centrales}: un Lindbladiano $\mcl M_Z$ para todo $Z \subset \ZZ^D$ finito;
  \item \textit{condiciones de contorno}: una familia de \textit{condiciones de contorno}
  $\{ \mcl B^{\partial \Lambda} \}_{\Lambda}$, para todo $\Lambda \subset \ZZ^D$ finito.
  \end{enumerate}
\end{defn}

Dada una familia uniforme de Lindbladianos así definida, para cada
\(\Lambda \subset \Gamma\) finito podemos definir dos generadores que
actúan en el:

\begin{align}
\mathcal L^\Lambda = \sum_{Z
\subset \Lambda} \mathcal M_Z &\quad \text{condiciones de contorno abiertas}
;
\\ \mathcal L^{\bar \Lambda} = \mathcal L^\Lambda + \mathcal
B^{\partial \Lambda} &\quad \text{condiciones de contorno cerradas}.
\end{align}

Al hablar de \(\lind^{\bar \Lambda}\), llamaremos a los términos
\(\mathcal M_Z\) interacciones centrales y a los
\(\mathcal B_d^{\partial \Lambda}\) interacciones de contorno.

Son necesarios unos comentarios sobre esta definición: no estamos
considerando una secuencia concreta de retículos crecientes
\(\Lambda_n\), sino que permitimos definir un Lindbladiano (y de hecho
dos) para todo \(\Lambda\) finito por el cual está dada una condición de
contorno.

Si elegimos dos \(\Lambda_1 \subset \Lambda_2 \subset \Gamma\) finitos,
y miramos a las interacciones que involucran las partículas en el centro
de \(\Lambda_1\), entendiendo por esto los vértices que están lejos de
\(\Gamma \setminus \Lambda_1\), entonces es facil ver que
\(\lind^{\overline \Lambda_1}\) y \(\lind^{\overline \Lambda_2}\) tienen
las mismas interacciones de corto alcance, y que sólo difieren en los
términos de largo alcance: o bien por el efecto de los términos de
\(\mathcal M_Z\) con \(Z\) que se extiende fuera de \(\Lambda_1\), o
bien por las diferencias entre \(\mathcal B^{\partial \Lambda_1}\) y
\(\mathcal B^{\partial \Lambda_2}\). Hablando informalmente, podemos
decir que los detalles microscópicos de las interacciones son los mismos
excepto por algún término de largo alcance. En la sección siguiente
asumiremos que la fuerza de las interacciones (la norma de los
operadores correspondientes), decrece en su alcance: por lo tanto, para
familias uniformes, la diferencia entre las interacciones centrales de
\(\lind^{\overline \Lambda_1}\) y \(\lind^{\overline \Lambda_2}\) será
pequeña. Esta es la propiedad fundamental y caracterizante de las
familias uniformes de Lindbladianos: a excepción de pequeños errores,
los detalles microscópicos de las interacciones centrales no dependen de
cómo de grande se ha elegido el sistema (con ser este lo suficientemente
grande para contenerlas).

\subsubsection{Hipótesis de
Lieb-Robinson}\label{hipuxf3tesis-de-lieb-robinson}

Hasta ahora, nuestra definición de Lindbladiano local es incompleta: si
no especificamos a qué ritmo las normas de las interacciones decaen,
siempre podemos descomponer un Lindbladiano en una suma de términos
locales todos nulos menos el último con soporte en todo el espacio. Si
en vez de esto imponemos que las normas decrezcan en función del
diámetro del soporte obtenemos una condición altamente no trivial. Dado
que el ritmo de decaimiento está relacionado con una propiedad que
presentaremos en seguida, llamada velocidad de Lieb-Robinson, llamaremos
estas condiciones \textbf{hipótesis de Lieb-Robinson}, y las daremos
sólo para las familias uniformes definidas anteriormente.

\begin{defn}[Hipótesis de Lieb-Robinson]
\label{es:def:lr-assumptions}
Existe una función \emph{creciente} $\nu(r)$ que verifica $\nu(x+y)\le \nu(x) \nu(y)$,
y constantes positivas $v$, $b$, y $c$, tales que:
\begin{equation}
  \label{es:eq:assumption-a1}
  \tag{A-1}
  \sup_{x\in \Gamma} \sum_{Z \ni x} \norm{\mcl M_Z}_\diamond \abs{Z} \nu(\diam Z) \le v < \infty,
\end{equation}
\begin{equation}
  \label{es:eq:assumption-a2}
  \tag{A-2}
  \sup_{x\in \Gamma} \sup_{r} \nu(r) \sum_{d = r}^N \norm{ B_d^{\partial b_x(N)}}_\diamond  \le c N^b.
\end{equation}
\end{defn}

Nótese que si \(\norm{\mcl M_Z}_\diamond\) decae exponencialmente en
\(\diam Z\) (o si es cero para todo \(Z\) de diámetro más grande que una
cierta constante, una situación llamada \textbf{interacciones de alcance
finito}) entonces podemos elegir \(\nu(r) = \exp(\mu r)\) para algún
\(\mu\) positivo. Si por el contrario decae polinomialmente, tenemos que
considerar funciones más lentas, como \(\nu(r) = (1+r)^\mu\). En
particular, si \(\norm{\mcl M_Z}_\diamond \sim (\diam Z)^{-\alpha}\),
entonces la \cref{es:eq:assumption-a2} se cumple si
\(\mu < \alpha - (2D+1)\), donde \(D\) es la dimensión geométrica de
\(\Gamma\) (lo que significa que \(\alpha\) tendrá que ser mayor que
\(2D+1\) para que la condición se cumpla).

La motivación tras estas hipótesis es que los sistemas que verifican
estas condiciones muestran una velocidad finita de propagación: el
soporte de un observable local crece en el tiempo, a menudo de manera
lineal, salvo por una cola exponencialmente pequeña. Esto implica que
las regiones que están separadas en el espacio, si están descorreladas a
tiempo cero, quedarán casi descorreladas por un tiempo finito, que
depende (a menudo de manera lineal) de su distancia. Daremos más
detalles sobre el tema en el \cref{cotas-de-lieb-robinson}

\subsubsection{Sistemas libres de
frustración}\label{sistemas-libres-de-frustraciuxf3n}

Otra condición que necesitaremos imponer en ciertos casos es la llamada
\textbf{ausencia de frustración}, parecida a una propiedad de
hamiltonianos con el mismo nombre. Merece la pena notar que es solamente
una condición sobre las interacciones centrales, y no sobre las de
contorno.

\begin{defn}
Diremos que una familia uniforme $\mcl L = \{ \mcl M, \mcl B\}$ es \emph{libre de frustración}
si para todo $\Lambda$ y todo punto fijo
$\rho_\infty$ de $\lind^{\bar \Lambda}$ se cumple
\begin{equation}
  \mcl M_Z (\rho_\infty) = 0 \quad \forall Z \subset \Lambda.
\end{equation}
\end{defn}

Hay muchos ejemplos interesantes y naturales de Lindbladianos que
verifican esta propiedad: entre ellos, los generadores de Davies y otros
tipos de muestreo de Gibbs para Hamiltoninanos conmutativos
\cite{arxiv1409.3435}, así como la preparación disipativa de estados
para PEPS.

\subsection{Herramientas}\label{herramientas}

Antes de presentar los resultados principales, haremos un resumen de
algunas herramientas usada en su demostración. Han sido desarrolladas
para este objetivo, pero pueden resultar igualmente interesantes en
otros contextos. Empezaremos con las cotas de Lieb-Robinson, una
herramienta estándar en los problemas de muchos cuerpos, y continuaremos
con unos resultados derivados de ellas, o simplemente inspirados.

En esta sección, daremos por hecho que el generador \(\lind\) verifica
las hipótesis \eqref{eq:assumption-a1} y \eqref{eq:assumption-a2}.

\subsubsection{Cotas de Lieb-Robinson}\label{cotas-de-lieb-robinson}

En los sistemas de muchos cuerpos se asume que las interacciones son
locales o quasi-locales: el espín en cada vértice del retículo sólo
puede interactuar directamente con sus vecinos (interacciones de alcance
finito), o si puede interactuar con espines mas alejados la fuerza de
esa interacción tiene que decaer rápidamente con la distancia. Por lo
tanto, la interacción entre espines distantes no es directa, sino
mediada por los espines intermedios que tienen que ``transmitir'' la
información. Es de esperar que dicha interacción no sea por lo tanto
inmediata, sino que tendrá un retraso, y más grande si la distancia
crece dado que más espines intermedios tendrán que verse involucrados.
Esto no es un efecto relativístico, dado que no hay velocidad finita de
la luz en nuestros modelos: una metáfora más exacta sería la de la
velocidad del sonido, que está dada por el medio en el cual la
información se propaga.

Esta visión intuitiva está formalizada por las cotas de Lieb-Robinson, y
la velocidad de propagación resultante se llama velocidad de
Lieb-Robinson. La primera demostración formal fue obtenida para sistemas
hamiltonianos y grupos de automorfismos, \cite{lieb1972,
robinson1968}, y por esta razón también se le llama velocidad de grupo.
Mas tarde fue generalizada a evoluciones disipativas
\cite{Nachtergaele12,Poulin10}.

Una consecuencia de las cotas de Lieb-Robinson y de la existencia de tal
velocidad es que, si consideramos una región finita \(A\) y modificamos
el generador de la evolución en unos vértices que están lejos de \(A\),
la evolución modificada será casi indistinguible de la original en
\(A\), por lo menos por un corto tiempo: antes de que la información
haya tenido tiempo de viajar desde los puntos donde se han hecho las
modificaciones hasta \(A\), los espines de \(A\) no ``saben'' que hubo
modificación alguna, y por lo tanto evolucionarán como si no hubiese
habido ninguna. Dado que la velocidad está dada por la hipótesis
\eqref{eq:assumption-a1}, será uniforme en el tamaño del sistema. Esto
ha permitido en \cite{Nachtergaele12} demostrar la existencia del límite
termodinámico.

El efecto de perturbar la dinámica en una región lejana viene dado por
el siguiente lema, que se deriva de la cota usual de Lieb-Robinson.

\begin{lemma}[{\cite[Lemma 5.4]{Stability-paper}}]
  \label{es:lemma:lieb-robinson-localization}
  Sean $\lind_1$ y $\lind_2$ dos Lindbladianos locales,
  y suponemos que $\lind_2$ verifica la hipótesis \eqref{eq:assumption-a1} con parámetros $v$ y $\nu(r)$.
  Consideramos un operador $O_X$ soportado en $X \subset \Lambda$, e indicamos con $O_i(t)$ su evolución bajo $\lind^*_i,\, i=1,2$.
  Suponemos que $\lind_1 - \lind_2 = \sum_{r\ge 0} M_r$,
  donde $M_r$ es un superoperador soportado en $Y_r$ y que se anula en $\identity$, con $\dist(X, Y_r) \ge r$.
  Entonces se verifica lo siguiente:
  \begin{equation}
    \norm{O_1(t) - O_2(t)} \le  \norm{O_X} \abs{X} \frac{e^{vt} -vt -1}{v} \sum_{r=0}^\infty  \norm{M_r}_{\diamond} \nu^{-1}(r) .
  \end{equation}
\end{lemma}

\subsubsection{Evoluciones abiertas y
cerradas}\label{evoluciones-abiertas-y-cerradas}

La definición de familia uniforme nos ha permitido definir dos
evoluciones, una con condiciones de contorno y una sin. La definición de
condiciones de contorno que hemos dado está justificada por el siguiente
resultado: si se verifican las hipótesis de la
\cref{es:eq:assumption-a1} y la \cref{es:eq:assumption-a2}, entonces el
efecto de la condición de contorno se difunde desde la frontera hacia el
centro del sistema con la misma velocidad finita de propagación de las
cotas de Lieb-Robinson. Por lo tanto, para tiempos cortos y observables
lejanos de la frontera, las dos evoluciones serán indistinguibles. Esto
ha sido probado usando el \cref{es:lemma:lieb-robinson-localization}.

\begin{lemma}[{\cite[Lemma 5.6]{Stability-paper}}]
\label{es:lemma:localizing-boundary}
Sea $O_A$ un observable soportado en $A\subset \Lambda$, y sea $O_A(t)$
(resp. $\bar O_A(t)$) su evolución bajo $\lind^{\Lambda *}$ (rep.,
$\lind^{\bar \Lambda *}$).
Sea $r = \dist(A,\Gamma\setminus\Lambda)$.  Entonces existen constantes
positivas $c$, $v$,
y $\beta$ tales que:
\begin{equation} \norm{ O_A(t) - \bar O_A(t)} \le
c \norm{O_A}\abs{A} \frac{e^{vt} - 1 -vt}{v} \nu^{-\beta}(r).
\end{equation} \end{lemma}

\subsubsection{Localización libre de
frustración}\label{localizaciuxf3n-libre-de-frustraciuxf3n}

Las cotas presentadas en esta sección son válidas para todo observable
localizado. El enunciado dual sobre la evolución de estados diría algo
sobre la evolución de cualquier estado que a tiempo cero se descompone
como un producto con respecto a esa región. Trabajando en el problema de
determinar una ley de área para la información mutua, hemos tenido que
afrontar un problema parecido pero distinto: ¿qué pasa para otros
estados que, aunque no tengan esta estructura de producto, satisfacen
alguna otra propiedad ``buena''? Mas exactamente, imaginemos preparar
nuestro sistema en \(\Lambda_n = b_0(n)\) en el estado \(\rho^\infty_n\)
que es el punto fijo de la evolución (con condiciones de contorno
cerradas) en \(\Lambda_n\), pero luego extendemos el sistema con un
estado arbitrario \(\tau\) en \(\Lambda_{n+1}\setminus \Lambda_n\) y
miramos la evolución del estado \(\rho^\infty_n\otimes \tau\) bajo el
generador definido en \(\Lambda_{n+1}\).

¿Qué podemos decir en este caso? Las cotas de Lieb-Robinson estándar no
nos dan ninguna información, dado que las regiones que estamos
considerando (\(\Lambda_n\) y \(\Lambda_{n+1}\setminus \Lambda_{n}\))
están a distancia cero, y no podemos asumir que \(\rho^\infty_n\) sea
producto (o casi producto) en ninguna otra bipartición del sistema. Por
otra parte, si asumimos que no haya frustración, la mayor parte de los
términos del generador serán cero sobre el estado considerado, y los
únicos que no son ceros son los que están cerca de la frontera de
\(\Lambda_n\). Entonces esperamos que la evolución sea aproximadamente
trivial en el centro de \(\Lambda_n\), y que la parte no trivial se
difunda de la frontera hacia el centro a la velocidad de Lieb-Robinson.

Esta idea intuitiva se formaliza rigurosa en el lema siguiente. Se debe
notar que, por lo que sabemos, esto no es una consecuencia directa de la
cota estándar de Lieb-Robinson. Para demostrarla hemos tenido que
reproducir las ideas y las técnicas de la demostración de las misma
cotas de Lieb-Robinson y adaptarlas a esta situación específica.

\begin{lemma}[{\cite[Lemma 12]{Area-Law-paper}}]
        \label{es:lemma:localization}
        Sea $ \mcl L = (\mcl M, \mcl B)$ una familia de Lindbladianos uniforme y
        libre de frustración.
        Sea $A \subset \Gamma$ una región finita. Fíjese un natural positivo $m$. Sea $B = A(m+1)$,
        $R = A(m+1)\setminus A(m)$ y
        $\rho_\infty^m$ el punto fijo de $T_t^{\bar A(m)}$ y $\tau$ un estado cualquiera en $R$. Entonces
        \begin{equation}
                \label{es:eq:localizing}
                  \norm{\left( T_t^{\bar B} - T_t^{ \bar B\setminus A} \right)(\rho_\infty^m \otimes \tau)}_1 \le
                  \poly(m) \nu^{-1}(m) \left[ e^{vt} -1 + t \right] ;
        \end{equation}
        donde $T_t^{\bar B \setminus A}$ es la evolución generada por
        \[ \mcl L^{\bar B \setminus A} = \sum_{Z \subset {B\setminus A}} \mcl M_Z + \sum_{d \le m+1} \mcl B^{\partial B}_d .\]
\end{lemma}

\subsection{Resultados principales}\label{resultados-principales}

Podemos ahora presentar los resultados principales obtenidos en
\cite{Stability-paper, Short-Stability-paper,Area-Law-paper}.

\subsubsection{Equilibración rápida
local}\label{equilibraciuxf3n-ruxe1pida-local}

\begin{defn}[Equilibración rápida local]
  \label{es:def:local-mixing}
Para $A \subset \Lambda$, definimos la \emph{contracción de $T_t$ con respecto a $A$} como
\begin{equation}
    \eta^A(T_t) := \sup_{\substack{\rho \ge 0 \\ \trace \rho\, =1 }}
    \norm{\trace_{A^c} \qty[ T_t(\rho) - T_{\infty}(\rho) ]}_1
    = \sup_{\substack{O_A \in \mathcal A_A \\ \norm{O_A}=1}}
    \norm{ T^*_t(O_A) - T^*_{\infty}(O_A) }.
\end{equation}
Diremos que $\mathcal L$ verifica \emph{equilibración rápida local} si, para todo $A \subset \Lambda$,
se cumple que
  \begin{equation}
    \label{es:eq:local-rapid-mixing}
    \eta^{A}(T_t) \le k(\abs{A}) e^{-\gamma t},
  \end{equation}
donde $k(r)$ crece polinomialmente en $r$, $\gamma >0$ y todas las contantes que aparecen
arriba son independientes del tamaño del sistema.
\end{defn}

Podemos igualmente definir un \textbf{tiempo de equilibración local}
como la inversa de \(\eta^A\):

\begin{equation}
    \tau^A(\epsilon) =\min \{ t>0 : \sup_\rho \norm{\trace_{A^c} [ T_t(\rho) - T_{\infty}(\rho) ]} \le \epsilon \}.
\end{equation}

La condición de equilibración rápida local implica que \(\tau^A\)
dependa de \(\abs{A}\) y \(\epsilon\), pero no de \(\Lambda\): esto
implica que, salvo por pequeños errores, los observables locales
convergen a su límite en una escala temporal que depende solamente del
soporte del observable y \textbf{no} del tamaño del sistema. Junto con
la cota de Lieb-Robinson, esto implica que los observables locales sólo
pueden interactuar con una región finita a su alrededor e independiente
del tamaño del sistema.

Parece sorprendente que solamente basándonos en una cota sobre el tiempo
de equilibración de una familia uniforme hayamos podido derivar una
propiedad tan fuerte, tal y como probamos en
\cite[Proposition 6.6]{Stability-paper}.

\begin{thm}
Para familias uniformes que satisfacen las hipótesis de Lieb-Robinson,
equilibración rápida implica equilibración rápida local.\end{thm}

\subsubsection{Indistinguibilidad local}\label{indistinguibilidad-local}

El resultado sobre equilibración local para observables puede en cierto
sentido ``dualizarse'' en una propiedad de la familia de los puntos
fijos. El razonamiento es el siguiente: el límite
\(O(\infty) = \lim_{t\to \infty} O(t)\) de la evolución de un observable
es el valor esperado con respecto al estado límite, es decir
\(O(\infty) = \trace(\rho_\infty O)\identity\), dado que:
\[ \trace \rho O(\infty) =
\lim_{t \to \infty} \trace \rho T_t^*(O) = \lim_{t \to \infty} \trace
T_t(\rho) O = \trace \rho_\infty O.\]

Por una parte las cotas de Lieb-Robinson implican que si \(O\) es un
observable local, entonces para tiempos cortos \(O(t)\) no depende de
las interacciones que están lejos de su soporte; por otra parte,
\(O(t)\) converge a \(O(\infty)\) en un tiempo independiente del tamaño
del sistema. Por lo tanto, \(O(\infty)\) sólo depende de las
interacciones que están cerca de su soporte a tiempo cero, y no de las
que están alejadas. Dado que \(O(\infty)\) es igual a
\(\tr \rho_\infty O\), esto implica que el observable \(O\) no puede
distinguir entre distintos puntos fijos de evoluciones definidas en
sistemas más grandes, dado que la única diferencia en las interacciones
está alejada del soporte de \(O\). El lema siguiente formaliza este
argumento \cite[Lemma 6.2]{Stability-paper}

\begin{lemma}
    \label{es:lemma:rapidmixing-localization}
    Sea $\mcl L = \{ \mcl M, \mcl B\}$ una familia uniforme de evoluciones disipativas
        con equilibración rápida, y asumamos que cada $T_t^{\bar \Lambda}$ tenga
        un punto fijo único y ningún otro punto periódico.
    Fíjese un $\Lambda$ y sea $\rho_\infty$ el único punto fijo de $T^{\bar \Lambda}_t$.
    Dado $A \subset \Lambda$, para todo $s \ge 0$ indicamos por $\rho^s_\infty$
        el único punto fijo de $T_t^{\bar A(s)}$.

    Entonces se cumple que:
    \begin{equation}
      \norm{ \trace_{A^c} ( \rho_\infty - \rho_\infty^s ) }_1 \le \abs{A}^\delta \Delta_0(s),
    \end{equation}
    donde $\Delta_0(s) = c (\abs{A(s)}/\abs{A})^{\delta v/(v+\gamma)} \nu^{-\beta \gamma /(v+\gamma)}$,
        y $c$ es una constante positiva mientras que $\beta$ y $v$ vienen dadas por el
        \cref{es:lemma:localizing-boundary} y $\delta$ y $\gamma$ por la \cref{es:defn:rapid-mixing}.
\end{lemma}

Esta propriedad es coherente con la idea que las familias uniformes
representen modelos en los que la dinámica microscópica está bien
definida y es independiente del tamaño del sistema.

\subsubsection{Estabilidad bajo
perturbaciones}\label{estabilidad-bajo-perturbaciones}

En el \cref{estabilidad-de-sistemas-cuuxe1nticos} hemos presentado la
importancia de la estabilidad bajo perturbaciones con el fin de
justificar a nivel teórico los modelos que consideramos. Vamos ahora a
dar una definición más formal de estabilidad. Dado que hay muchas
maneras distintas de definirla que pueden tener sentido en contextos
distintos, haremos una elección muy conservadora e impondremos las
mínimas hipótesis posibles en la perturbación, mientras que pediremos la
noción más fuerte de estabilidad. Hay naturalmente muchas maneras de
relajar este resultado.

Dada una familia uniforme de Lindbladianos \(\lind\), definida por sus
interacciones centrales \(\mcl M_Z\) y por sus condiciones de contorno
\(\mcl B_d\), consideraremos una perturbación al mismo tiempo de los
términos centrales \(\mcl M_Z^\prime = \mcl M_Z + E_Z\) y de las
condiciones de contorno \(\mcl B^\prime_d = \mcl B_d + E_d\). La
perturbación debería ser pequeña comparada con la norma del los
Lindbladianos originales, así que vamos a asumir que para todo \(E_Z\) y
\(E_d\) se cumpla
\(\norm{E_Z}_\diamond \le \epsilon \norm{M_Z}_\diamond\) y
\(\norm{E_d}_\diamond \le \epsilon \norm{B_d}_\diamond\), por algún
\(\epsilon >0\).

Nótese que una perturbación de este tipo es pequeña a nivel
microscópico, pero dado que actúa en cada término local, la suma
\(E^{\bar \Lambda} = \sum_{Z \subset \Lambda} E_Z + \sum_d E_d\) tiene
una norma que es divergente con el tamaño del sistema, y por lo tanto es
una perturbación no acotada una vez que olvidamos la estructura local.
Esto implica que no podemos simplemente aplicar la teoría de
perturbaciones estándar.

Todavía necesitamos alguna condición sobre la perturbación para que sea
``físicamente realista''. Sería suficiente pedir que \(\mcl M_Z^\prime\)
y \(\mcl B^\prime_d\) sean Lindbladianos, pero podemos relajar esta
condición y solamente pedir las condiciones siguientes (que se cumplen
de manera automática si los generadores perturbados son Lindbladianos):

\begin{itemize} \item $\du E_Z(\identity) = \du
E_d(\identity) = 0$; \item $S_t =
\exp[t(\lind^{\bar \Lambda} + E^{\bar \Lambda})]$ is a contraction for
each $t\ge 0$.  \end{itemize}

Con este modelo de perturbación, y bajo la hipótesis de equilibración
rápida, hemos podido probar el siguiente resultado de estabilidad
\cite[Theorem
6.7]{Stability-paper}.

\begin{thm}
  \label{es:thm:stability}
  Sea $\mathcal L$ una familia uniforme de Lindbladianos locales con un único punto fijo
  y con equilibración rápida.
  Sea $S_t$ definido como arriba.
  Para un observable $O_A$ soportado en $A  \subset \Lambda$, se cumple que para todo $t \ge 0$:
  \begin{equation}\label{es:eq:stability}
    \norm{T_t^{*}(O_A) - S_t^{*}(O_A)} \le c(\abs{A})\, \norm{O_A} \left( \epsilon + \abs{\Lambda} \nu^{-\eta}(d_A) \right),
  \end{equation}
  donde $d_A = \dist(A, \Lambda^c)$;
  $\eta$ es positivo e independiente de $\Lambda$;
  $c(|A|)$ es independiente de $\Lambda$ y $t$, y está acotado por un polinomio en $\abs{A}$.
\end{thm}

Comentamos el término de la derecha de la \cref{es:eq:stability}. El
factor multiplicativo \(\norm{O_A}\) es una normalización esperada, dado
que el término de la izquierda es 1-homogéneo. La constante
\(c(\abs{A})\) sólo depende del soporte de \(A\) y no de \(\Lambda\):
por lo tanto, si fijamos \(A\) y hacemos crecer \(\Lambda\), esta será
simplemente una constante fija. Discutiremos más adelante qué pasa
cuando no es este el caso.

La norma local de la perturbación es \(\epsilon\), así que no es
inesperado que aparezca en el lado derecho de la cota. Más sorprendente
es que sólo aparezca como un factor lineal. Recordamos nuevamente que
aunque la perturbación sea una suma de términos locales actuando en cada
vértice, \(\epsilon\) es la norma (diamante) de uno sólo de ellos y no
de toda la perturbación, y que por lo tanto es independiente del tamaño
del sistema.

El otro término \(\abs{\Lambda} \nu^{-\eta}(d_A)\) es un factor de
corrección de la frontera, que tiene en consideración el efecto de la
condición de contorno perturbada sobre el observable. Como es de
esperar, decae con la distancia entre \(A\) y la frontera, y tiende a
cero con \(\Lambda\) que tiende al retículo infinito \(\Gamma\). Por lo
tanto, para \(\Lambda\) suficientemente grande, será más pequeño que
\(\epsilon\), y por lo tanto despreciable. En algún caso intermedio este
factor puede parecer lejos de ser óptimo: por ejemplo, no se esperaría
un factor de este tipo en el caso de interacciones invariantes por
translaciones, aunque los observables estén localizados cerca de la
frontera. Al fin y al cabo, en este caso, la frontera es sólo una
necesidad matemática, y no corresponde a ninguna diferencia en las
interacciones físicas entre los espines. De hecho, podemos trasladar el
sistema para mover la frontera lo más lejos posible de \(A\), y por lo
tanto en esos casos se debería considerar \(d_A\) como la mitad del
diámetro del complementario de \(A\) en \(\Lambda\). Con esta
observación, el decaimiento del termino \(\nu^{-\eta}(d_A)\) cancelará
rápidamente la contribución de \(\abs{\Lambda}\) y el resultado será
despreciable con respecto a \(\epsilon\).

Hay que observar que si consideramos observables no-locales (es decir,
observables cuyo soporte crece con \(\Lambda\)), entonces el factor
\(c(\abs{A})\) no puede ser más pequeño que lineal, y en particular esto
implica que no se puede mejorar la cota para que no sea divergente en
\(\Lambda\). De hecho, es sencillo construir ejemplos simples de espines
sin interacciones tales que existen observables soportados en todo en
retículo y tales que el término de la izquierda de la
\cref{es:eq:stability} crece de manera lineal en el tamaño del sistema.
Esto implica en particular que el factor \(c(\abs{A})\) tiene que ser
por lo menos lineal. Véase \cite[Example 4.8]{Stability-paper} para
dicha construcción.

\subsubsection{Ley de área con corrección
logarítmica}\label{ley-de-uxe1rea-con-correcciuxf3n-logaruxedtmica}

Sobre el problema de decaimiento de correlaciones para el punto fijo de
la evolución, hemos obtenido el siguiente resultado \cite[Theorem
14]{Area-Law-paper}:

\begin{thm} Sea $\mathcal L$ una familia uniforme
de Lindbladianos locales con único punto fijo y equilibración local.
Entonces el punto fijo de cada $\lind^{\bar \Lambda}$
verifica: \begin{equation} \label{es:eq:correlation-decay} T(A : B) \le
3 (\abs{A} + \abs{B})^\delta \Delta_0\left(\frac{d_{AB}}{2}\right),
\end{equation} donde $\Delta_0$ es la función de decaimiento rápido dada en el
\cref{es:lemma:rapidmixing-localization}.  \end{thm}

A causa del \cref{es:thm:fannes-mutual}, \(I(A:B)\) tendrá el mismo
decaimiento.

Consideremos ahora la cuestión de si \(\rho_\infty\) verifica una ley de
área. No hemos podido dar una respuesta definitiva, pero hemos obtenido
un resultado ligeramente menos fuerte: \(I(A:A^c)\) escala como
\(\abs{\partial A} \log \abs{A}\), que en términos del radio de \(A\) va
como \(r^{D-1}\log r\). Para obtener tal resultado, tuvimos que hacer
alguna hipótesis más \cite[Proposition 16, Theorem 17]{Area-Law-paper}.

\begin{thm} \label{es:thm:area-law} Sea $\lind$ una familia uniforme
de Lindbladianos locales con único punto fijo y equilibración
rápida. Además, asumimos que $\lind$ verifique una de las condiciones
siguientes: \begin{itemize} \item $\rho_\infty$ es un estado puro para
todo $\Lambda$; \item $\lind$ es libre de frustración; \end{itemize}
Entonces se cumple lo siguiente para todo punto fijo de $\lind^{\bar
\Lambda}$, para alguna constante $c$ independiente de $\Lambda$:
\begin{equation} I(A:A^c) \le c \abs{\partial A} \log \abs{A}
\end{equation} \end{thm}

Es interesante notar que las dos condiciones alternativas en el
\cref{es:thm:area-law} son independientes entre sí. Por lo tanto,
sospechamos que su necesidad sea un efecto de la demostración, y que se
pueda probar el resultado sin ninguna de ellas.

\section{Perspectivas y trabajos
futuros}\label{perspectivas-y-trabajos-futuros}

Hemos visto que una condición sobre el crecimiento del tiempo de
equilibración de un sistema dinámico cuántico puede tener un gran
impacto en las propiedades que este presenta: sean estas dinámicas (como
la estabilidad) o estáticas (como la ley de área y la indistinguibilidad
de los puntos fijos).

Esto nos lleva a reconsiderar el tiempo de equilibración como una
propiedad fundamental y característica de estos modelos, de la misma
manera en la cual lo es el gap espectral para los sistemas
hamiltonianos. Esta idea ya había sido propuesta en
\cite{PhysRevB.90.045101}. En esta tesis hemos identificado una clase de
sistemas que, dadas las propiedades mencionadas arriba, claramente
tienen un papel especial en una posible clasificación de evoluciones
disipativas. Todavía no nos resulta claro si esta clase de modelos surge
de considerar el caso ``bueno'' o más bien el caso ``trivial'': será
importante ver qué propiedades se pueden recuperar y cuáles no al
sustituir la condición de equilibración rápida por algo menos
restrictivo, y cómo de amplia y variada es la clase de modelos que sí
satisfacen equilibración rápida. Es una línea de investigación
interesante, y las observaciones siguientes son los primeros pasos en su
desarrollo.

\subsection{Tiempo de equilibración
polinomial}\label{tiempo-de-equilibraciuxf3n-polinomial}

Si consideramos a los Lindbladianos como ``máquinas disipativas'' o
procedimientos de preparación de estados cuánticos con cierta utilidad,
entonces desde un punto de vista algorítmico un tiempo de equilibración
polinomial es perfectamente aceptable, y constituye una hipótesis más
natural que la de equilibración rápida. En el mismo espíritu, para
efectuar computaciones, a veces es útil considerar hamiltonianos cuyo
gap se cierra pero solamente de manera \emph{polinómica} en el tamaño
del sistema. En los dos casos, tenemos una situación que es claramente
una desventaja en el límite termodinámico, pero que para sistemas
finitos todavía puede ser manejada en un tiempo razonable y con una
cantidad razonable de recursos.

En esta situación, la conexión con la teoría de la materia condensada
(con sus resultados matemáticos pero al mismo tiempo con su
``filosofía'') se vuelve menos útil: sería parecido a estudiar la
eficiencia de un algoritmo de ordenamiento mirando cómo se comporta en
un conjunto infinito de elementos. Necesitamos herramientas distintas.

Describir cuál es el poder computacional y las propiedades de esta clase
de evoluciones es un problema muy interesante y sería una manera de
clarificar si las propuestas de preparación disipativa de estados puedan
ser implementadas en experimentos a grandes escalas. En este contexto la
estabilidad, no necesariamente del mismo tipo que hemos considerado en
este trabajo, sería la propiedad fundamental a considerar.

\subsection{Preparación de modelos
topológicos}\label{preparaciuxf3n-de-modelos-topoluxf3gicos}

El resultado sobre la ley de área puede también verse como una propiedad
negativa de los modelos con equilibración rápida: no pueden generar
estados que no tengan una ley de área. Por otra parte sabemos que
algunos estados topológicos no pueden ser obtenidos en tiempo sub-lineal
\cite{PhysRevB.90.045101}, al menos con hipótesis realistas sobre los
generadores, aunque sí cumplan con una ley de área. Dado que los estados
topológicos se ven como ingredientes cruciales para una memoria cuántica
robusta, sería interesante estudiar si:

\begin{enumerate}
\def\labelenumi{\roman{enumi}.}
\tightlist
\item
  pueden ser preparados con un proceso disipativo de manera que este sea
  estable;
\item
  existen ``buenos'' procesos disipativos que mantengan el estado, una
  vez que este haya sido preparado de otra forma.
\end{enumerate}

La diferencia entre los dos casos es que en el segundo no estamos
pidiendo que el proceso prepare el estado topológico de manera robusta y
rápida al partir de cualquier estado inicial, sino que solamente pedimos
que estas propiedades se cumplan cuando el proceso arranca justamente en
el estado que queremos preservar. Si el sistema es robusto, entonces
todo estado que esté lo suficientemente cerca del estado topológico
convergerá hacia él, y toda pequeña perturbación de los generadores
solamente cambiará poco este punto fijo. En principio, no haría falta
pedir otras propiedades fuera de estas: podría haber otros punto fijos,
estables o también inestables.

En esta línea de investigación, debemos mencionar los trabajos
siguientes: la propuesta de un proceso de ``codificación''
\cite{Dengis2014}, que no solamente prepara un estado fundamental del
Toric Code en 2D, sino que también permite codificar información lógica,
en el sentido de que un par específico de qubits del sistema será
copiado en los qubits virtuales del código. El proceso requiere tiempo
lineal, y no es invariante por traslaciones. Cuál es el efecto del ruido
en este modelo es todavía un problema abierto.

En \cite{arxiv1409.3435} se estudian dos familias de modelos
Lindbladianos derivados de hamiltonianos conmutativos, de lo cuales
preparan el estado de Gibbs. Se demuestra que para temperaturas
suficientemente altas, ambos procesos tienen un gap espectral y que por
lo tanto tienen un tiempo de equilibración polinomial. En el caso
clásico correspondiente es posible demostrar que además cumplen la
condición más fuerte de la desigualdad de log-Sobolev y equilibración
rápida, por lo tanto es posible que también el resultado cuántico se
pueda mejorar. Nótese que esto no contradiría el resultado sobre la
preparación de modelos topológicos mencionado antes: dado que estamos
considerando el estado de Gibbs a temperatura finita, y sabemos que para
el Toric Code en 2D no hay propiedades topológicas en este caso, no hay
nada en principio en contra de que puedan existir procesos disipativos
con equilibración rápida que preparen el estado de Gibbs del Toric Code
en 2D.

Sabemos que al mandar la temperatura a cero, el estado de Gibbs converge
al estado fundamental, y bajo alguna hipótesis sobre la densidad de
estados en los distintos niveles de energía, para tener una aproximación
buena es suficiente que la temperatura escale como el inverso del
logaritmo del número de partículas \cite{Hastings2007}. Por lo tanto
sería interesante estudiar el comportamiento de los modelos considerados
en \cite{arxiv1409.3435} cuando la temperatura tiende a cero, para el
caso específico del hamiltoniano del Toric Code. No está claro si la
aplicación resultante tendrá algún punto fijo distinto del subespacio
fundamental, ni si tendrá un gap espectral. En \cite{1602.01108v1} se
considera un problema similar de preparar un estado de Gibbs para un
hamiltoniano con temperatura crítica, y se demuestra que en el régimen
de coexistencia de fases no puede haber un único punto fijo. Esto no
resuelve el problema del Toric Code, dado que este último no tiene
temperatura crítica, pero es ciertamente una comparación interesante.

\subsection{Demostrar equilibración
rápida}\label{demostrar-equilibraciuxf3n-ruxe1pida}

En esta tesis hemos demostrado cómo la equilibración rápida implica
propiedades del punto fijo de la evolución, y hemos presentados la
desigualdad de log-Sobolev como una manera de probar esta condición en
el caso de Lindbladianos reversibles. Sería extremadamente útil tener
herramientas y condiciones que nos permitan probar la condición de
equilibración rápida, a través de la desigualdad de log-Sobolev o no. El
trabajo hecho en \cite{arxiv1409.3435} puede también verse como ir en
esta dirección: determinar condiciones sobre el punto fijo que implican
cotas en el tiempo de equilibración. La inspiración viene de los
resultados obtenidos para modelos clásicos (véase
\cite{martinelli1993finite,martinelli1999lectures} para una reseña),
para los cuales se ha probado que una condición sobre el decaimiento de
correlaciones del punto fijo de una dinámica de Glauber implica una
desigualdad de log-Sobolev y a su vez esta implica equilibración rápida.
Además, este tipo de decaimiento de correlaciones suele estar presente
en estados de Gibbs a temperatura alta.

En \cite{arxiv1409.3435} los autores elegieron un enfoque parecido, y
una definición especifica de decaimiento de correlaciones fue usada para
probar que la dinámica correspondiente tiene un gap espectral.
Igualmente demostraron que esta condición se cumple para temperaturas
altas, y además en el caso en 1D. Sería muy interesante ver si se puede
mejorar el resultado y probar una desigualdad de log-Sobolev, lo que
correspondería al resultado clásico.

Los resultados presentado en el \cref{hipercontractividad} muestran que
al generalizar la desigualdad de log-Sobolev del caso clásico al caso
cuántico, no obtenemos una única desigualdad, sino una familia (indexada
por \(p \in [1,\infty)\)) de desigualdades. Clásicamente son todas
equivalentes, pero en el caso cuántico no se sabe, y por lo tanto varios
autores han tenido que asumir la regularidad \(L_p\): la hipótesis que
la constante de \(L_2\) sea una cota inferior de todas las otras. Para
recuperar la conexión con la hipercontractividad se necesita
efectivamente la familia entera de desigualdades, pero como hemos
mostrado en el \cref{entropuxeda-y-desigualdad-de-log-sobolev} es sólo
el caso \(L_1\) el que interesa para demostrar equilibración rápida.

No sabemos cómo de genérica es la condición de regularidad \(L_p\): el
único resultado general que tenemos es que clases importantes de
Lindbladianos, como los generadores de Davies \cite{Quantum-Log-Sobolev}
y los procesos que preservan la unidad \cite{hypercontractivitylp} sí la
verifican. En \cite{Quantum-Log-Sobolev} se ha conjeturado que cada
Lindbladiano reversible sea \(L_p\) regular, y que cada Lindbladiano
primitivo verifique una versión débil de dicha condición. Si la
conjetura es falsa, entonces uno podría restringir su estudio a la
desigualdad de \(L_1\), dado que es la que nos provee la cota en el
tiempo de equilibración. Mostrar qué condiciones tienen que imponerse en
el punto fijo para probar esa desigualdad de log-Sobolev nos ayudará a
entender si la clase de sistemas con equilibración rápida es ``pequeña''
o ``grande''.

\subsection{Otros trabajos en distintas lineas de
investigación}\label{otros-trabajos-en-distintas-lineas-de-investigaciuxf3n}

Otra línea de investigación que ha sido desarrollada a lo largo del
Doctorado, aparte del estudio de las dinámicas disipativas abiertas, ha
sido el estudio de qué propiedades del límite termodinámico de una
sucesión de hamiltonianos se pueden inferir del estudio de algunos casos
finitos. En particular, nos hemos interesado en posibles problemas que
puedan surgir al estudiar, como es costumbre, el límite a través de una
secuencia de casos finitos, de manera numérica o experimental, para
luego extrapolar información sobre el límite. En un número de casos muy
relevantes esta manera de afrontar el problema ha tenido éxito
\cite{march1992electron, LandauLifshitzVolume5,domb1983phase,Tagliacozzo2008,Pirvu2012}
y ha proporcionado importante información sobre propiedades del modelo
físico en la escala de sistemas grandes. Por otra parte, se ha probado
un resultado general negativo: el problema de decidir si una sucesión de
hamiltonianos locales invariantes por traslaciones en 2D tiene un gap
espectral en el límite es un problema indecidible
\cite{Spectralgapundecidability}. Esto implica que estos modelos pueden
exhibir comportamientos impredecibles, y nos ha llevado a explorar las
posibilidades de construir ejemplos exóticos.

El resultado de esta investigación ha sido presentado en un artículo
(todavía no publicado) \cite{size-driven}, en el cual presentamos dos
familias de modelos que presentan una propiedad sorprendente: para toda
región finita más pequeña que un umbral dado, el estado fundamental y
los estados de bajas energías son estados clásicos (estados productos en
la base computacional); por encima del umbral estos muestran por el
contrario propiedades topológicas, que son características de ciertos
modelos cuánticos. Si consideramos espines con dimensión local más
grande, podemos hacer crecer el umbral de manera espectacular, y ya sólo
con considerar dimensión local \(d=10\) este se vuelve más grande que el
número estimado de partículas en el universo. Hemos llamado a este
fenómeno \textbf{transición de fase inducida por el tamaño}, dado que
puede considerarse como un cambio brusco entre un modelo clásico y uno
cuántico al incrementar el parámetro del tamaño del sistema.

Las dos construcciones están basadas en ideas distintas, y tienen
umbrales distintos. Ambas se basan en \textbf{problemas de teselación}:
una teselación es un recubrimiento de una región del plano con cuadrados
de lado uno y bordes coloreados, de manera que los colores de los bordes
de cuadrados adyacentes se correspondan. Se ha demostrado que el
problema de decidir si, dado un conjunto finito de teselas, es posible
teselar el plano entero es un problema indecidible
\cite{berger1966undecidability,robinson1971undecidability}. Este
resultado está en la base del resultado de indecidiblidad del gap de
\cite{Spectralgapundecidability}. Hemos modificado su construcción con
el uso de interacciones de plaquetas y estrellas (en vez de usar
simplemente plaquetas como en la construcción original), y hemos
construido dos familias de modelos. La primera se basa en la idea de
construir patrones periódicos de periodo muy grande (comparado con el
número de colores usados en la teselación), de manera que un patrón
específico sólo ocurra una vez cada periodo. Al penalizar tal patrón
podemos inducir una frustración de la energía para todo retículo más
grande que el periodo: esto nos permite implementar la transición entre
el modelo clásico y el cuántico.

La otra construcción se basa en la idea, ya presente en los resultados
previos de indecidibilidad, de codificar la historia de una máquina de
Turing en el estado fundamental del hamiltoniano. De esta manera podemos
dar una penalización energética si la máquina termina su computación, y
de la misma manera que en el caso anterior introducir una frustración de
la energía si el sistema es suficientemente grande para que la máquina
termine. Dado que el problema de determinar si (y cuándo) una máquina de
Turing termina es un problema indecidible, este ha sido el ingrediente
fundamental para probar la indecidibilidad del gap espectral. Hemos sido
capaces de optimizar mucho el coste de la codificación, en el sentido de
necesitar un espacio de Hilbert local de dimensión mucho más pequeña
para inscribir la historia de una máquina de Turing en el modelo de
espines. Con este código optimizado, hemos considerados las llamadas
máquinas \emph{Busy Beavers}: una máquina de dimensión muy pequeña, pero
cuyo tiempo de terminación es increíblemente largo - de hecho crece más
rápidamente que cualquier función computable. De esta manera hemos
obtenido modelos que tienen una dimensión local relativamente pequeña,
pero para los cuales hay una frustración en la energía sólo para
sistemas extremamente grandes. Nuevamente, esta frustración nos permite
implementar la transición de fase.

\selectlanguage{english}
\printbibliography[heading=bibintoc]

\end{document}